\documentclass{pasj00}
\draft

\begin{document}
\SetRunningHead{K. Fujisawa et al.}
{Methanol Masers with EAVN}
\Received{}
\Accepted{}

\title{Observations of 6.7~GHz Methanol Masers with EAVN I
: VLBI Images of the first Epoch of Observations}


%
\author{Kenta \textsc{Fujisawa},\altaffilmark{1,2}
Koichiro \textsc{Sugiyama},\altaffilmark{2}
Kazuhito \textsc{Motogi},\altaffilmark{2}
Kazuya \textsc{Hachisuka},\altaffilmark{3}
Yoshinori \textsc{Yonekura},\altaffilmark{4}
Satoko \textsc{Sawada-Satoh},\altaffilmark{5}
Naoko \textsc{Matsumoto},\altaffilmark{6}
Kazuo \textsc{Sorai},\altaffilmark{7}
Munetake \textsc{Momose},\altaffilmark{8}
Yu \textsc{Saito},\altaffilmark{8}
Hiroshi \textsc{Takaba},\altaffilmark{9}
Hideo \textsc{Ogawa},\altaffilmark{10}
Kimihiro \textsc{Kimura},\altaffilmark{10}
Kotaro \textsc{Niinuma},\altaffilmark{2}
Daiki \textsc{Hirano},\altaffilmark{2}
Toshihiro \textsc{Omodaka},\altaffilmark{11}
Hideyuki \textsc{Kobayashi},\altaffilmark{6}
Noriyuki \textsc{Kawaguchi},\altaffilmark{6}
Katsunori M. \textsc{Shibata},\altaffilmark{6}
Mareki \textsc{Honma},\altaffilmark{6}
Tomoya \textsc{Hirota},\altaffilmark{6}
Yasuhiro \textsc{Murata},\altaffilmark{12, 13}
Akihiro \textsc{Doi},\altaffilmark{12, 13}
Nanako \textsc{Mochizuki},\altaffilmark{12}
Zhiqiang \textsc{Shen},\altaffilmark{3,14}
Xi \textsc{Chen},\altaffilmark{3,14}
Bo \textsc{Xia},\altaffilmark{3,14}
Bin \textsc{Li},\altaffilmark{3,14} and
Kee-Tae \textsc{Kim}\altaffilmark{15}
}
  \altaffiltext{1}{The Research Institute for Time Studies, Yamaguchi University, 1677-1 Yoshida, Yamaguchi, Yamaguchi 753-8511, Japan}
  \altaffiltext{2}{Graduate school of Science and Engineering, Yamaguchi University, 1677-1 Yoshida, Yamaguchi, Yamaguchi 753-8512, Japan}
  \altaffiltext{3}{Shanghai Astronomical Observatory, Chinese Academy of Sciences, China}
  \altaffiltext{4}{Center for Astronomy, Ibaraki University, 2-1-1 Bunkyo, Mito, Ibaraki 310-8512, Japan}
  \altaffiltext{5}{Mizusawa VLBI Observatory, National Astronomical Observatory of Japan (NAOJ), 2-12 Hoshigaoka-cho, Mizusawa-ku, Oshu, Iwate 023-0861, Japan}
  \altaffiltext{6}{Mizusawa VLBI Observatory, NAOJ, 2-21-1 Osawa, Mitaka, Tokyo 181-8588, Japan}
  \altaffiltext{7}{Department of Physics / Department of Cosmosciences, Hokkaido University, Kita 10, Nishi 8, Kita-ku, Sapporo, Hokkaido 060-0810, Japan}
  \altaffiltext{8}{College of Science, Ibaraki University, 2-1-1 Bunkyo, Mito, Ibaraki 310-8512, Japan}
  \altaffiltext{9}{Faculty of Engineering, Gifu University, 1-1 Yanagido, Gifu, Gifu 501-1193, Japan}
  \altaffiltext{10}{Department of Physical Science, Osaka Prefecture University, 1-1 Gakuen-cho, Naka-ku, Sakai, Osaka 599-8531, Japan}
  \altaffiltext{11}{Department of Physics and Astronomy, Graduate School of Science and Engineering, Kagoshima University, 1-21-35 Korimoto, Kagoshima, Kagoshima 890-0065, Japan}
  \altaffiltext{12}{The Institute of Space and Astronautical Science, Japan Aerospace Exploration Agency, 3-1-1 Yoshinodai, Chuou-ku, Sagamihara, Kanagawa 229-8510, Japan}
  \altaffiltext{13}{Department of Space and Astronautical Science, The Graduate University for Advanced Studies, 3-1-1 Yoshinodai, Chuou-ku, Sagamihara, Kanagawa 229-8510, Japan}
  \altaffiltext{14}{Key Laboratory of Radio Astronomy, Chinese Academy of Sciences, China}
  \altaffiltext{15}{Korea Astronomy and Space Science Institute, 776 Daedeokdae-ro, Yuseong-gu, Daejeon 305-348, Republic of Korea}

\email{kenta@yamaguchi-u.ac.jp}


\maketitle

\begin{abstract}

Very long baseline interferometry (VLBI) monitoring of the 6.7~GHz methanol maser allows us
to measure the internal proper motions of the maser spots
and therefore study the gas motion around high-mass young stellar objects.
To this end,
we have begun monitoring observations with the East-Asian VLBI Network.
In this paper we present the results of the first epoch observation for 36 sources,
including 35 VLBI images of the methanol maser.
Since two independent sources were found in three images, respectively,
images of 38 sources were obtained.
In 34 sources, more than or equal to 10 spots were detected.
The observed spatial scale of the maser distribution was from 9 to 4900 astronomical units,
and the following morphological categories were observed:
elliptical, arched, linear, paired, and complex.
The position of the maser spot was determined to an accuracy of approximately 0.1 mas,
sufficiently high to measure the internal proper motion from two years of monitoring
observations.
The VLBI observation, however, detected only approximately 20{\%} of all maser emission,
suggesting that the remaining 80{\%} of the total flux was spread into an undetectable extended
distribution.
Therefore, in addition to high-resolution observations,
it is important to observe the whole structure of the maser emission including
extended low-brightness structures, to reveal the associated site of the maser
and gas motion.

\end{abstract}


\section{Introduction}\label{section:introduction}
Although high-mass star formation has been intensively studied,
it remains poorly understood because of the large distance and high obscuration
of the high-mass star-forming regions and short duration of critical evolutionary phases
(\cite{2007ARA&A..45..481Z}, and references therein).
Star forming regions are associated with maser emissions of
high brightness temperature, and high transparency in the radio band,
which are suitable for probing young stellar objects (YSOs).
Maser emissions are particularly useful for tracing circumstellar
gas motions close to the central star. The 6.7~GHz methanol maser transition,
which is the brightest among the methanol masers, is observed only
from high-mass star-forming regions (e.g., \cite{1991ApJ...380L..75M}; \cite{1995MNRAS.274.1126C}; \cite{2003A&A...403.1095M}; \cite{2008A&A...485..729X})
and considered one of the best tracers of gas dynamics around high-mass YSOs.

Some 6.7~GHz methanol masers show linearly elongated morphology
with a linear velocity gradient
(\cite{1993ApJ...412..222N}; \cite{1998MNRAS.300.1131P};
\cite{1998MNRAS.301..640W}; \cite{2000A&A...362.1093M}).
These masers can be interpreted as circumstellar disks viewed edge-on.
\citet{2009A&A...502..155B} analyzed samples surveyed by very-long-baseline interferometry (VLBI)
imaging using the European VLBI Network (EVN).
They found elliptical morphology in 30{\%} of the sampled methanol masers
and deduced that this morphology arises from inclined rotating disks
with expansion or infall motions.
In fact, rotational motions consistent with circumstellar disks have been
identified as internal proper motions in a few 6.7~GHz methanol maser sources
(G~16.59$-$0.05, G~23.01$-$0.41, and IRAS~20126$+$4104; \cite{2010A&A...517A..71S},
\yearcite{2010A&A...517A..78S}; \cite{2011A&A...526A..66M}).
In addition, the 6.7~GHz methanol maser source AFGL~5142 exhibits infall proper motion
\citep{2011A&A...535L...8G}, while a rotation and infall motion
is observed in Cepheus A (Sugiyama et al. 2013).

On the other hand, the associating sites of 6.7~GHz methanol masers remain obscure.
\citet{2003MNRAS.341..277D} and \citet{2009A&A...493..127D} reported that, in 60{\%} of their samples,
the methanol maser spots were distributed along the elongated direction of
the H$_2$~$v = 1-0$~S(1) at 2.12~$\mu$m and SiO thermal line emissions, which are shock diagnostic.
According to these authors, such a parallel distribution suggests that methanol masers are
directly associated with outflows. Supporting this inference, outward proper motions
have been found in a few 6.7~GHz methanol maser sources
(\cite{2010A&A...511A...2R}; \cite{2011PASJ...63...53S};
\cite{2011PASJ...63.1345M}; Sawada-Satoh et al. 2013).
\citet{2011ApJ...730...55P} observed linear/arched morphology in only
nine out of 50 sources with the Multi-Element Radio-Linked Interferometric Network (MERLIN)
and the Karl G. Jansky Very Large Array (JVLA), and they did not detect any source
with a clear elliptical morphology, in contrast to the results of \citet{2009A&A...502..155B}.

If masers are to be used for studying high-mass star formation,
the origin of the 6.7~GHz methanol maser must be elucidated.
This can be achieved with VLBI monitoring of numerous unbiased sources,
from which the spatial distributions and three-dimensional velocity field
(radial velocity and proper motions in RA and Dec) of the maser can be statistically investigated.

To date, VLBI images of 6.7~GHz methanol masers have been reported
for approximately 60 sources
(e.g., \cite{2000A&A...362.1093M};
\cite{2004MNRAS.351..779D}; \cite{2008PASJ...60...23S};
\cite{2009A&A...502..155B}),
while the masers have been detected in more than 900 high-mass star-forming regions
(\cite{2005A&A...432..737P}, and references therein;
\cite{2007MNRAS.377..571E}; \cite{2007ApJ...656..255P};
\cite{2009A&A...507.1117X};
\cite{2010MNRAS.404.1029C}, \yearcite{2011MNRAS.417.1964C};
\cite{2010MNRAS.409..913G}, \yearcite{2012MNRAS.420.3108G}).
As mentioned above, the internal proper motion of the 6.7~GHz methanol maser has been measured
in only a fraction of cases.
Therefore, we have started a VLBI monitoring project of the 6.7~GHz methanol maser sources
with the East-Asian VLBI Network (EAVN) to systematically investigate their internal proper motions.
This study presents the initial results of this project, namely,
the spatial distributions of the 6.7~GHz methanol maser spots.

Section~2 describes the criteria for target source selection and provides
details of the observations and data reduction methods.
Section~3 presents the EAVN images, while individual sources are discussed in Section~4.
Section~5 focuses on the spatial morphology and feasibility of measuring the internal proper motion.
Throughout this paper, sources are named by their Galactic coordinates,
expressed in the form \textit{xxx.xx$+$xx.xx}
following the IAU recommendation for nomenclature,
unless the source has been previously named (e.g., G~9.621$+$0.196).


\section{Observations and Data Reduction}\label{section:observation}

\subsection{Source Selection}\label{selection}
The target sources were selected from the methanol maser catalog of Pestalozzi et al. (2005)
and the Methanol Multibeam Survey catalog (Caswell et al.\ 2010; Green et al.\ 2010)
using the following criteria:
1) source declination $\delta >$~$- 40^{\circ}$,
2) catalogued peak flux density $F_{p} >$~65~Jy, and
3) no previous VLBI observation.
These criteria were satisfied by 34 sources. Two additional 6.7~GHz methanol
maser sources 031.28$+$00.06 and 049.49$-$00.38 were included in the target sources,
despite having been previously observed by EVN
(Minier et al. 2000; \cite{2005ASPC..340..342P}; \cite{2012A&A...541A..47S}).
These were used to compare the imaging capabilities of EAVN with those of EVN.
The selected 36 sources, together with their properties
(Galactic and IRAS names (if any), coordinates, peak velocity, peak flux density, distance, and the references)
are summarized in Table~\ref{table1}. Most of these sources
(34/36, 94{\%}) are located in the southern hemisphere ($\delta <$~0$^{\circ}$).

\subsection{EAVN Array}\label{eavnarray}
Observations were conducted with EAVN (\cite{shen2004}),
which consists of the following three VLBI networks:
the Japanese VLBI Network (JVN; Doi et al. 2006), Korean VLBI Network (KVN; Minh et al. 2003),
and Chinese VLBI Network (CVN; Ye et al. 1991).
Due to the location of the EAVN stations at latitudes below 40$^{\circ}$ N,
the facility is suitable to observing sources in the southern hemisphere.
There are three main frequency bands of EAVN observations: 6.7, 8 and 22~GHz.
The two Japanese telescopes Yamaguchi and Hitachi participating in this project are described in
Fujisawa et al. (2002) and Yonekura et al. (2013), respectively.

\subsection{Observations and Data Reduction}\label{reduction}
\begin{figure*}[htbp]
\begin{center}
\includegraphics[width=140mm,clip]{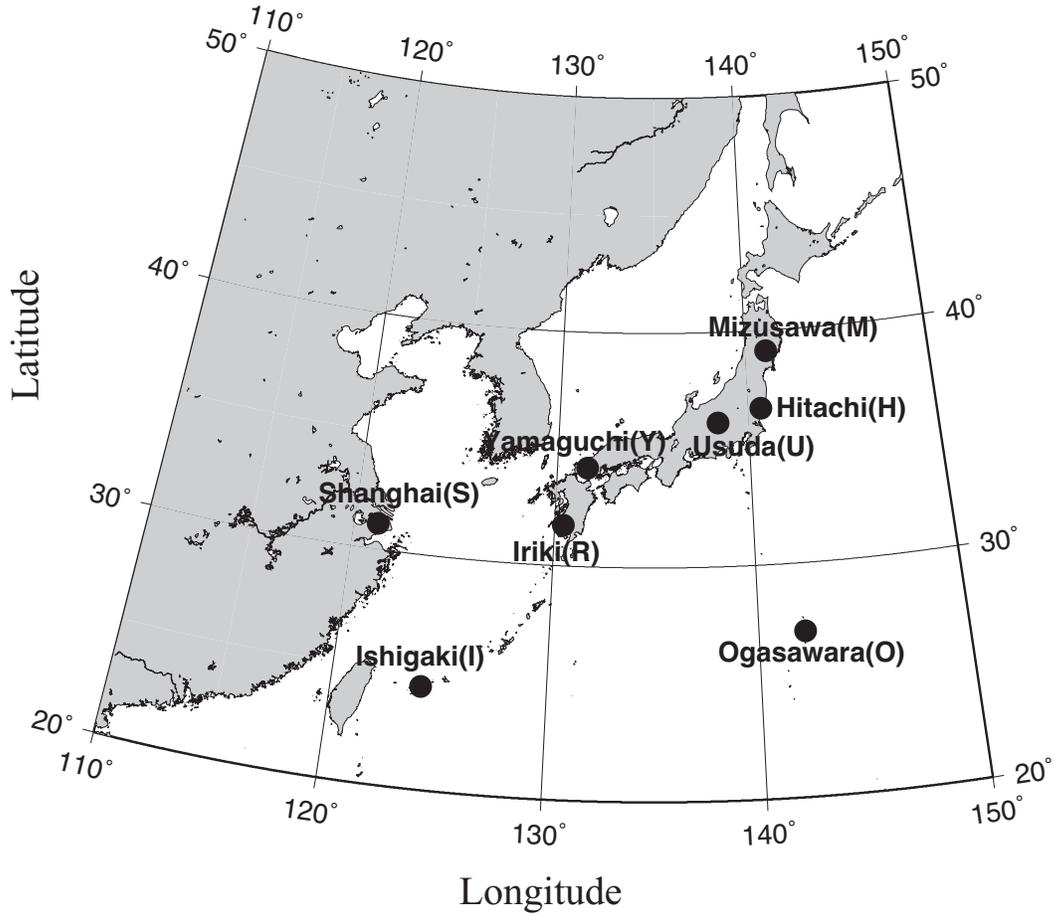}
\end{center}
\caption{{\footnotesize
The location of the participating telescopes.
}}	
\label{fig0}	
\end{figure*}

The first epoch observations of this monitoring project were conducted
during six sessions between 2010 and 2011. Table~\ref{table2} lists the observational
parameters of each session, including the date, time, and participating telescopes.
The location of the telescopes is shown in Figure~\ref{fig0}.
The projected baselines were from  6~M$\lambda$ (Yamaguchi--Iriki) to 50~M$\lambda$ (Mizusawa--Ishigaki)
corresponding to fringe spacings of 34.4~mas and 4.1~mas, respectively, at 6.7~GHz.
The typical size of the minor axis of the synthesized beam was 5~mas,
although it varied depending on the uv-coverage.

The continuum sources 3C454.3 and NRAO530 were used as the fringe finder
and bandpass calibrator, respectively. Continuum sources located adjacent
to target maser sources, J1700$-$2610, J1743$-$0350, J1845$-$2852, J1824$+$0119,
and J1930$+$1532, were used for delay calibration.
For one source, 232.62+00.99, located apart from the other sources,
J0607$-$0834, J0609$-$1542, and J0730$-$1141 were used for delay calibration.
Five to seven maser sources were observed in each session.
The integration time of a single scan of each maser source was 15 min,
and scanning was repeated
three or four times with an interval of 1 or 2 h,
yielding a total integration time of approximately 1 h for each source.
The data were recorded on magnetic tapes using the VSOP-terminal system
at a data rate of 128~Mbps with 2-bit sampling and correlated at
the Mitaka FX correlator (Shibata et al. 1998). We selected 4~MHz
including spectral lines from the recorded 32~MHz at the correlation.
The selected 4~MHz bandwidths were then divided into 1024 channels,
yielding a velocity coverage of 180~km s$^{-1}$ and channel spacing of 0.178~km s$^{-1}$.

The data were reduced using the Astronomical Image Processing System (AIPS; Greisen 2003).
Correlator digitization errors were corrected using the task ACCOR.
The clock and clock-rate offsets were corrected and bandpass was calibrated
using the strong continuum calibrators. Next, the delay was calibrated and
Doppler corrections were performed. Amplitude calibration parameters were
derived from the total-power spectra of maser lines using the template method
in the task ACFIT. Template spectra at or near the VLBI observation date were
obtained from single-dish observations of each target source with the Yamaguchi 32 m telescope.
Fringe fitting was performed using one spectral channel of the strongest maser feature,
followed by self-calibration. The fringe-fitted solutions were
poor for some sources at longer-baseline because the maser components of these sources,
including the strongest component, were heavily resolved out. At the amplitude calibration
stage using the ACFIT, we flagged antennas for which calibration failed.
Following calibration, uniformly weighted channel maps were made every 0.178 km~s$^{-1}$,
and maser components were searched for within the image cubes.
Maser components are considered real if detected with signal-to-noise ratio (SNR) $\geq 5$
at similar (within the beam FWHM) positions in two or more consecutive channels.
Once a maser component was found, it was fitted to an elliptical Gaussian using the task JMFIT.
The identified maser spots are shown as VLBI images in the following sections.

We also used the task FRMAP for fringe-rate mapping, from which we obtained
the absolute coordinates of each source with an accuracy of 200~mas.
The fringe-rate map was made for selected bright maser features
in each source, and the coordinates were estimated from the average of multiple solutions.
The absolute coordinates estimated by fringe-rate mapping are shown in Table~\ref{table3}.
Due to the extended north-south size of the synthesized beam for sources near the equator, 
nine sources suffered positional uncertainties of $\ge 1$~arcsec, and so were excluded.
The positions of the other four sources were not determined mainly due to the weak flux,
and the sources are also excluded from Table~\ref{table3}.

\begin{table*}
\caption{Summary of the 6.7~GHz methanol maser sources observed by VLBI}
\label{table1}
\begin{tabular}{clcllrrcc}
\hline
No. & G-Name          &  IRAS      &  \multicolumn{2}{c}{Coordinates (J2000)} & \multicolumn{1}{c}{$V_{\mathrm{lsr}}$} &  \multicolumn{1}{c}{$F_{\mathrm{p}}$}  & \multicolumn{1}{c}{$D$} & Ref. \\ \cline{4-5}
    &                 &            &  \multicolumn{1}{c}{R.A.}  & \multicolumn{1}{c}{Dec.}  &        &        &       &       \\
    &                 &            &  \multicolumn{1}{c}{($^\mathrm{h}~^\mathrm{m}~^\mathrm{s}$)} & \multicolumn{1}{c}{($^{\circ}~^{\prime}~^{\prime \prime}$)}
    & \multicolumn{1}{c}{(km~s$^{-1}$)} & \multicolumn{1}{c}{(Jy)}  & \multicolumn{1}{c}{(kpc)} & \\
\hline
1  & 000.54$-$00.85   &17470$-$2853& 17 50 14.35 & $-$28 54 31.1 &  11.8   &  68   & \hspace{2mm}7.2                & cas10  \\ 
2  & 000.64$-$00.04   &17441$-$2822& 17 47 18.65 & $-$28 24 25.0 &  49.1   &  69   & \hspace{2mm}7.9$^{\dagger}$    & cas10  \\ 
3  & 002.53$+$00.19   &17476$-$2638& 17 50 46.47 & $-$26 39 45.3 &   3.1   &  88   & \hspace{2mm}4.2                & cas10  \\ 
4  & 006.18$-$00.35   &            & 18 01 02.16 & $-$23 47 10.8 & $-$30.2 & 228.57& \hspace{2mm}5.1                & gre10  \\ 
5  & 006.79$-$00.25   &17589$-$2312& 18 01 57.75 & $-$23 12 34.9 &  16.3   &  91.07& \hspace{2mm}3.8                & gre10  \\ 
6  & 008.68$-$00.36   &18032$-$2137& 18 06 23.49 & $-$21 37 10.2 &  43.2   & 102.0 & \hspace{2mm}4.5                & gre10  \\ 
7  & 008.83$-$00.02   &18024$-$2119& 18 05 25.67 & $-$21 19 25.1 &  $-$3.8 & 159.08& \hspace{2mm}5.2                & gre10  \\ 
8  & 009.61$+$00.19   &18032$-$2032& 18 06 14.92 & $-$20 31 44.3 &   5.5   &  70.00& \hspace{2mm}5.2$^{\dagger}$    & gre10  \\ 
9  & 009.98$-$00.02   &18048$-$2019& 18 07 50.12 & $-$20 18 56.5 &  42.2   &  67.58&  12.0                          & gre10  \\ 
10 & 010.32$-$00.16   &18060$-$2005& 18 09 01.46 & $-$20 05 07.8 &  11.5   &  90.05& \hspace{3.5mm}2.39             & gre10  \\ 
11 & 011.49$-$01.48   &18134$-$1942& 18 16 22.13 & $-$19 41 27.1 &   6.6   &  68.40& \hspace{2mm}1.6                & gre10  \\ 
12 & 011.90$-$00.14   &18092$-$1842& 18 12 11.44 & $-$18 41 28.6 &  42.9   &  64.89& \hspace{2mm}4.0                & gre10  \\ 
13 & 012.02$-$00.03   &18090$-$1832& 18 12 01.86 & $-$18 31 55.7 & 108.3   &  96.26&    11.1                        & gre10  \\ 
14 & 012.68$-$00.18   &            & 18 13 54.75 & $-$18 01 46.6 &  57.5   & 544.0 & \hspace{3.5mm}2.40$^{\dagger}$ & imm13  \\ 
15 & 012.88$+$00.48   &18089$-$1732& 18 11 51.40 & $-$17 31 29.6 &  39.3   &  68.88& \hspace{3.5mm}2.34$^{\dagger}$ & gre10  \\ 
16 & 014.10$+$00.08   &18128$-$1640& 18 15 45.81 & $-$16 39 09.4 &  15.4   &  87.26& \hspace{2mm}5.4                & gre10  \\ 
17 & 020.23$+$00.06   &18249$-$1116& 18 27 44.56 & $-$11 14 54.2 &  71.8   &  77   & \hspace{2mm}4.4                & cas09  \\ 
18 & 023.43$-$00.18   &18319$-$0834& 18 34 39.25 & $-$08 31 38.5 & 103     &  45   & \hspace{2mm}5.9$^{\dagger}$    & cas09  \\ 
19 & 025.65$+$01.05   &18316$-$0602& 18 34 20.91 & $-$05 59 40.5 &  41.9   & 178   &   12.5                         & xu09   \\ 
20 & 025.71$+$00.04   &18353$-$0628& 18 38 03.15 & $-$06 24 15.0 &  92.8   & 364   &    11.8                        & xu09   \\ 
21 & 025.82$-$00.17   &18361$-$0627& 18 39 03.63 & $-$06 24 09.5 &  91.2   &  70   & \hspace{2mm}5.0                & xu09   \\ 
22 & 028.83$-$00.25   &18421$-$0348& 18 44 51.08 & $-$03 45 48.5 &  83.5   &  73   & \hspace{2mm}4.6                & cyg09  \\ 
23 & 029.86$-$00.04   &            & 18 45 59.57 & $-$02 45 04.4 & 101.4   &  67   & \hspace{2mm}9.3                & xu09   \\ 
24 & 030.70$-$00.06   &18450$-$0205& 18 47 36.9  & $-$02 01 05   &  88     &  87   & \hspace{2mm}5.9                & xu09   \\ 
25 & 030.76$-$00.05   &18450$-$0200& 18 47 39.73 & $-$01 57 22.0 &  92     &  68   & \hspace{2mm}4.8                & xu09   \\ 
26 & 030.91$+$00.14   &18448$-$0146& 18 47 15.0  & $-$01 44 07   & 104     &  95.2 & \hspace{2mm}5.6                & xu09   \\ 
27 & 031.28$+$00.06   &18456$-$0129& 18 48 12.39 & $-$01 26 22.6 & 110     &  71   & \hspace{2mm}5.8                & xu09   \\ 
28 & 032.03$+$00.06   &18470$-$0049& 18 49 37.3  & $-$00 45 47   &  92.8   &  93   & \hspace{2mm}7.2                & xu09   \\ 
29 & 037.40$+$01.52   &18517$+$0437& 18 54 10.5  & $+$04 40 49   &  41.1   & 279   & \hspace{2mm}2.1                & xu09   \\ 
30 & 049.49$-$00.38   &19213$+$1424& 19 23 43.949& $+$14 30 34.44&  59.2   & 850   & \hspace{3.5mm}5.41$^{\dagger}$ & xu09   \\ 
31 & 232.62$+$00.99   &07299$-$1651& 07 32 09.79 & $-$16 58 12.4 &  23     & 162   & \hspace{3.5mm}1.68$^{\dagger}$ & cas09  \\ 
32 & 351.77$-$00.53   &17233$-$3606& 17 26 42.57 & $-$36 09 17.6 &   1.3   & 231   & \hspace{2mm}0.4                & cas10  \\ 
33 & 352.63$-$01.06   &17278$-$3541& 17 31 13.91 & $-$35 44 08.7 &  $-$2.9 & 183   & \hspace{2mm}0.9                & cas10  \\ 
34 & 353.41$-$00.36   &17271$-$3439& 17 30 26.18 & $-$34 41 45.6 & $-$20.3 & 116   & \hspace{2mm}3.8                & cas10  \\ 
35 & 354.61$+$00.47   &17269$-$3312& 17 30 17.13 & $-$33 13 55.1 & $-$24.4 & 166   & \hspace{2mm}3.8                & cas10  \\ 
36 & 359.43$-$00.10   &            & 17 44 40.60 & $-$29 28 16.0 & $-$47.8 &  73.50& \hspace{2mm}8.2                & cas10  \\ 
\hline 
\multicolumn{8}{l}{\hbox to 0pt{\parbox{160mm}{\footnotesize
Column~1: ID number;
Columns~2, 3: Galactic and IRAS names (if any), respectively;
Columns~4, 5: Absolute coordinates (referenced in Column 9);
Columns~6, and 7: Radial velocity and peak flux density of the brightest maser feature, respectively;
Column~8: Source distance (referenced in section~\ref{comments});
Column~9: Absolute coordinate reference.\\
Reference -- cas09: \citet{2009PASA...26..454C}; cyg09: \citet{2009ApJ...702.1615C};
xu09: \citet{2009A&A...507.1117X}; cas10: \citet{2010MNRAS.404.1029C};
gre10: \citet{2010MNRAS.409..913G}; imm13: \citet{2013arXiv1304.2041I}.\\
$\dagger$: distances determined by trigonometric parallax.
}\hss}}
\end{tabular}
\end{table*}

\begin{table*}[htbp]
\begin{center}
\caption{Summary of the first epoch of EAVN observations}
\label{table2}
\begin{tabular}{cccl}
\hline
Session & Date       & Time        & Telescopes$^{\ast}$                    \\
        & (y/m/d)    & (UT)        &                                               \\
\hline 
1       & 2010/08/28 & 07:00-16:00 & M, R, O, I, H, S                              \\
2       & 2010/08/29 & 07:00-28:00 & M, R, O, I, H, S                              \\
3       & 2010/08/30 & 08:00-17:00 & M, R, O, I, H                                 \\
4       & 2011/10/27 & 03:00-10:00 & M, R, O, I, Y, H                              \\
5       & 2011/10/28 & 03:00-10:00 & M, R, O, I, Y, H, S                           \\
6       & 2011/11/26 & 01:30-09:00 & M, R, O, I, Y, U, H, S                        \\
\hline 
\multicolumn{4}{l}{\hbox to 0pt{\parbox{90mm}{\footnotesize
$\ast$: M: Mizusawa, R: Iriki, O: Ogasawara, I: Ishigaki, 
Y: Yamaguchi, U: Usuda, H: Hitachi, S: Shanghai25.
}\hss}}
\end{tabular}
\end{center}
\end{table*} 

\section{Results}\label{section:result}
The spatial distributions of the 6.7~GHz methanol maser spots were successfully
obtained for all sources except 014.10$+$00.08, whose fringe was detected only
in the Mizusawa$-$Hitachi baseline. Among the 35 VLBI images, 33 were obtained for the first time.
This study has increased the number of reported VLBI images of 6.7~GHz methanol masers by a factor of 1.5.

\begin{figure*}[htbp]
\begin{center}
\includegraphics[width=140mm,clip]{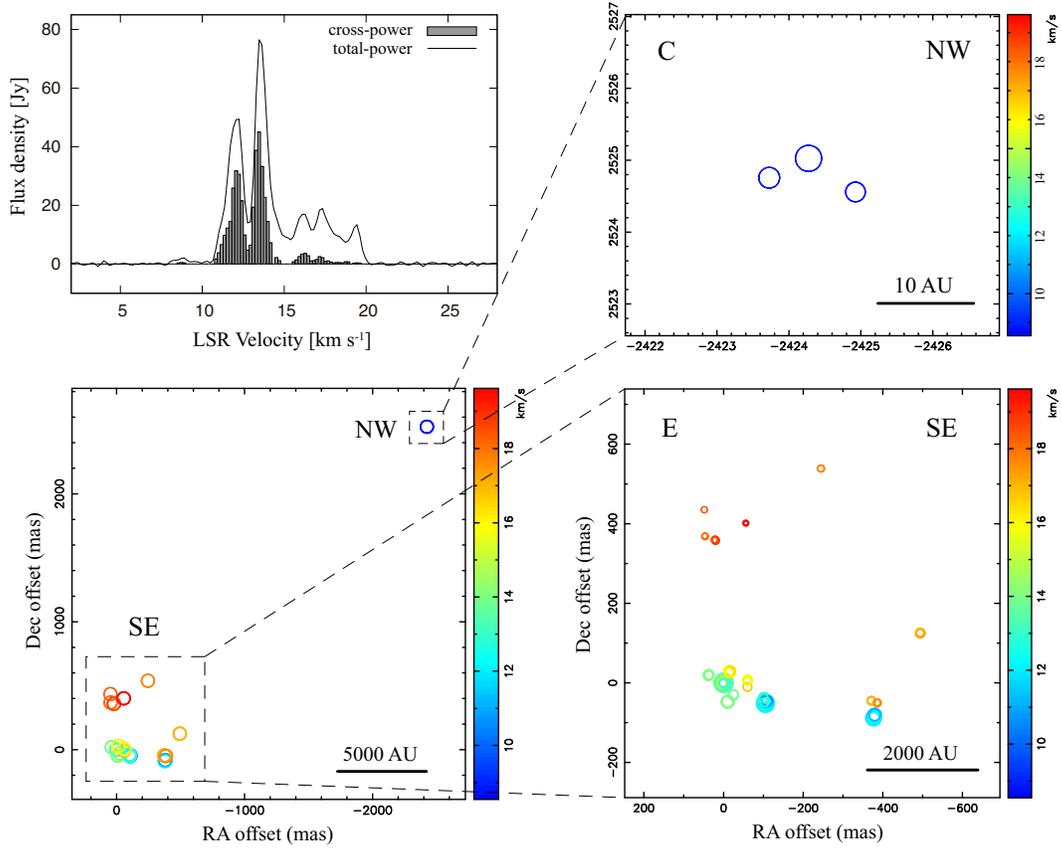}
\end{center}
\caption{{\footnotesize
The 6.7~GHz methanol maser emissions of source 000.54$-$00.85.
\textit{Upper-left-panel}: Total-power (solid line) and cross-power (hatched box) spectra.
\textit{Lower- and upper-right-panels}: Spatial distributions of the methanol maser spots obtained
from the EAVN observations. Detail are provided in Section~4.
}}	
\label{fig1}		
\end{figure*}

\begin{figure*}[htbp]
\begin{center}
\includegraphics[width=140mm,clip]{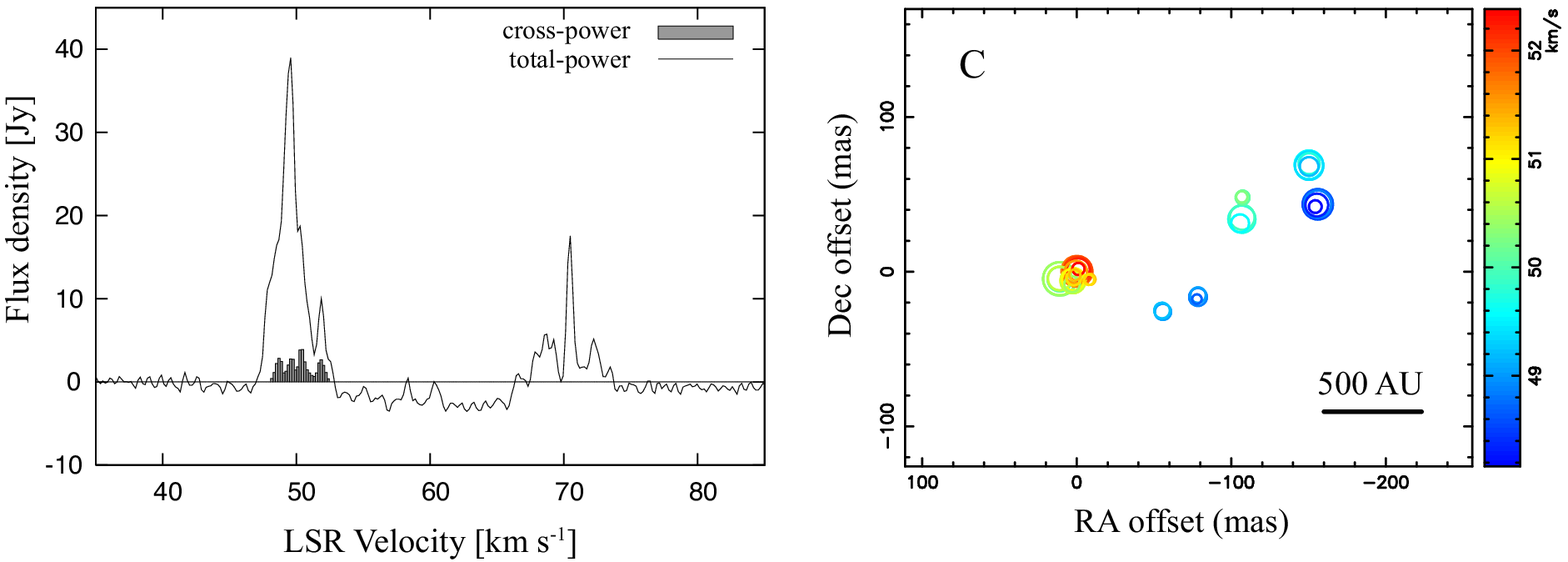}
\end{center}
\caption{{\footnotesize
Data for source 000.64$-$00.04, plotted as for Figure~\ref{fig1}.
}}	
\label{fig2}		
\end{figure*}

\begin{figure*}[htbp]
\begin{center}
\includegraphics[width=140mm,clip]{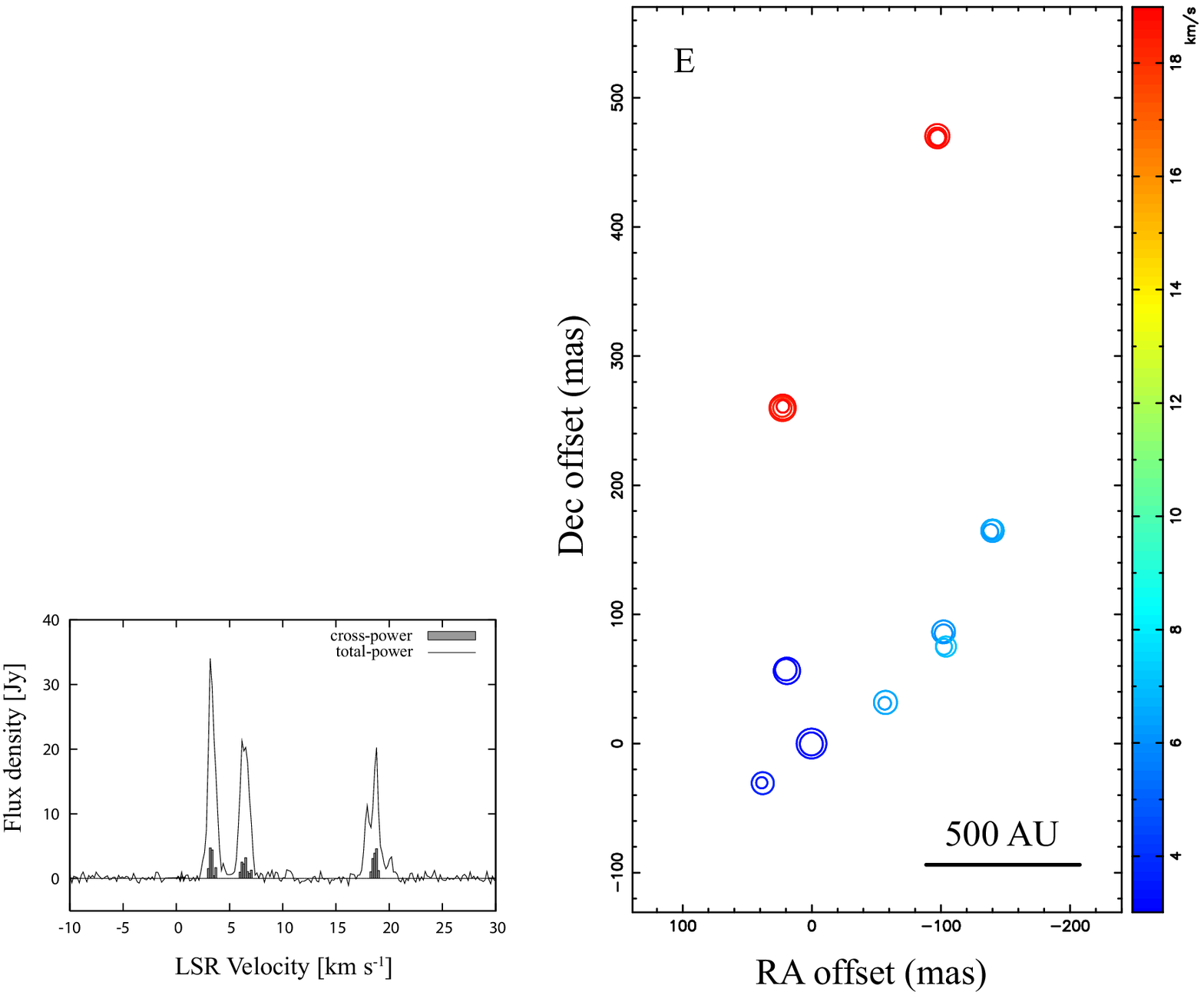}
\end{center}
\caption{{\footnotesize
Data for source 002.53$+$00.19, plotted as for Figure~\ref{fig1}.
}}	
\label{fig3}		
\end{figure*}

\begin{figure*}[htbp]
\begin{center}
\includegraphics[width=140mm,clip]{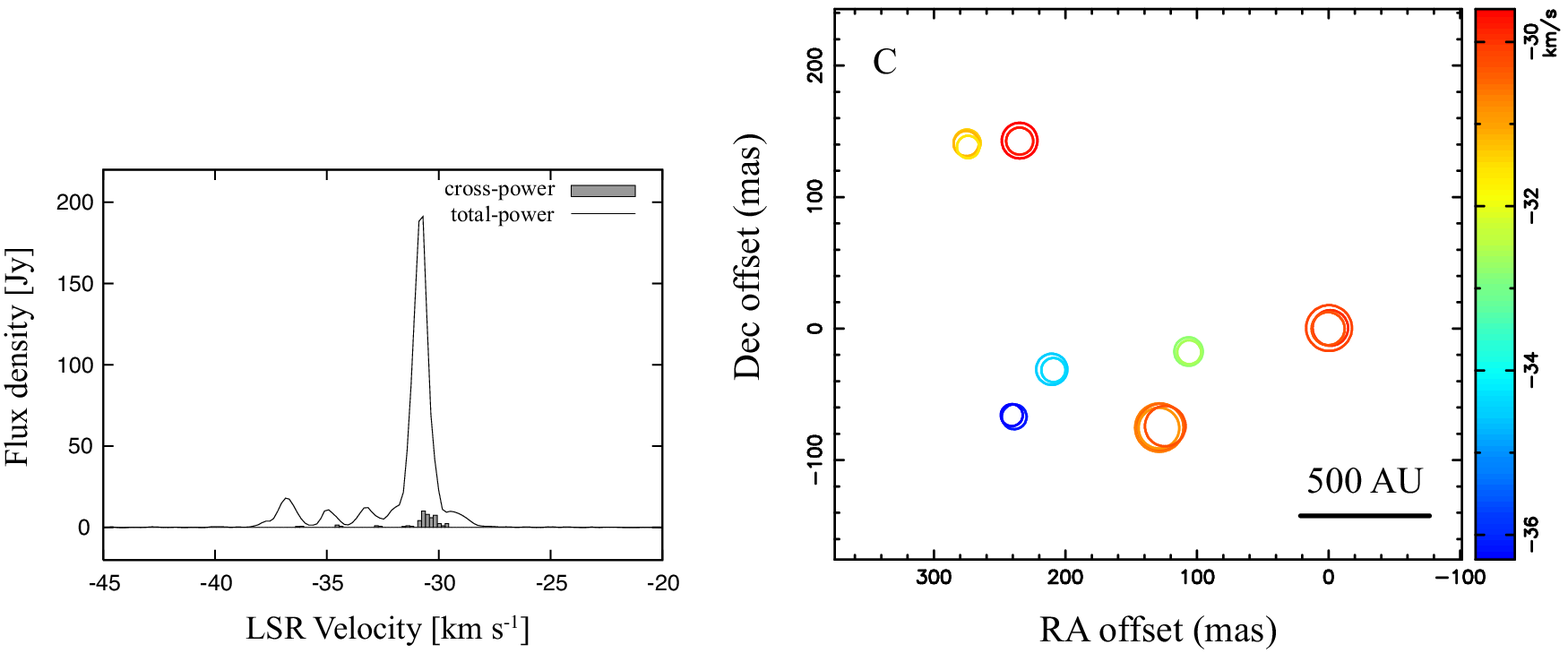}
\end{center}
\caption{{\footnotesize
Data for source 006.18$-$00.35, plotted as for Figure~\ref{fig1}.
}}	
\label{fig4}		
\end{figure*}

\begin{figure*}[htbp]
\begin{center}
\includegraphics[width=140mm,clip]{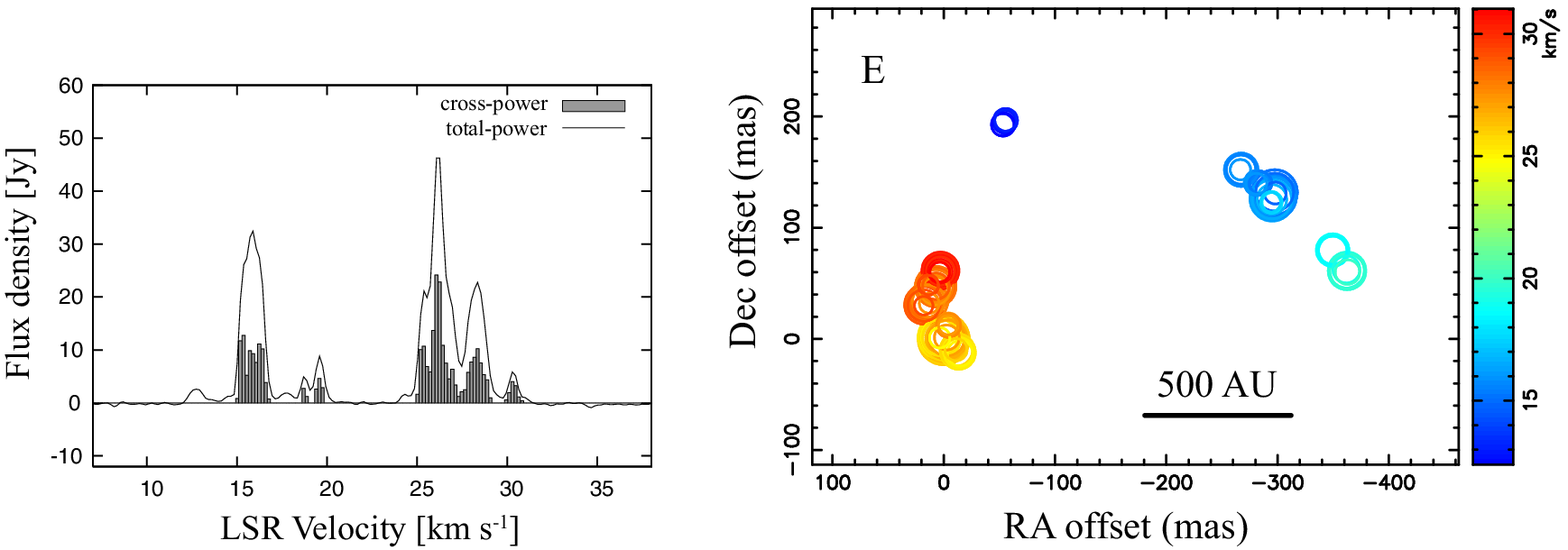}
\end{center}
\caption{{\footnotesize
Data for source 006.79$-$00.25, plotted as for Figure~\ref{fig1}.
}}	
\label{fig5}		
\end{figure*}

\begin{figure*}[htbp]
\begin{center}
\includegraphics[width=140mm,clip]{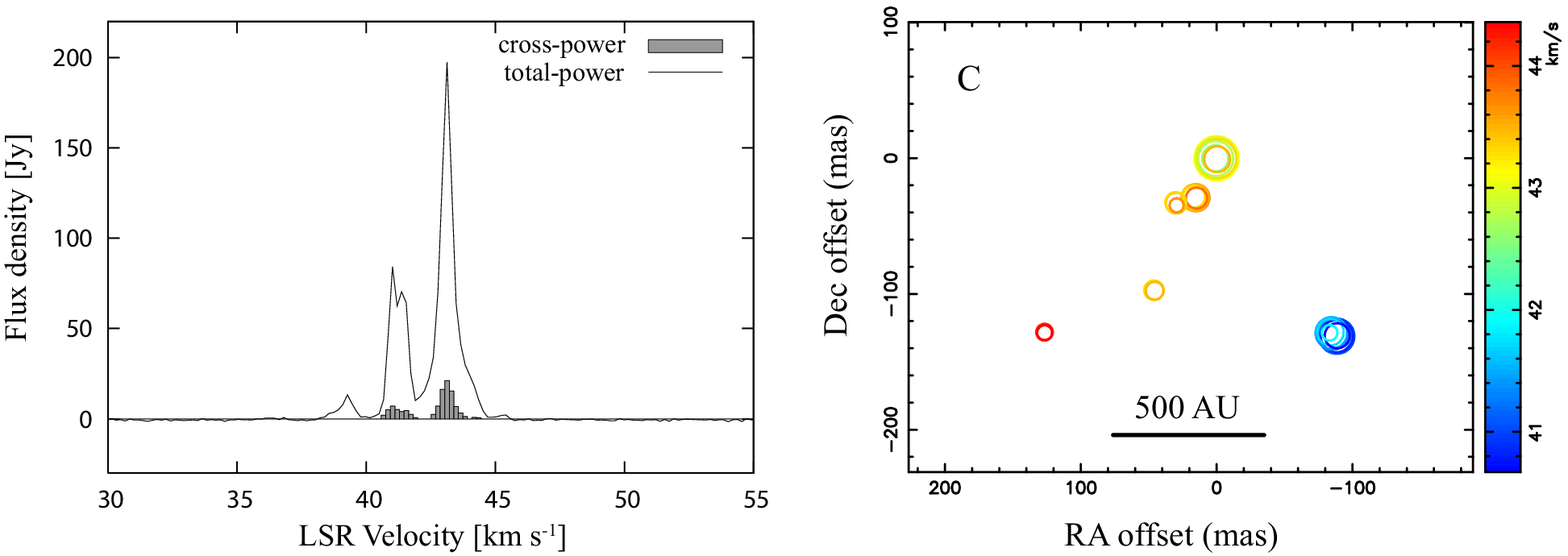}
\end{center}
\caption{{\footnotesize
Data for source 008.68$-$00.36, plotted as for Figure~\ref{fig1}.
}}	
\label{fig6}		
\end{figure*}

\begin{figure*}[htbp]
\begin{center}
\includegraphics[width=140mm,clip]{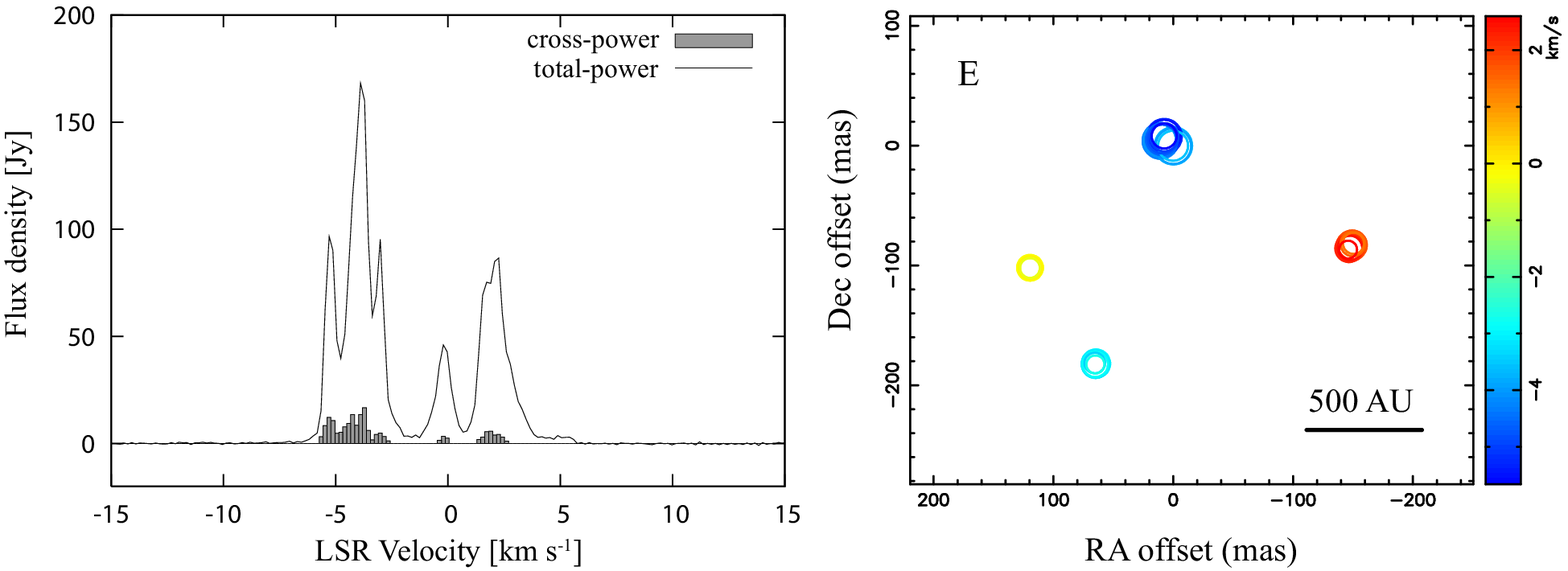}
\end{center}
\caption{{\footnotesize
Data for source 008.83$-$00.02, plotted as for Figure~\ref{fig1}.
}}	
\label{fig7}
\end{figure*}

\begin{figure*}[htbp]
\begin{center}
\includegraphics[width=140mm,clip]{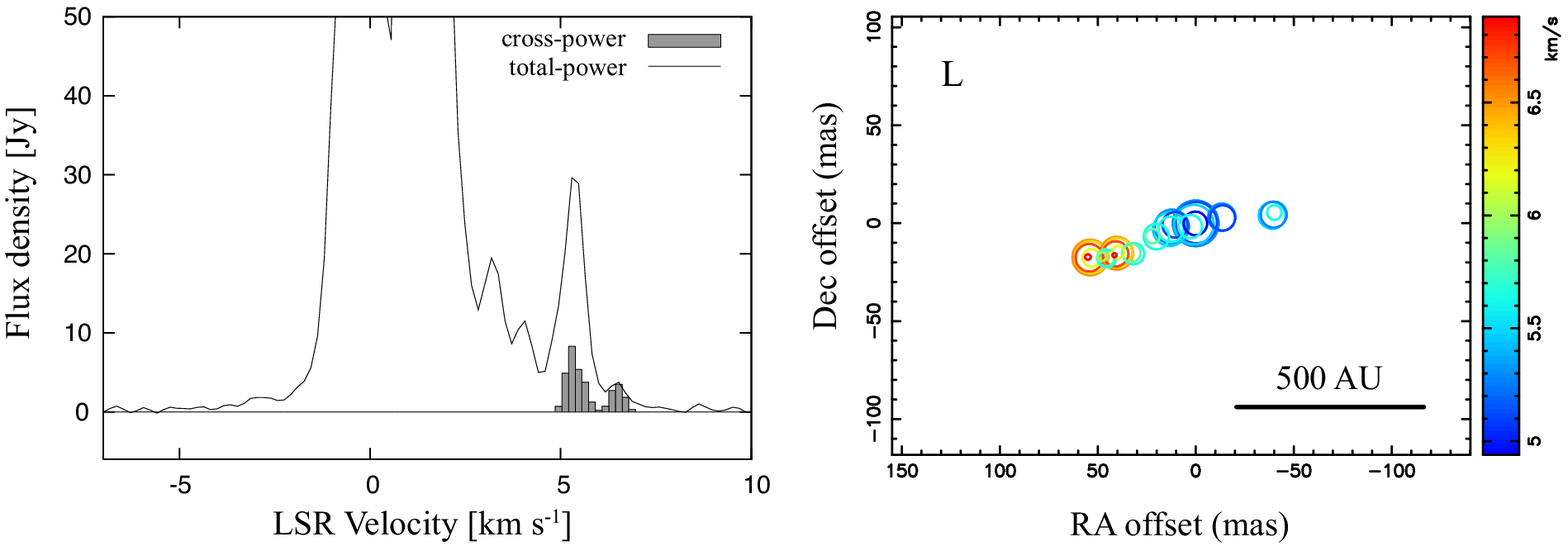}
\end{center}
\caption{{\footnotesize
Data for source 009.61$+$00.19, plotted as for Figure~\ref{fig1}.
}}	
\label{fig8}		
\end{figure*}

\begin{figure*}[htbp]
\begin{center}
\includegraphics[width=140mm,clip]{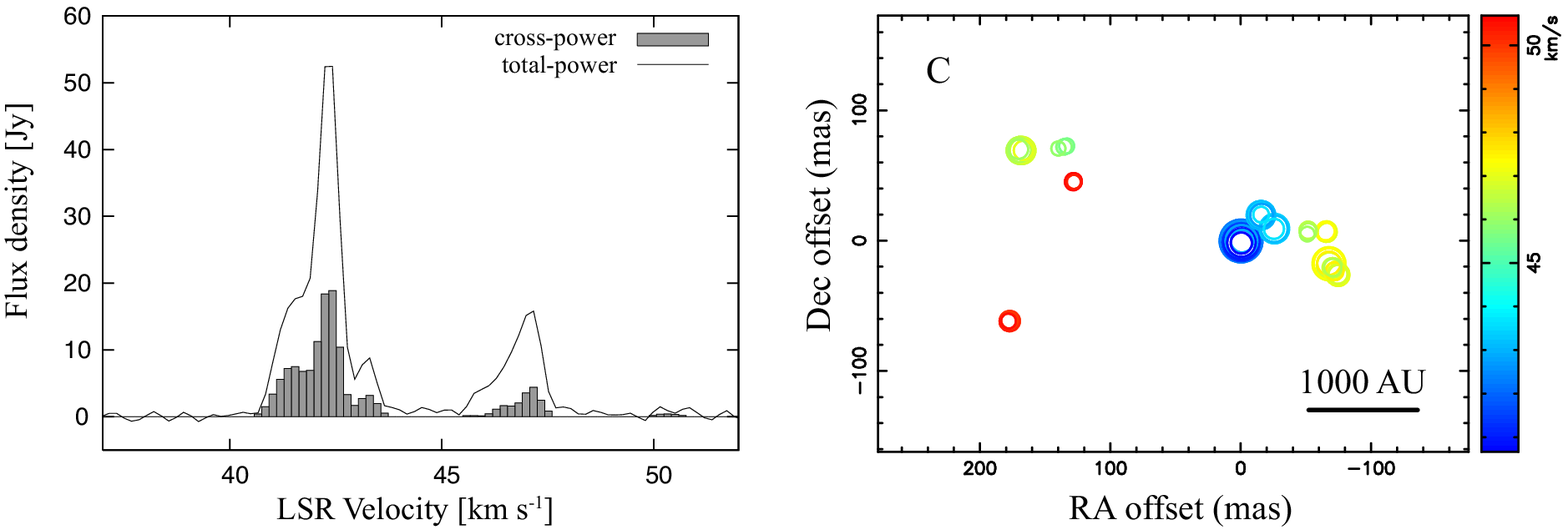}
\end{center}
\caption{{\footnotesize
Data for source 009.98$-$00.02, plotted as for Figure~\ref{fig1}.
}}	
\label{fig9}		
\end{figure*}

\begin{figure*}[htbp]
\begin{center}
\includegraphics[width=140mm,clip]{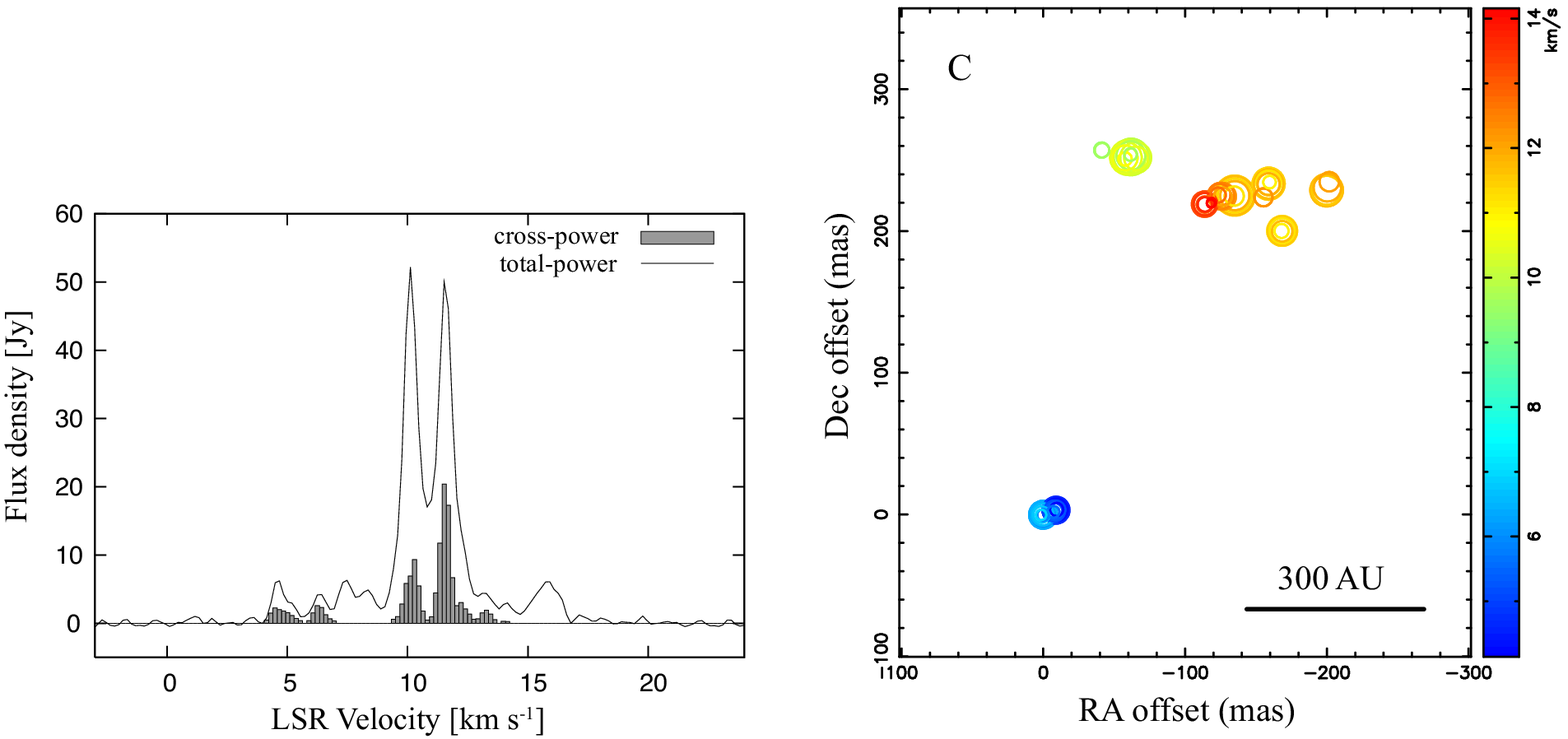}
\end{center}
\caption{{\footnotesize
Data for source 010.32$-$00.16, plotted as for Figure~\ref{fig1}.
}}	
\label{fig10}		
\end{figure*}

\begin{figure*}[htbp]
\begin{center}
\includegraphics[width=140mm,clip]{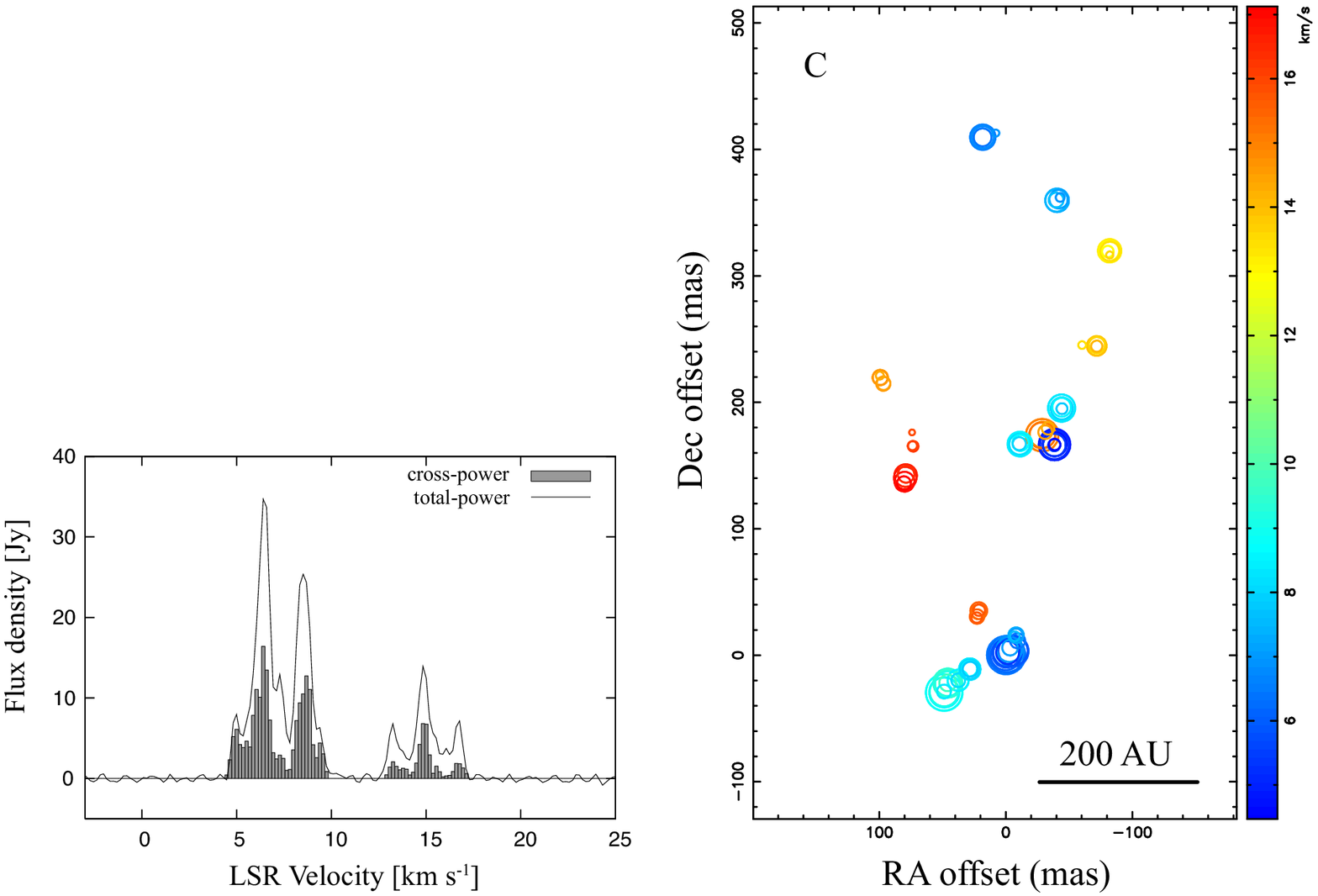}
\end{center}
\caption{{\footnotesize
Data for source 011.49$-$01.48, plotted as for Figure~\ref{fig1}.
}}	
\label{fig11}		
\end{figure*}

\begin{figure*}[htbp]
\begin{center}
\includegraphics[width=140mm,clip]{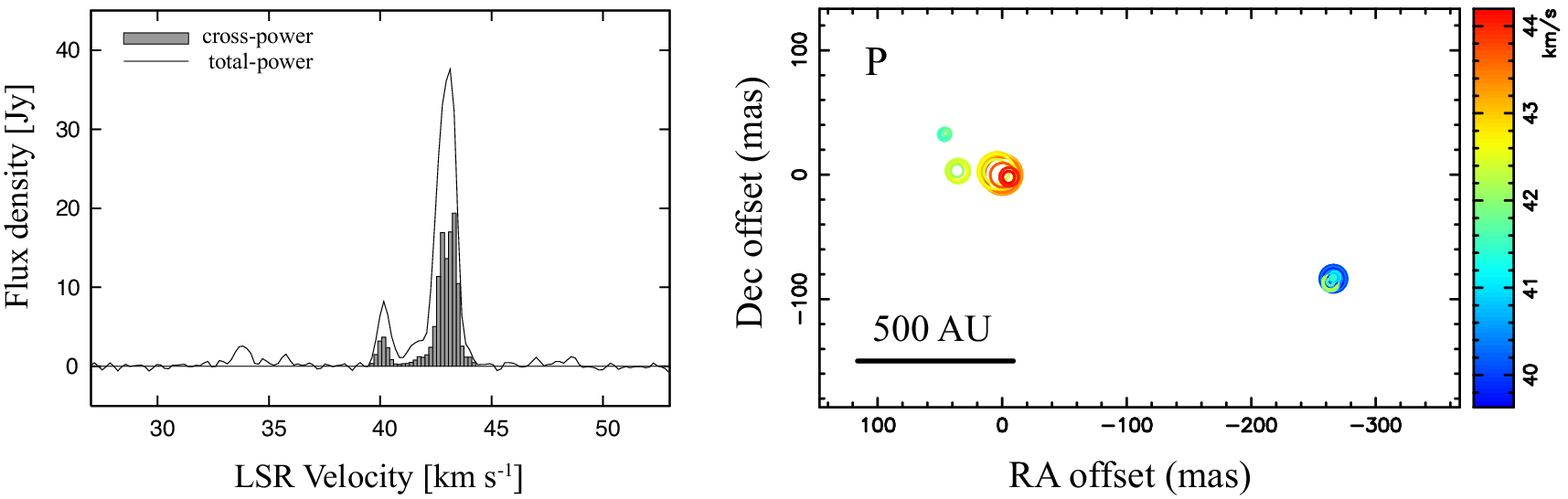}
\end{center}
\caption{{\footnotesize
Data for source 011.90$-$00.14, plotted as for Figure~\ref{fig1}.
}}	
\label{fig12}		
\end{figure*}

\begin{figure*}[htbp]
\begin{center}
\includegraphics[width=140mm,clip]{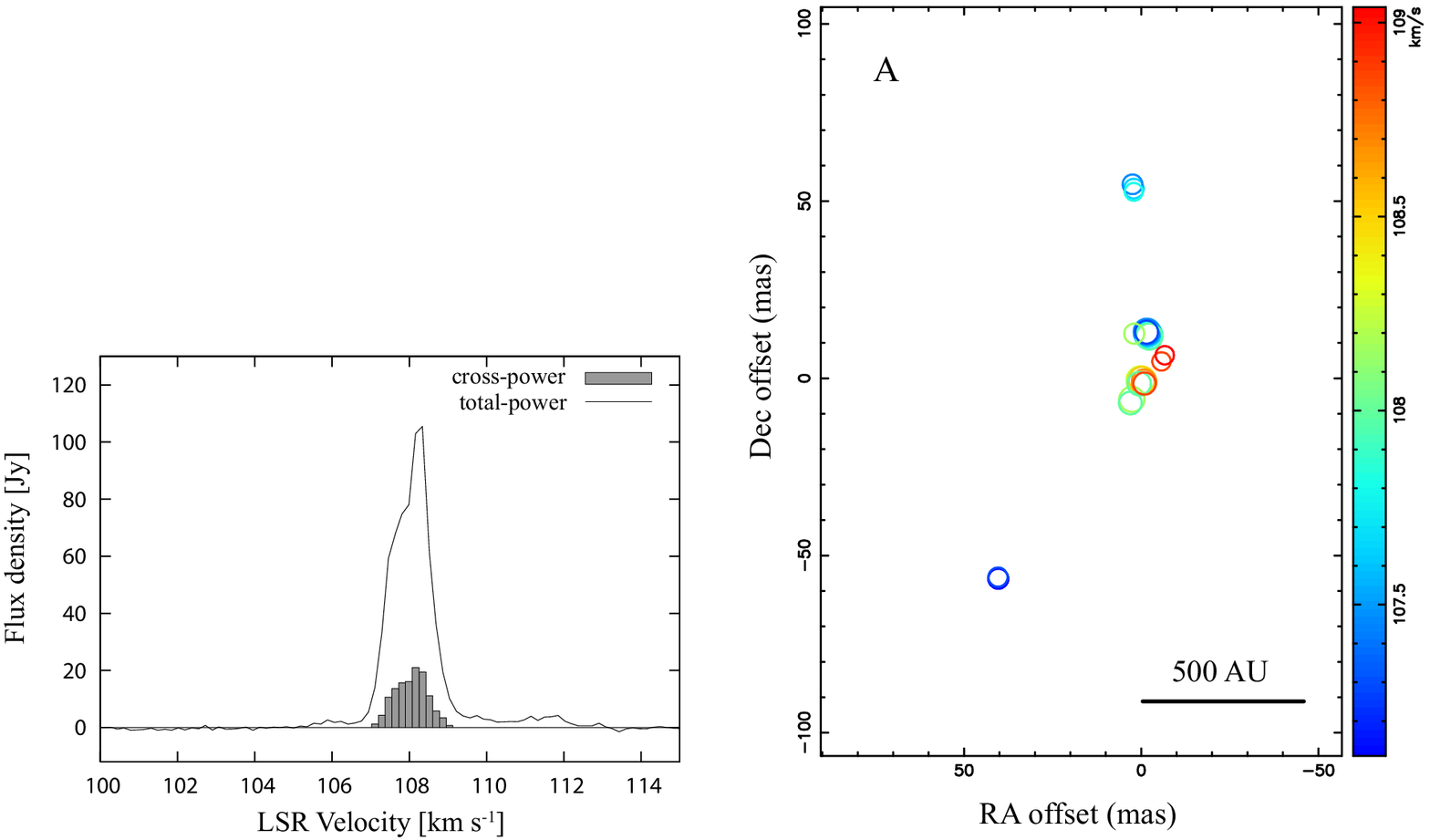}
\end{center}
\caption{{\footnotesize
Data for source 012.02$-$00.03, plotted as for Figure~\ref{fig1}.
}}	
\label{fig13}		
\end{figure*}

\begin{figure*}[htbp]
\begin{center}
\includegraphics[width=140mm,clip]{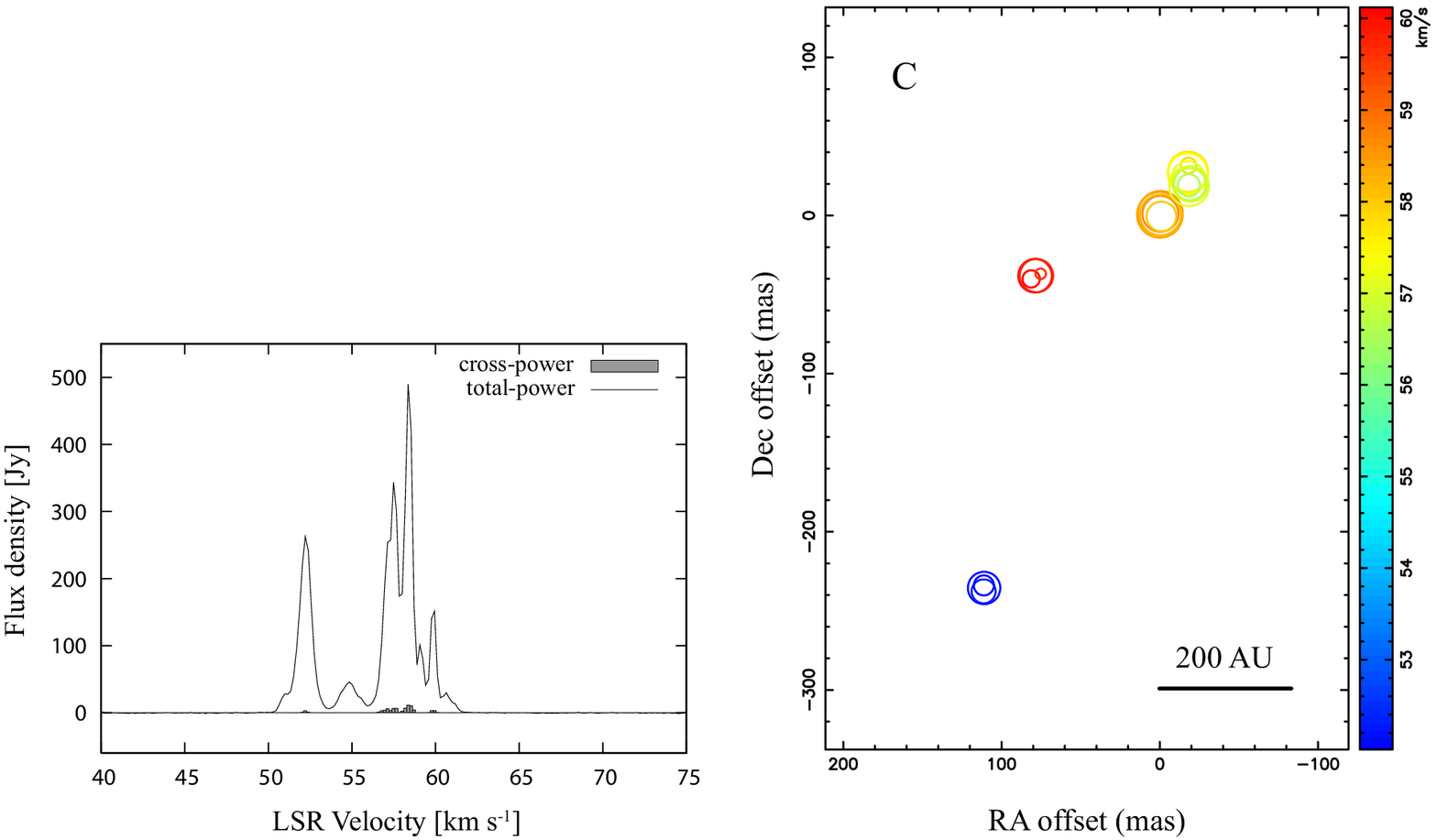}
\end{center}
\caption{{\footnotesize
Data for source 012.68$-$00.18, plotted as for Figure~\ref{fig1}.
}}	
\label{fig14}		
\end{figure*}

\begin{figure*}[htbp]
\begin{center}
\includegraphics[width=140mm,clip]{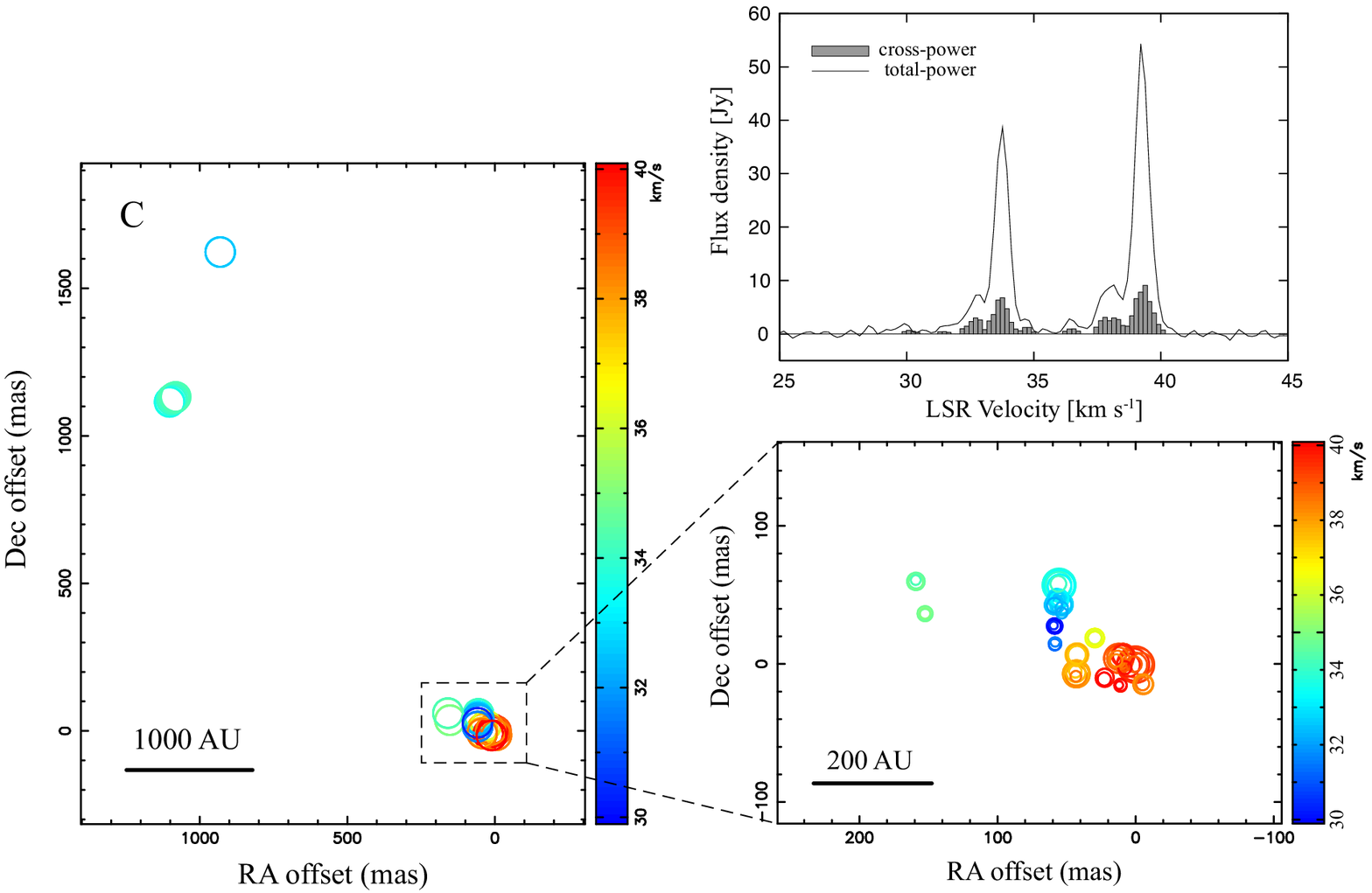}
\end{center}
\caption{{\footnotesize
Data for source 012.88$+$00.48, plotted as for Figure~\ref{fig1}.
}}	
\label{fig15}		
\end{figure*}

\begin{figure*}[htbp]
\begin{center}
\includegraphics[width=140mm,clip]{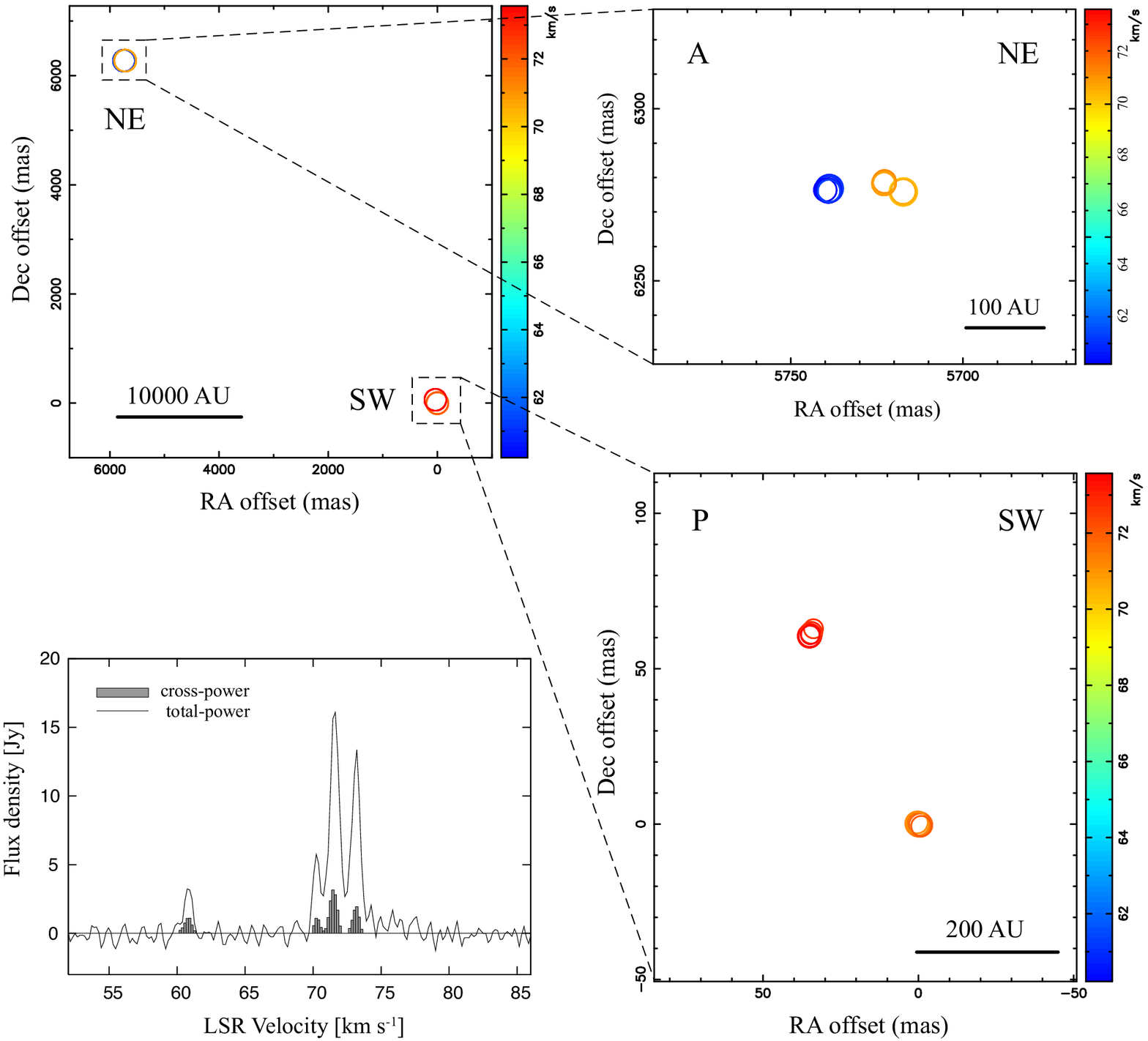}
\end{center}
\caption{{\footnotesize
Data for source 020.23$+$00.06, plotted as for Figure~\ref{fig1}.
}}	
\label{fig16}		
\end{figure*}

\begin{figure*}[htbp]
\begin{center}
\includegraphics[width=120mm,clip]{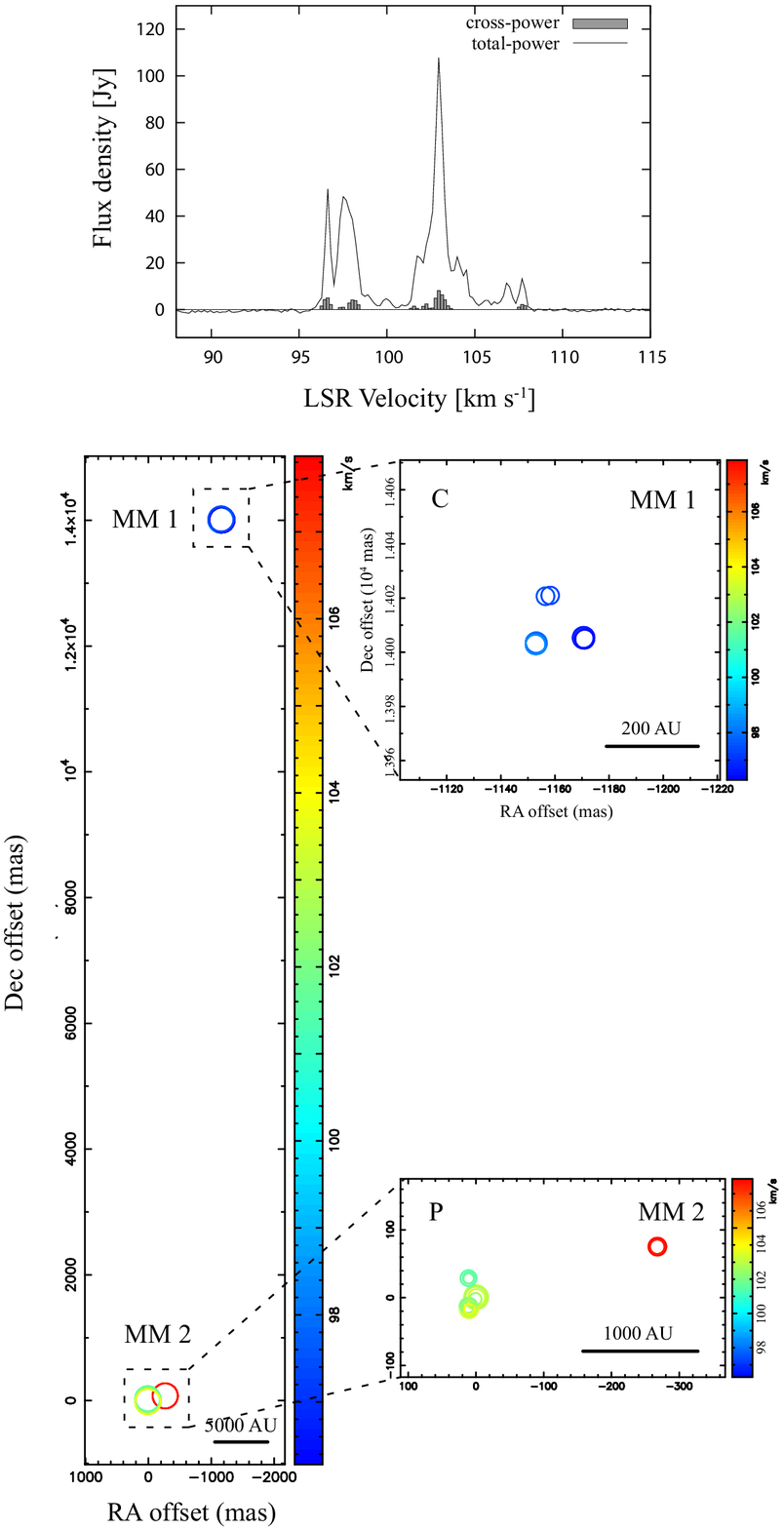}
\end{center}
\caption{{\footnotesize
Data for source 023.43$-$00.18, plotted as for Figure~\ref{fig1}.
}}	
\label{fig17}		
\end{figure*}

\begin{figure*}[htbp]
\begin{center}
\includegraphics[width=140mm,clip]{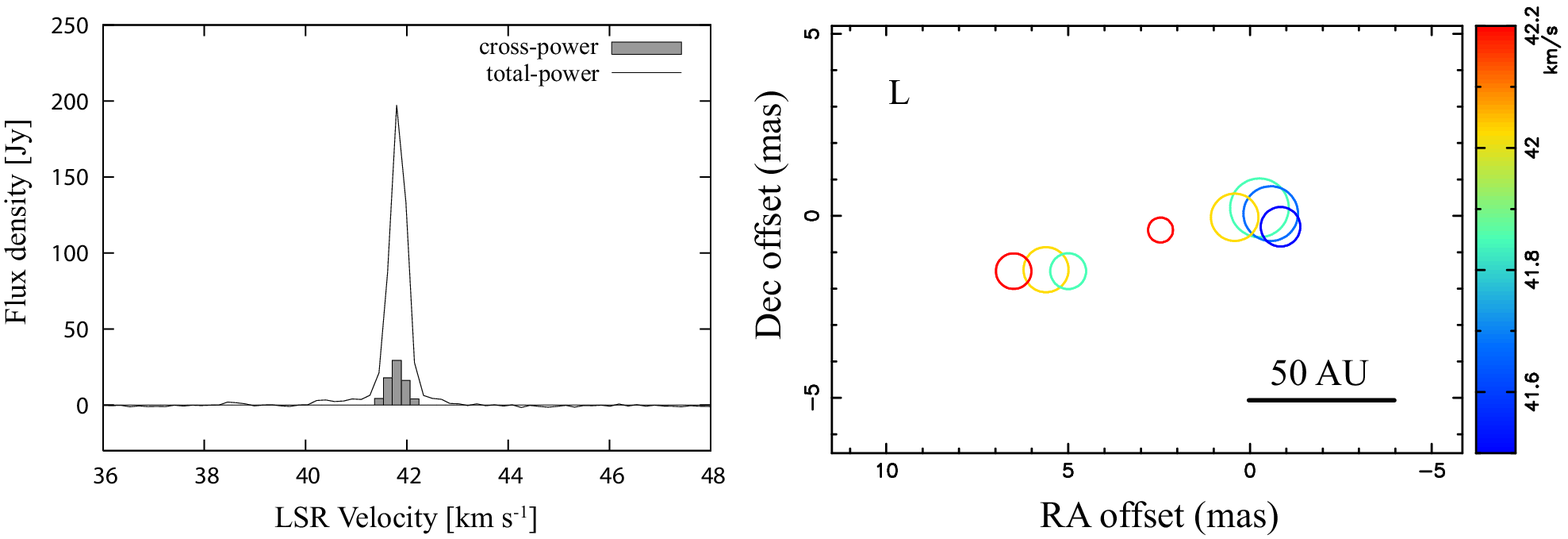}
\end{center}
\caption{{\footnotesize
Data for source 025.65$+$01.05, plotted as for Figure~\ref{fig1}.
}}	
\label{fig18}		
\end{figure*}

\begin{figure*}[htbp]
\begin{center}
\includegraphics[width=140mm,clip]{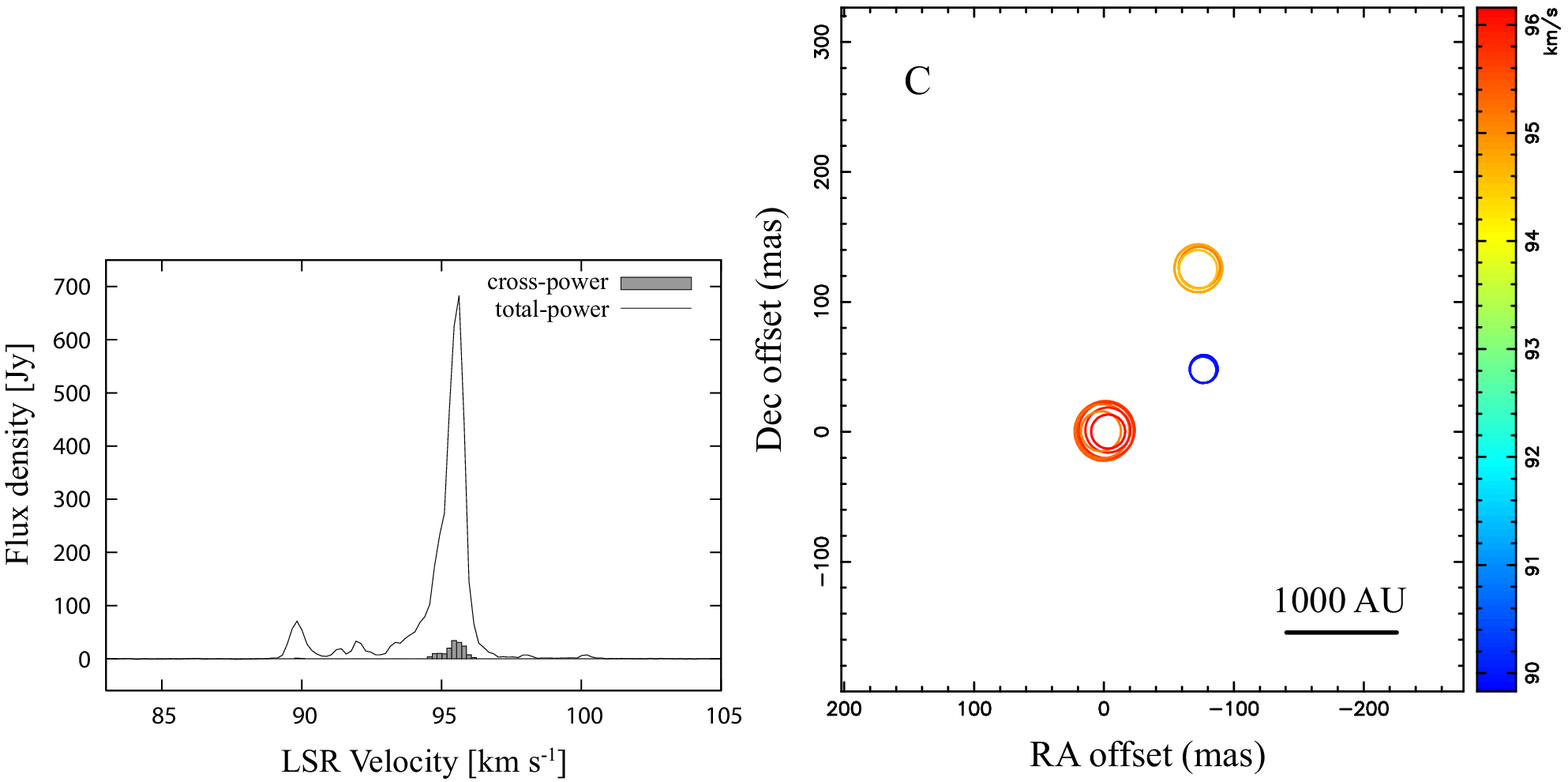}
\end{center}
\caption{{\footnotesize
Data for source 025.71$+$00.04, plotted as for Figure~\ref{fig1}.
}}	
\label{fig19}		
\end{figure*}

\begin{figure*}[htbp]
\begin{center}
\includegraphics[width=140mm,clip]{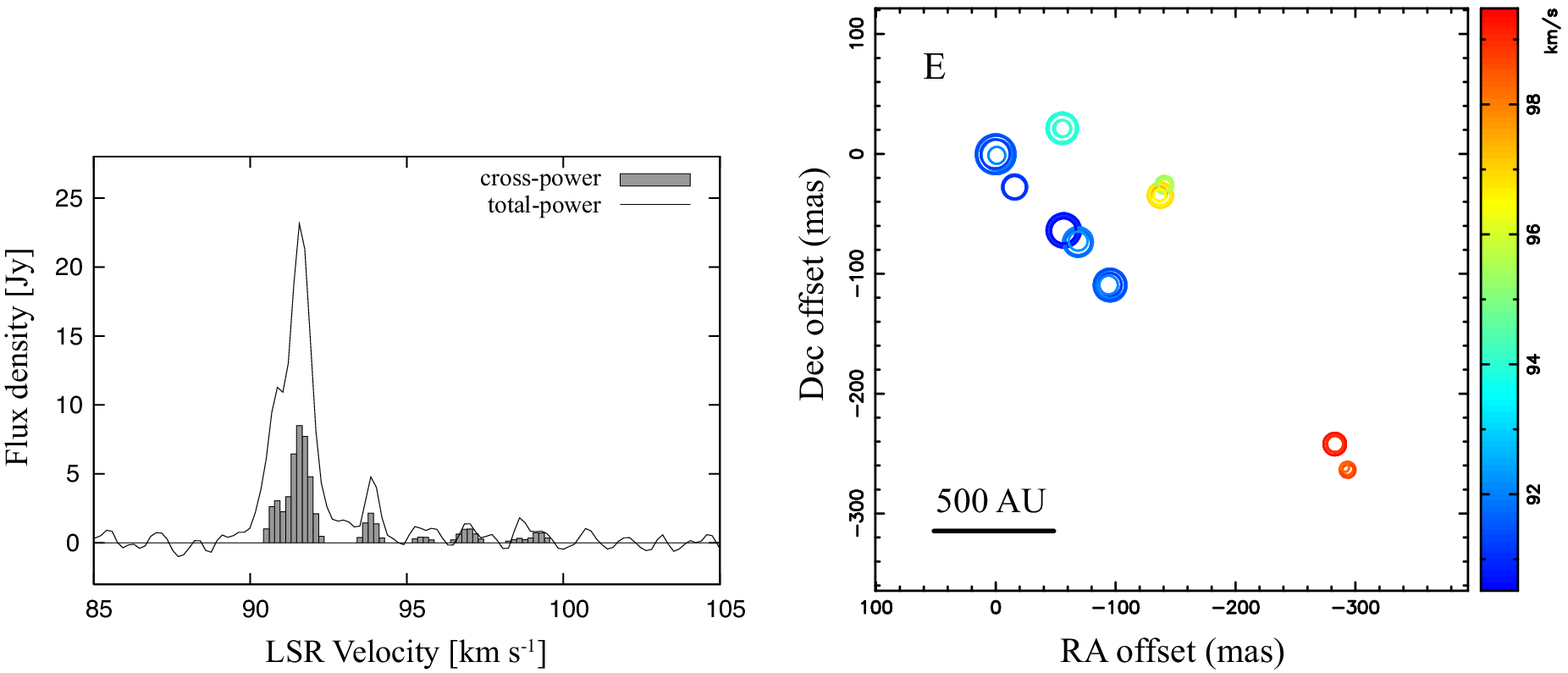}
\end{center}
\caption{{\footnotesize
Data for source 025.82$-$00.17, plotted as for Figure~\ref{fig1}.
}}	
\label{fig20}		
\end{figure*}

\begin{figure*}[htbp]
\begin{center}
\includegraphics[width=140mm,clip]{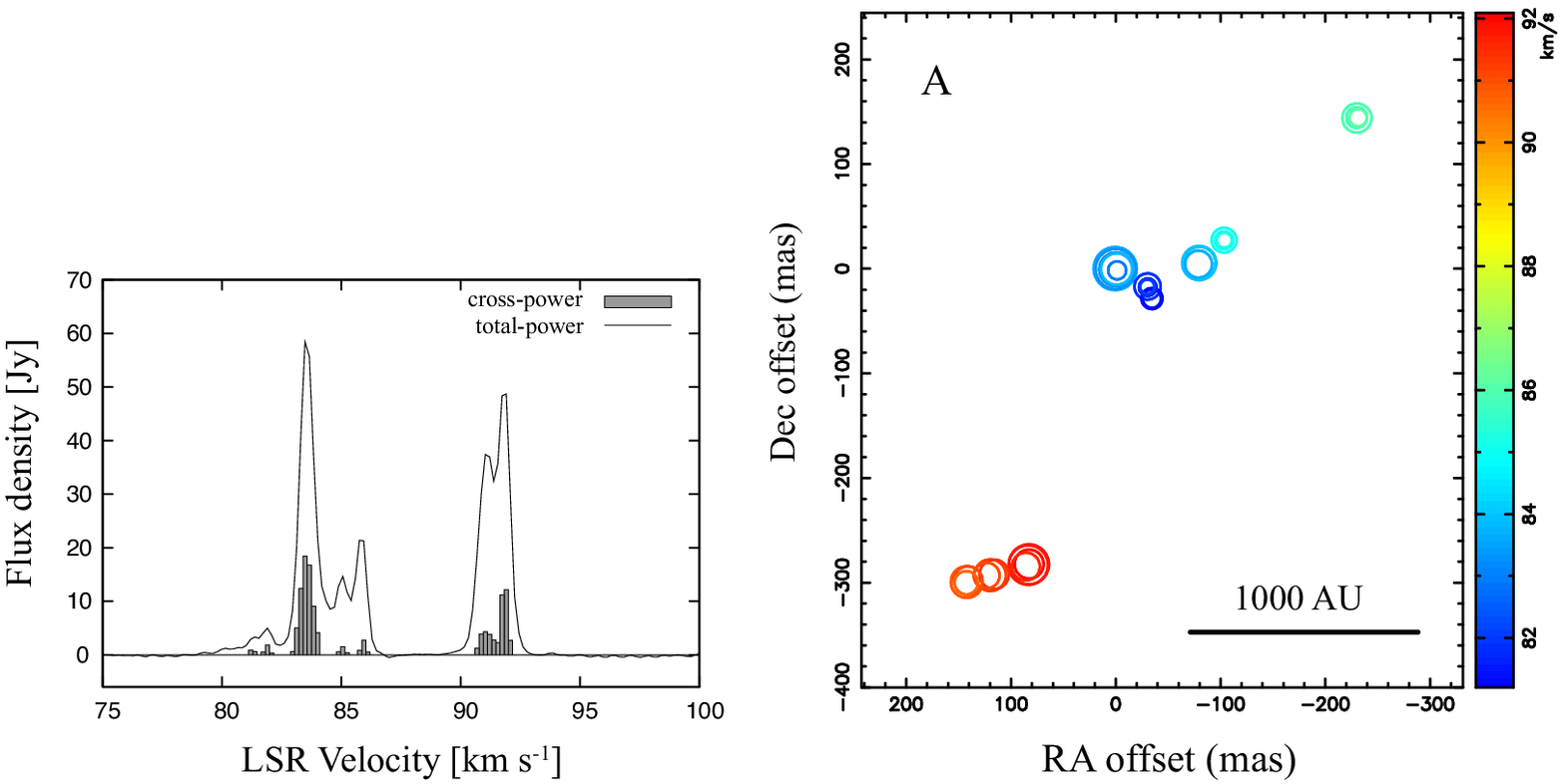}
\end{center}
\caption{{\footnotesize
Data for source 028.83$-$00.25, plotted as for Figure~\ref{fig1}.
}}	
\label{fig21}		
\end{figure*}

\begin{figure*}[htbp]
\begin{center}
\includegraphics[width=140mm,clip]{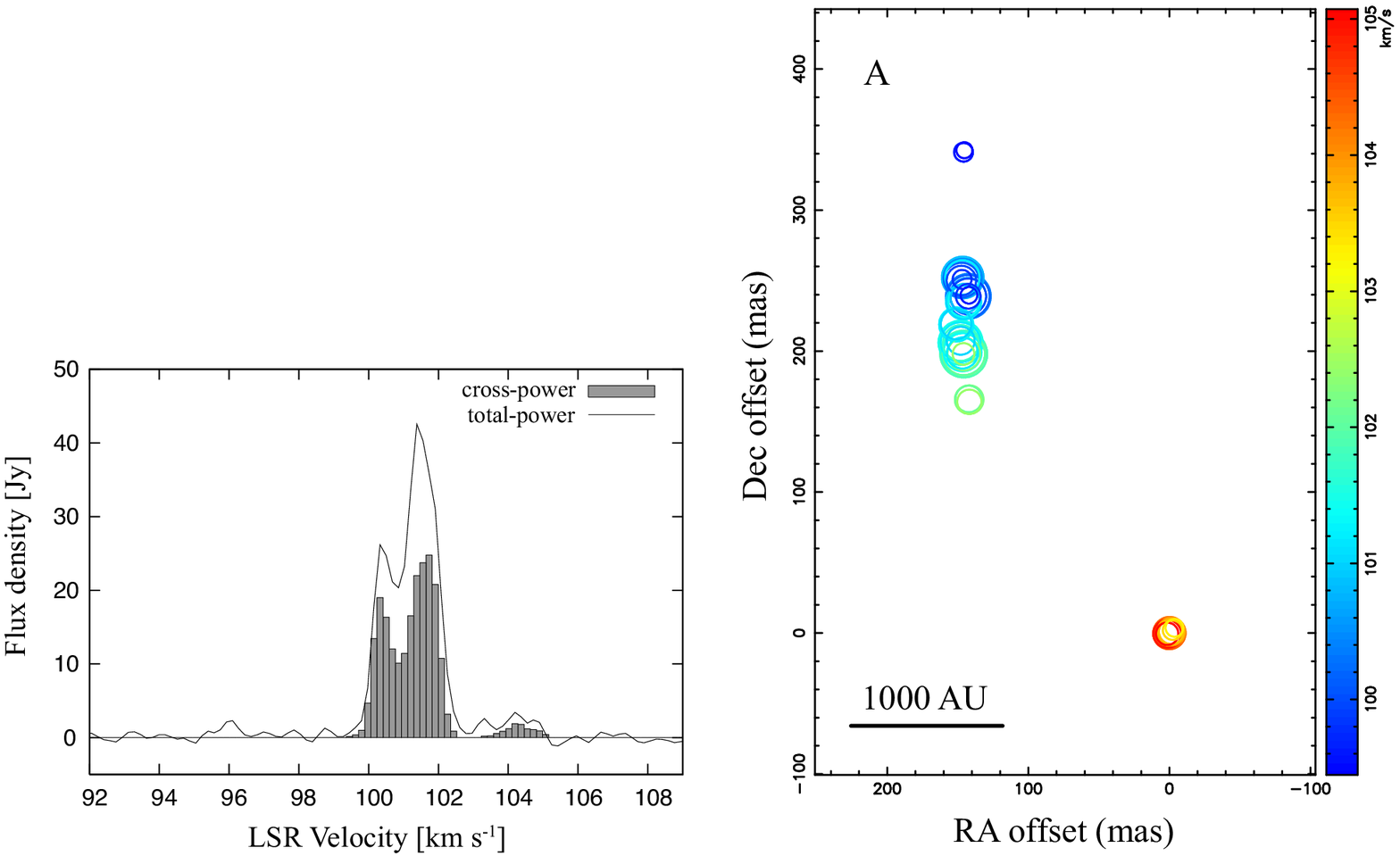}
\end{center}
\caption{{\footnotesize
Data for source 029.86$-$00.04, plotted as for Figure~\ref{fig1}.
}}	
\label{fig22}		
\end{figure*}

\begin{figure*}[htbp]
\begin{center}
\includegraphics[width=140mm,clip]{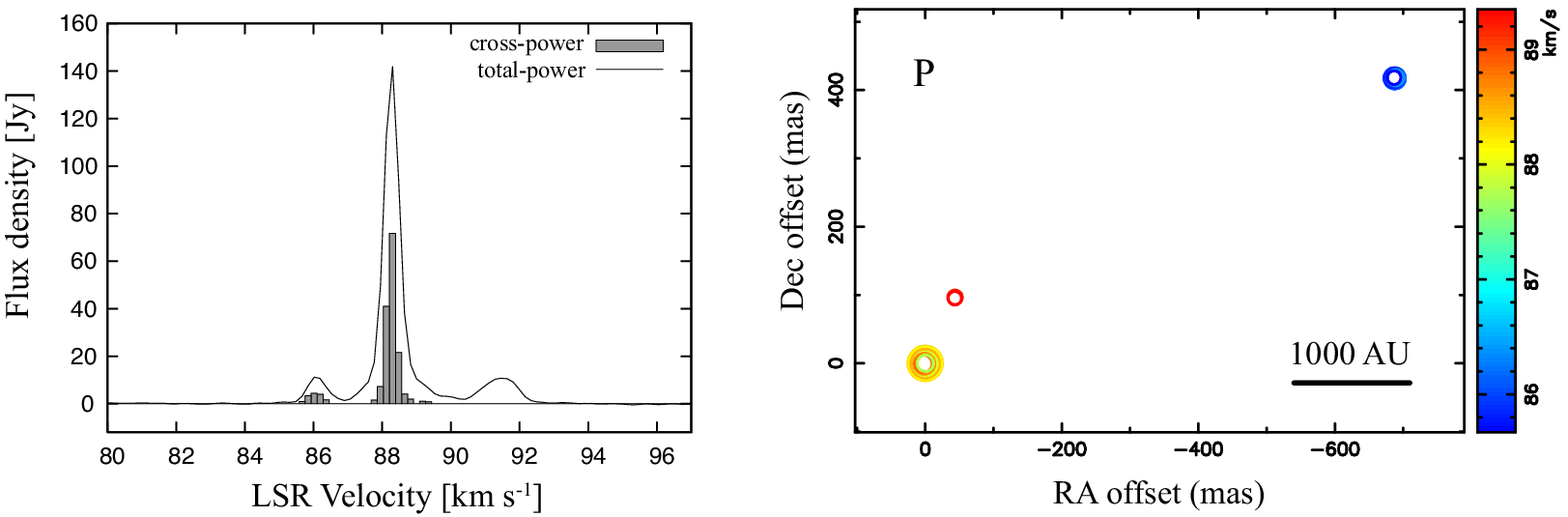}
\end{center}
\caption{{\footnotesize
Data for source 030.70$-$00.06, plotted as for Figure~\ref{fig1}.
}}	
\label{fig23}		
\end{figure*}

\begin{figure*}[htbp]
\begin{center}
\includegraphics[width=140mm,clip]{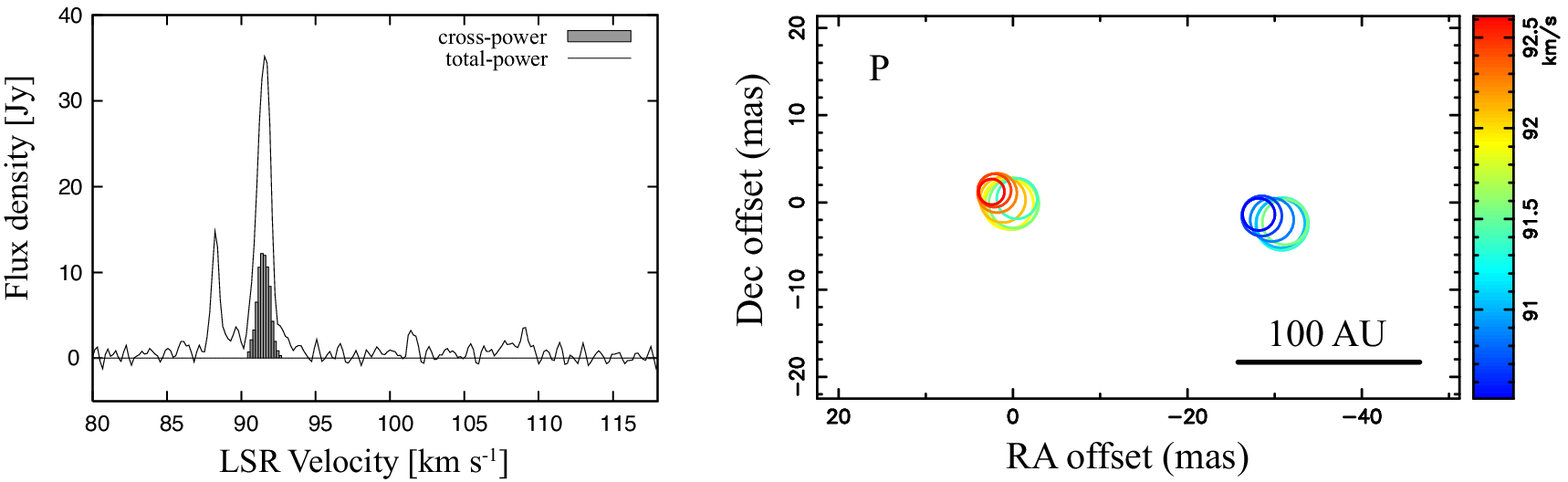}
\end{center}
\caption{{\footnotesize
Data for source 030.76$-$00.05, plotted as for Figure~\ref{fig1}.
}}	
\label{fig24}		
\end{figure*}

\begin{figure*}[htbp]
\begin{center}
\includegraphics[width=140mm,clip]{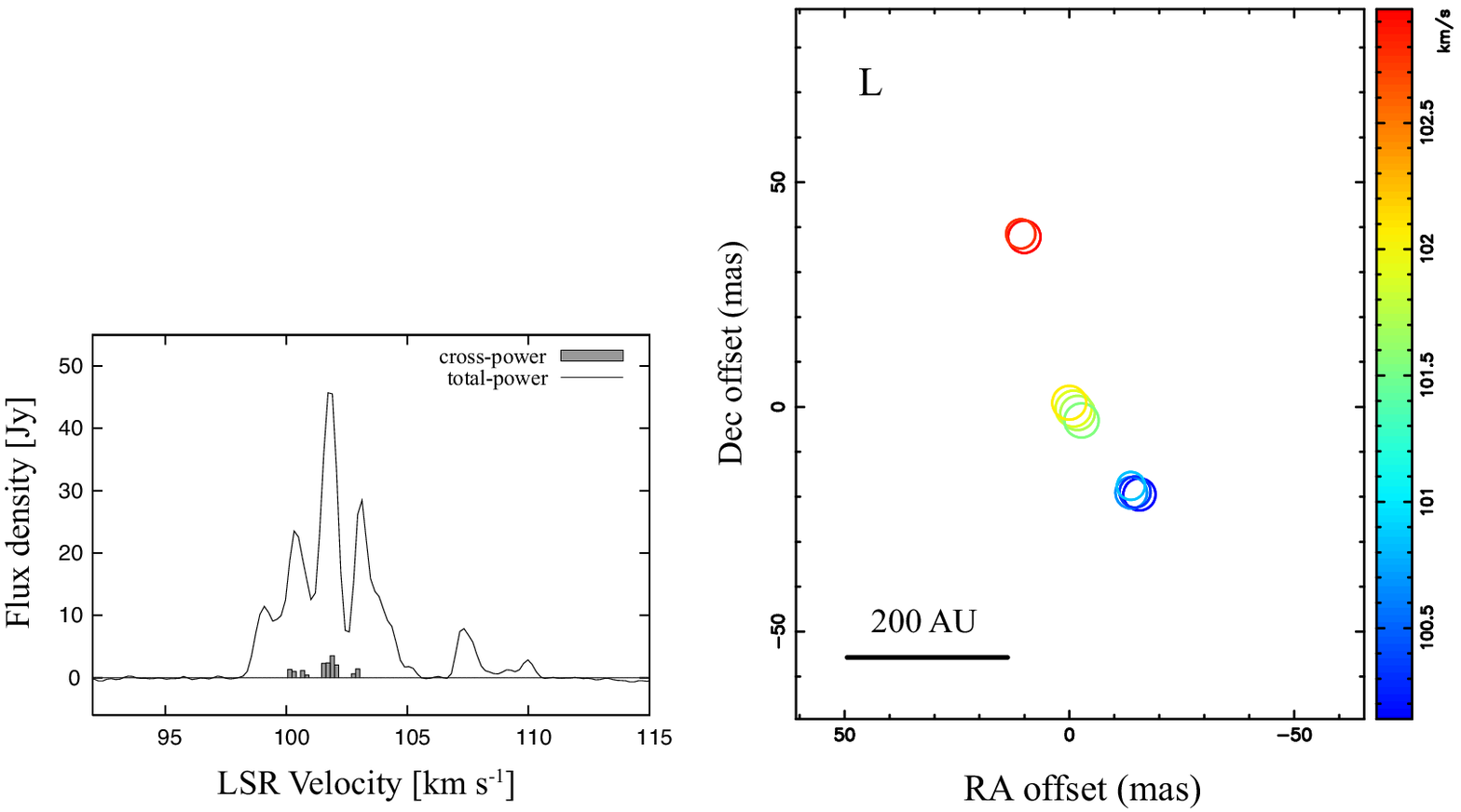}
\end{center}
\caption{{\footnotesize
Data for source 030.91$+$00.14, plotted as for Figure~\ref{fig1}.
}}	
\label{fig25}		
\end{figure*}

\begin{figure*}[htbp]
\begin{center}
\includegraphics[width=140mm,clip]{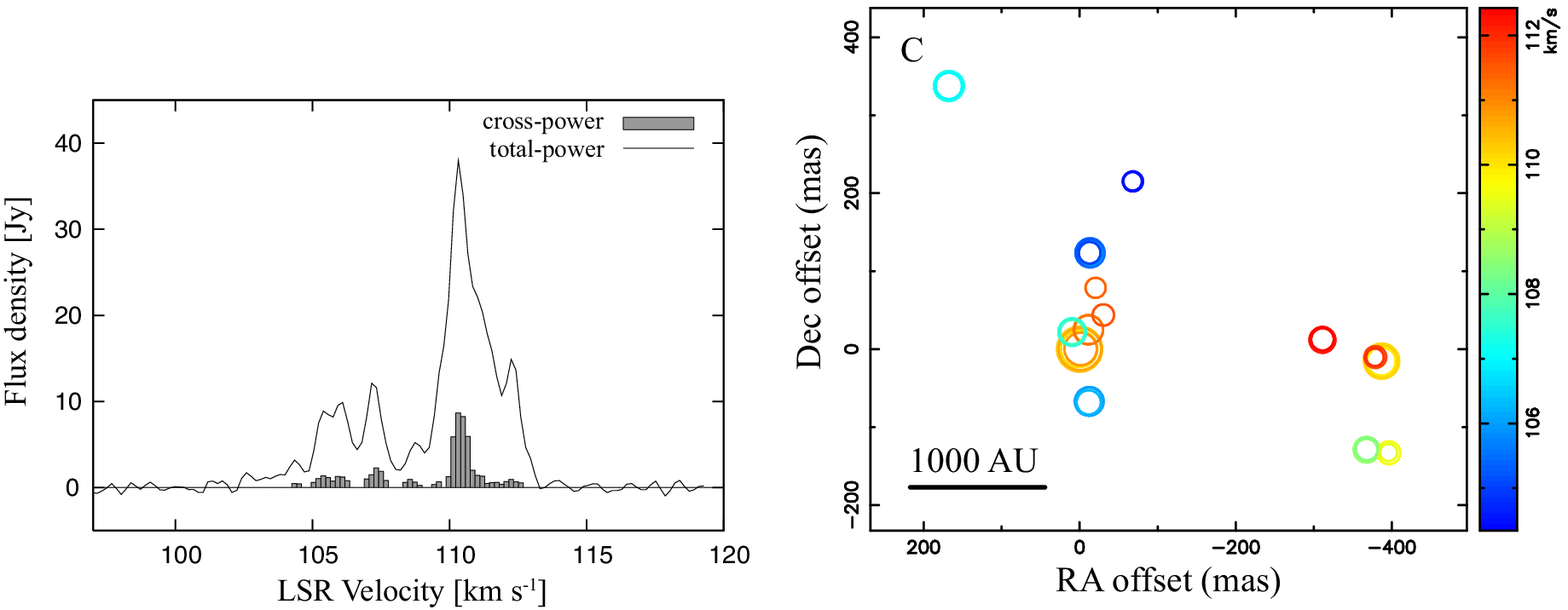}
\end{center}
\caption{{\footnotesize
Data for source 031.28$+$00.06, plotted as for Figure~\ref{fig1}.
}}	
\label{fig26}		
\end{figure*}

\begin{figure*}[htbp]
\begin{center}
\includegraphics[width=140mm,clip]{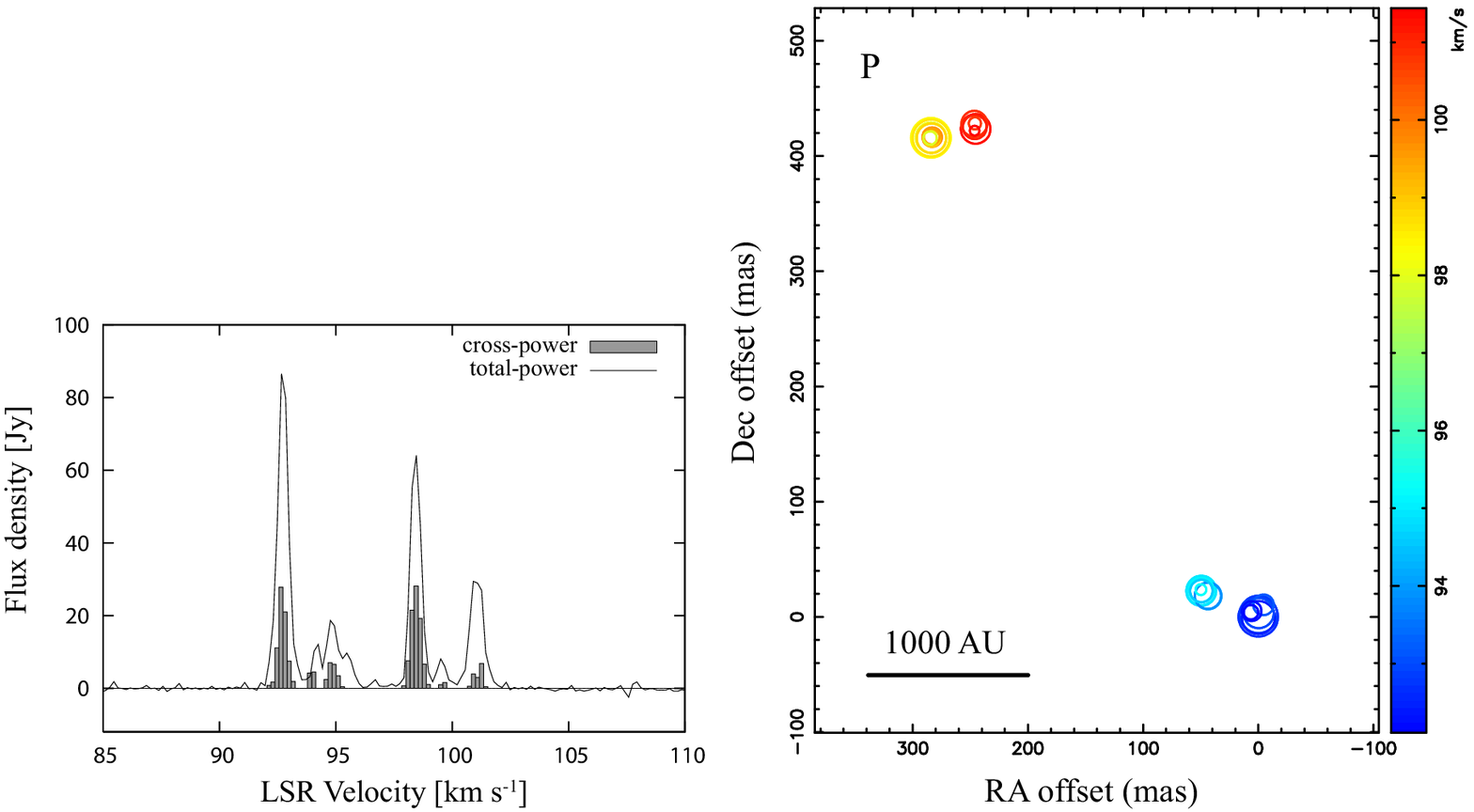}
\end{center}
\caption{{\footnotesize
Data for source 032.03$+$00.06, plotted as for Figure~\ref{fig1}.
}}	
\label{fig27}		
\end{figure*}

\begin{figure*}[htbp]
\begin{center}
\includegraphics[width=140mm,clip]{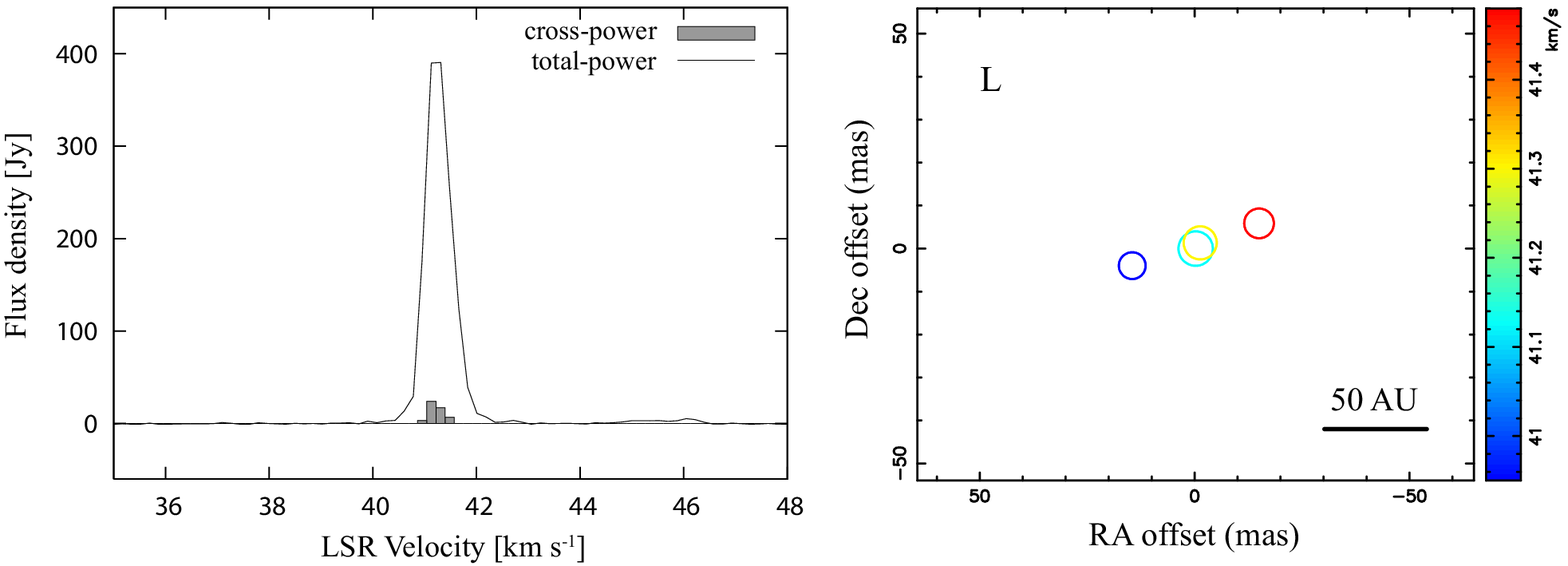}
\end{center}
\caption{{\footnotesize
Data for source 037.40$+$01.52, plotted as for Figure~\ref{fig1}.
}}	
\label{fig28}		
\end{figure*}

\begin{figure*}[htbp]
\begin{center}
\includegraphics[width=140mm,clip]{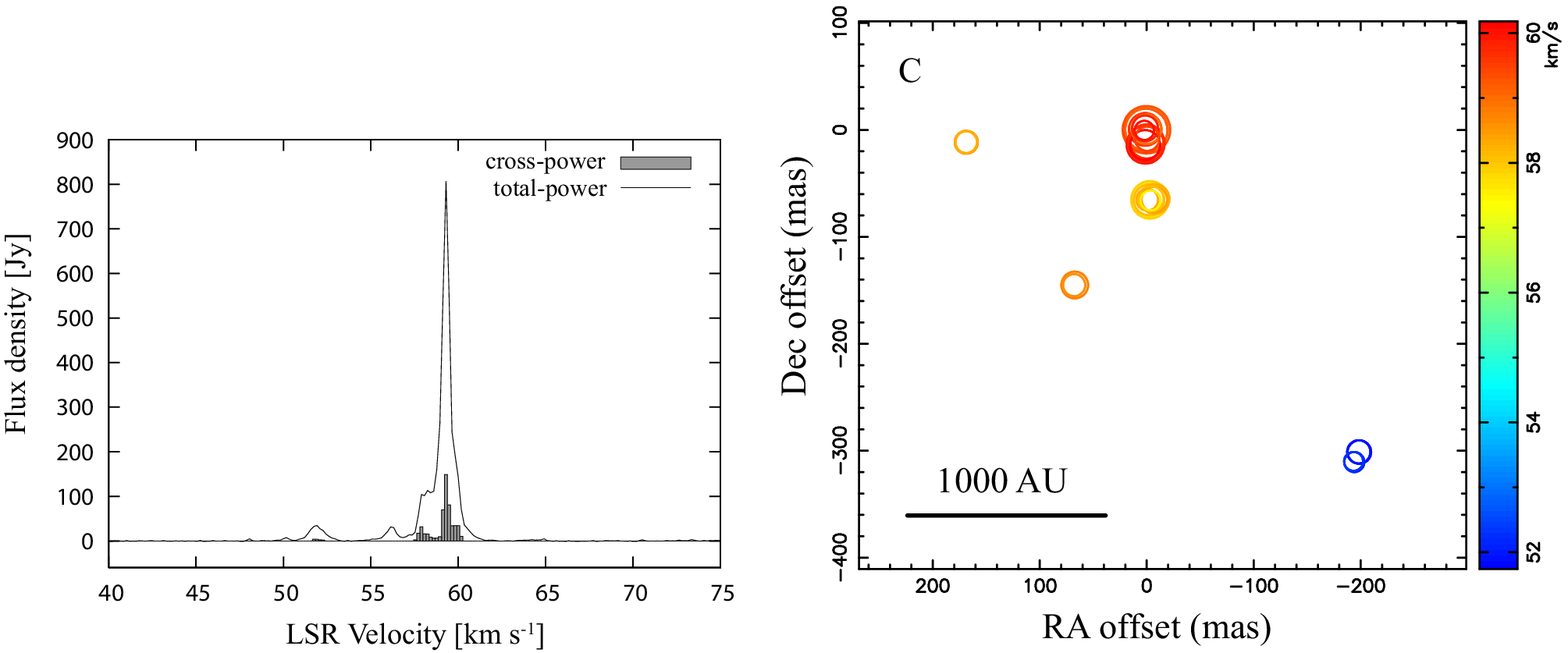}
\end{center}
\caption{{\footnotesize
Data for source 049.49$-$00.38, plotted as for Figure~\ref{fig1}.
}}	
\label{fig29}		
\end{figure*}

\begin{figure*}[htbp]
\begin{center}
\includegraphics[width=140mm,clip]{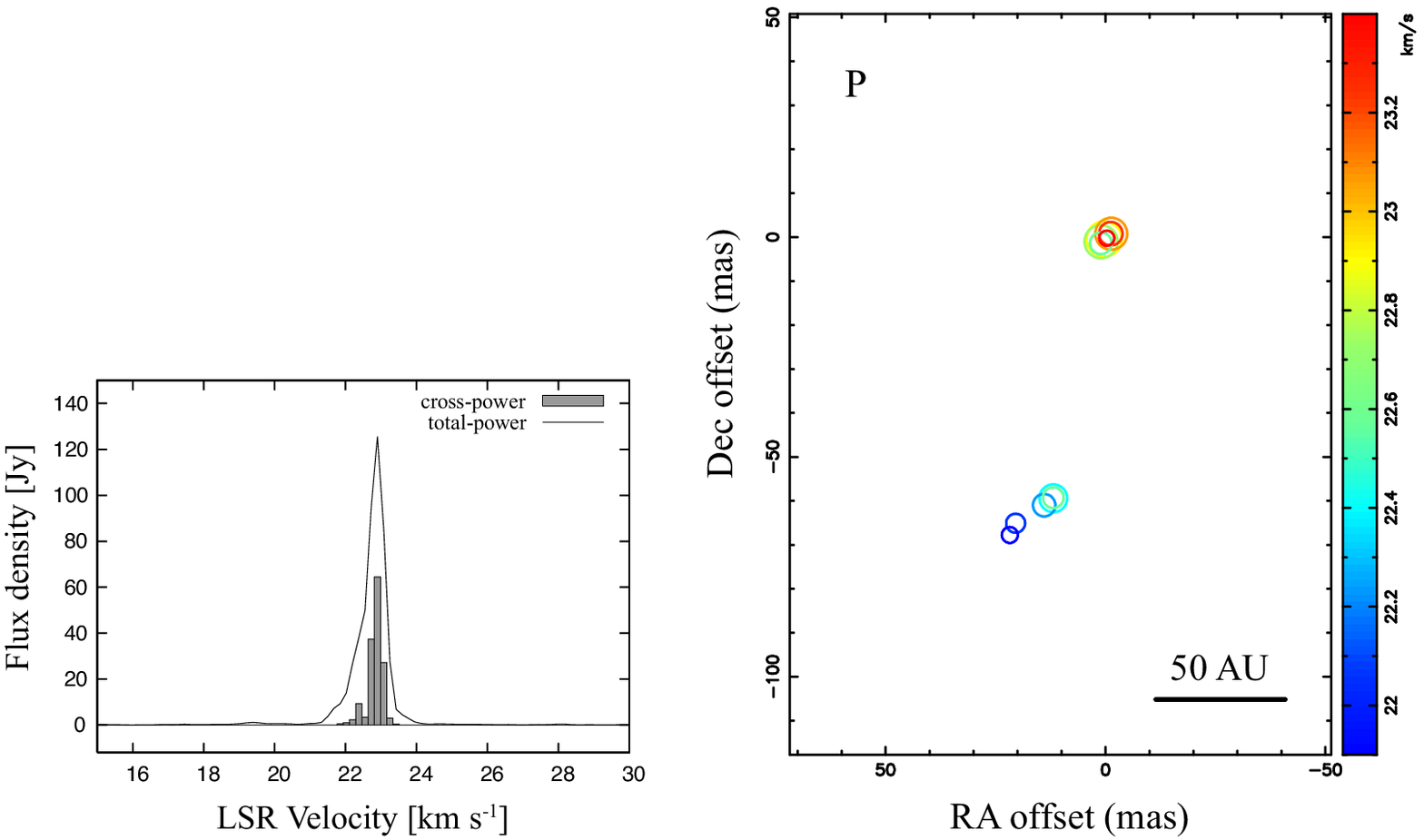}
\end{center}
\caption{{\footnotesize
Data for source 232.62$+$00.99, plotted as for Figure~\ref{fig1}.
}}	
\label{fig30}		
\end{figure*}

\begin{figure*}[htbp]
\begin{center}
\includegraphics[width=140mm,clip]{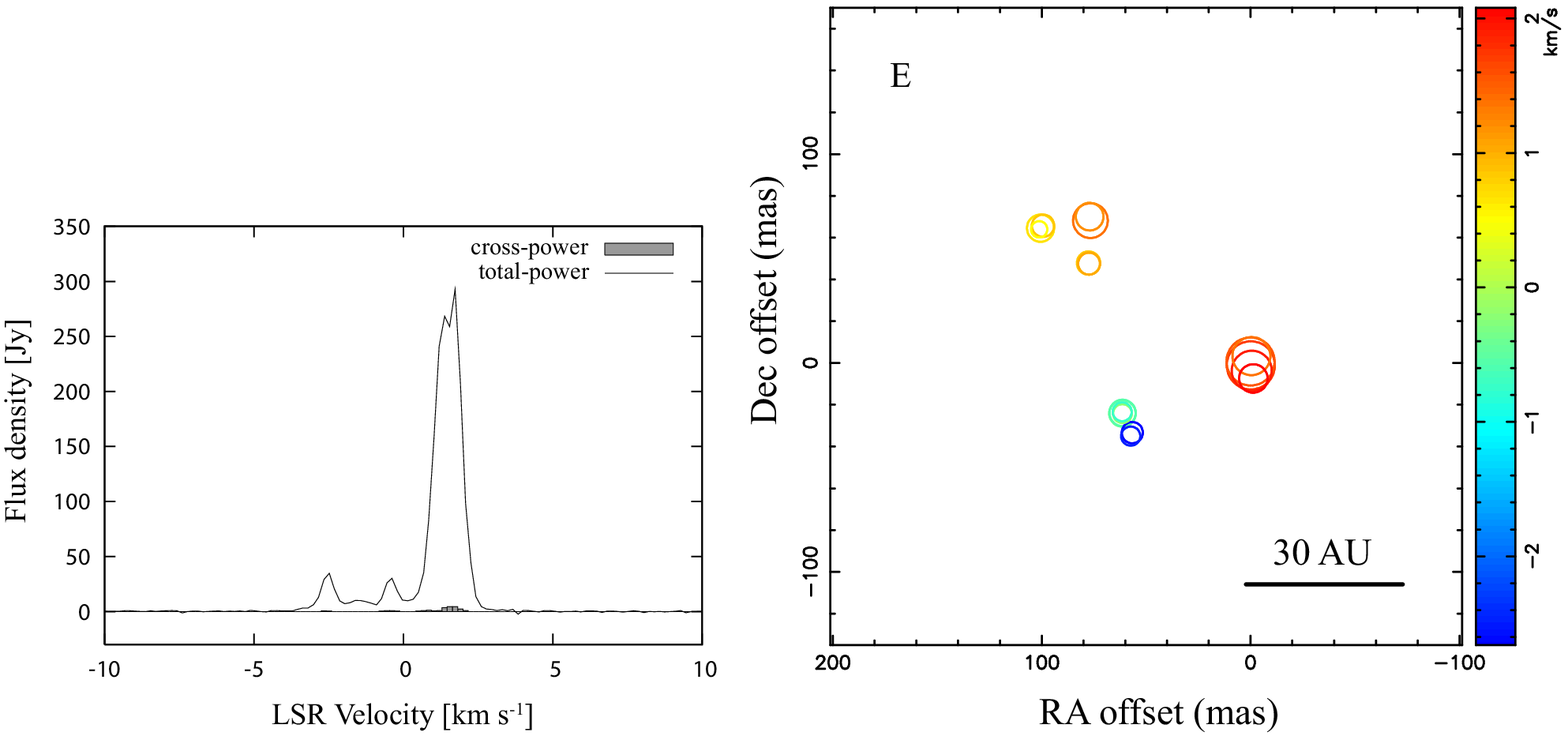}
\end{center}
\caption{{\footnotesize
Data for source 351.77$-$00.53, plotted as for Figure~\ref{fig1}.
}}	
\label{fig31}		
\end{figure*}

\begin{figure*}[htbp]
\begin{center}
\includegraphics[width=140mm,clip]{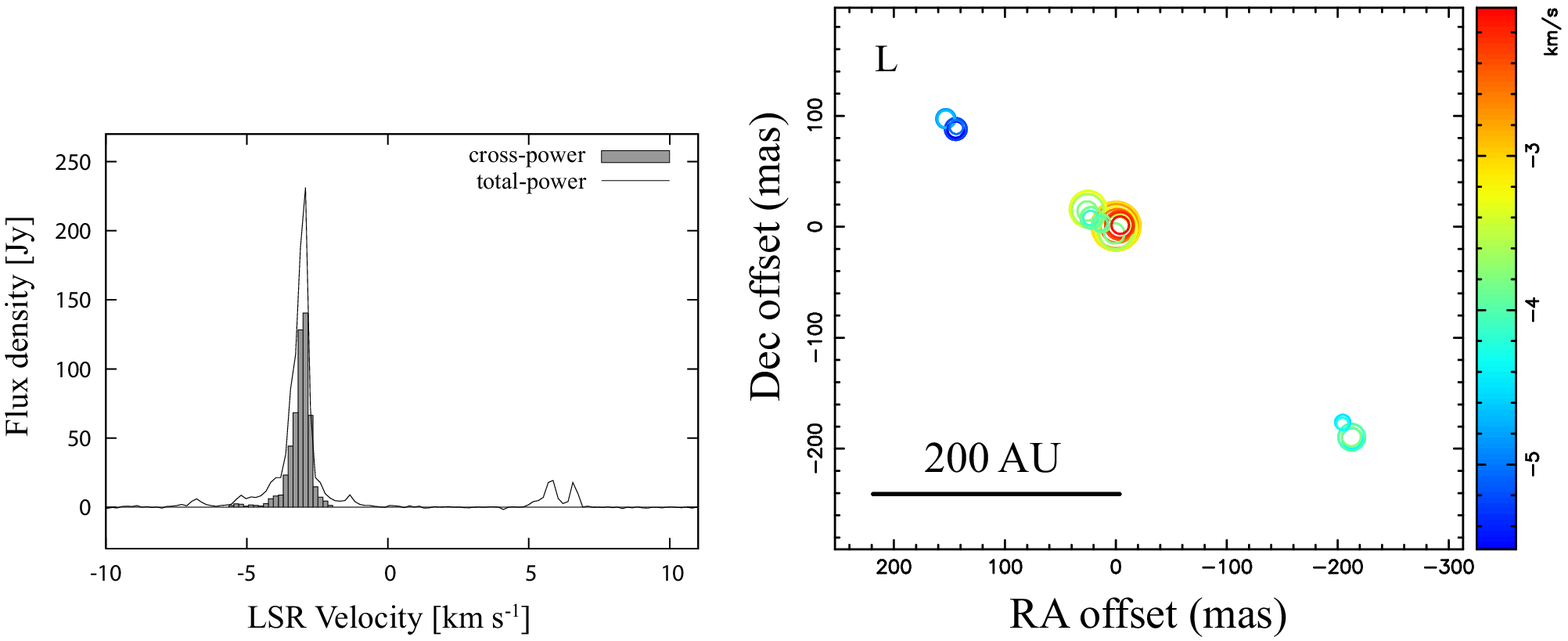}
\end{center}
\caption{{\footnotesize
Data for source 352.63$-$01.06, plotted as for Figure~\ref{fig1}.
}}	
\label{fig32}		
\end{figure*}

\begin{figure*}[htbp]
\begin{center}
\includegraphics[width=140mm,clip]{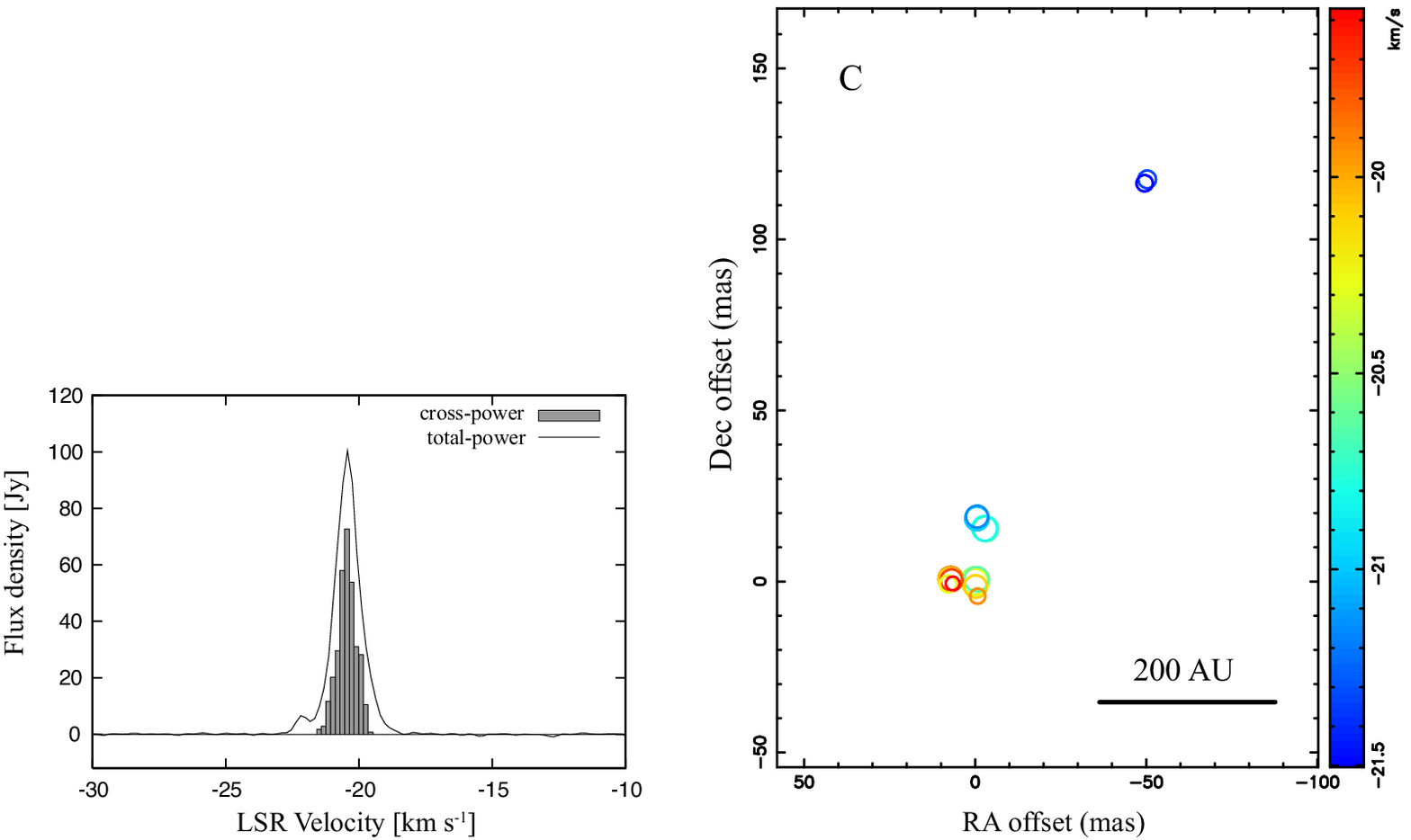}
\end{center}
\caption{{\footnotesize
Data for source 353.41$-$00.36, plotted as for Figure~\ref{fig1}.
}}	
\label{fig33}		
\end{figure*}

\begin{figure*}[htbp]
\begin{center}
\includegraphics[width=140mm,clip]{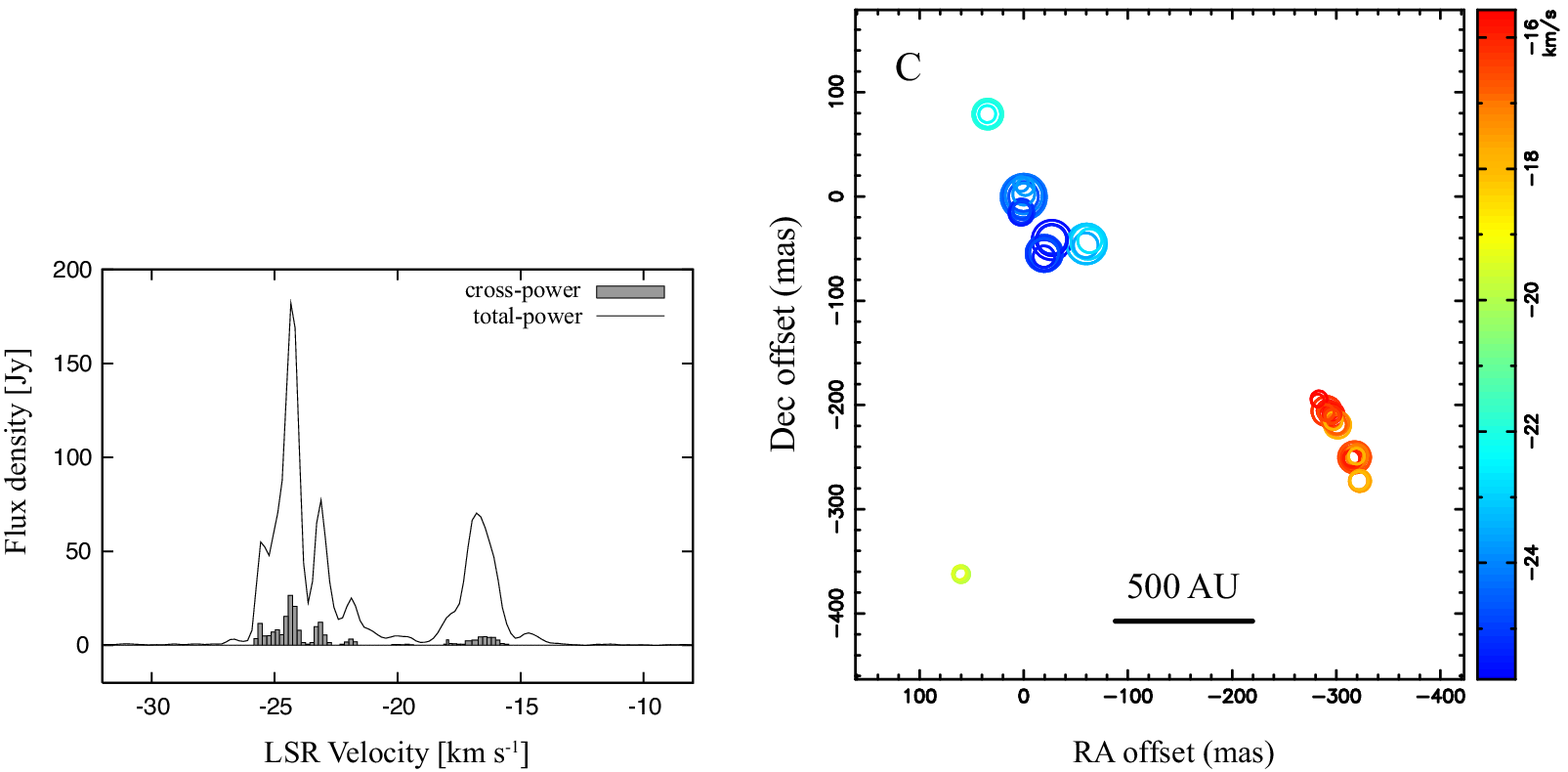}
\end{center}
\caption{{\footnotesize
Data for source 354.61$+$00.47, plotted as for Figure~\ref{fig1}.
}}	
\label{fig34}		
\end{figure*}

\begin{figure*}[htbp]
\begin{center}
\includegraphics[width=140mm,clip]{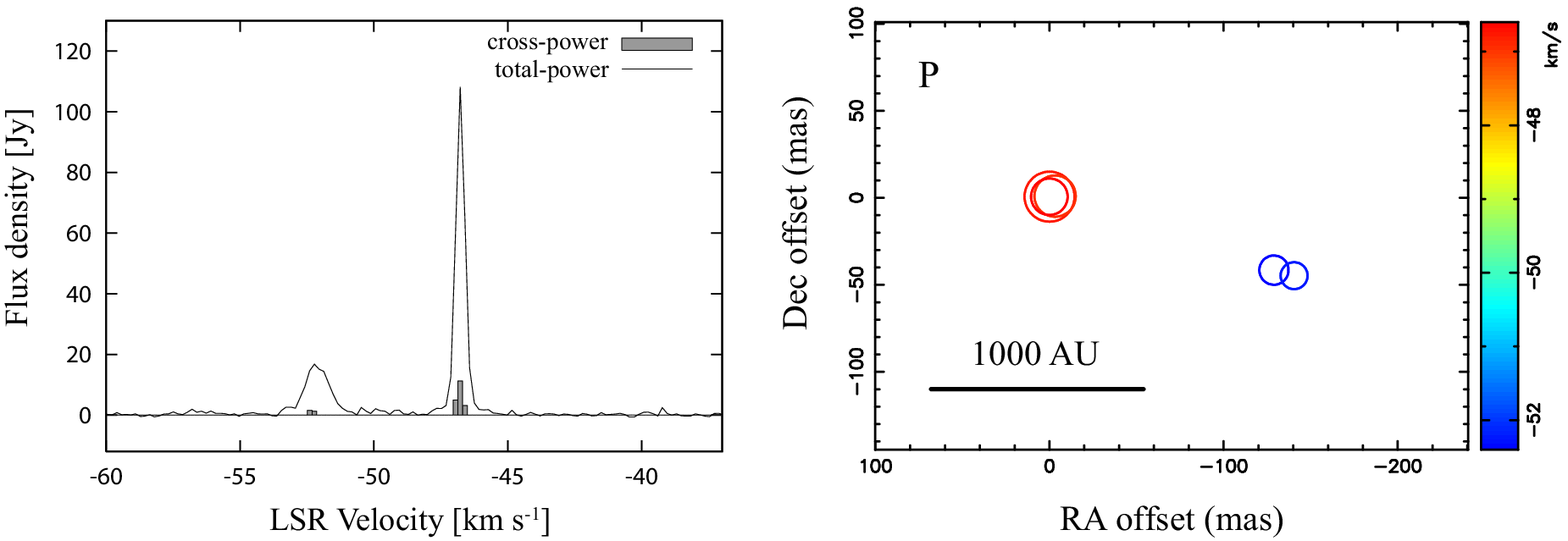}
\end{center}
\caption{{\footnotesize
Data for source 359.43$-$00.10, plotted as for Figure~\ref{fig1}.
}}	
\label{fig35}		
\end{figure*}

The VLBI images of the 35 sources in addition to their spectra are shown
in Figures~\ref{fig1}-\ref{fig35}.
In the source spectra, the solid line and hatched box represent the total-
(autocorrelation of the Hitachi 32 m data) and cross-power spectra (integrated over all baselines),
respectively. The spot sizes in the VLBI images indicate the peak intensity of the spots
on the logarithmic scale. Radial velocities are indicated by the color index of the color bar
displayed to the right of each figure.
The origin of each map is the absolute source coordinates listed in Table~\ref{table3}.
The spatial scale bar is displayed at the bottom corner of each figure.

Besides the coordinates, Table~\ref{table3} lists the following observational parameters:
the radial velocity and flux density of the peak maser spot, number of detected
maser spots, spatial scale (in right ascension and declination coordinates),
velocity range of the detected spectrum, morphological type, and integrated flux ratio of
the correlated to the total-power spectrum of each source. As mentioned in the following subsection,
we consider three sources contain two separated star-forming regions in each image,
yielding 38 imaged sources. The number of detected spots in each source varies from 3 to 72,
with 34 sources (89{\%}) displaying larger than or equal to 10 spots.
The spatial scale of the maser distribution is from 9 to 4900 astronomical units (AU).

Following \citet{2009A&A...502..155B}, we classified the observed morphologies
into the following five types: Elliptical, Arched, Linear, Paired, and Complex, as shown in Table~\ref{table3}.
Since the classification was performed by eye, it is not strict, but
provides an indication of the morphological structure. In the classification
process, we only used the spatial distribution of the spots, the velocity
distribution was not considered. The classification process is as follows:
First, we recognized that almost all spots form small clusters, where we
define a cluster here as a spot group including one or more spots, and the
size is roughly one tenth of the total extent.
When only two clusters exist in a source, the source is classified as \textit{Paired}.
When three or more clusters exist and all of them are successively and linearly distributed, the source is \textit{Linear}.
When three or more clusters exist and all are distributed successively, but loosely curved ($< 45$ degree), the source is \textit{Arched}. 
When four or more clusters exist and all are applicable to an ellipse, the source is \textit{Elliptical}. 
The remaining sources are classified as \textit{Complex}.

\begin{table*}[htbp]
\scriptsize
\begin{center}
\caption{Observed parameters}\label{table3}
\begin{tabular}{clcccccccccr}
\hline\hline
No.& Source & Session & \multicolumn{2}{c}{Coordinates (J2000.0)}                                           & $V_{\mathrm{p}}$ & $F_{\mathrm{p}}$ & $N_{\mathrm{s}}$ & Scale      & $V_{\mathrm{range}}$ & Spatial  & $F_{\mathrm{VLBI}} / F$ \\ \cline{4-5}
   &        &         &               RA            & Dec                                                   &                  &                  &                  &            &                      &  Morph.  &   \\
   &        &         & ($^\mathrm{h}~^\mathrm{m}~^\mathrm{s}$)   & ($^{\circ}~^{\prime}~^{\prime \prime}$) & (km~s$^{-1}$)    & (Jy)             &                  & (AU$^{2}$) &  (km~s$^{-1}$)       &          & (\%) \\
\hline
1  & 000.54$-$00.85 NW          & 4 & 17 50 14.38   & $-$28 54 28.9  &     8.7 &   0.5 &   3 &    9~$\times$~3    & [8.5, 8.9]         & C        & 14 \\
   & 000.54$-$00.85 SE          &   & 17 50 14.56   & $-$28 54 31.4  &    13.3 &  63.2 &  55 & 3900~$\times$~4500 & [10.8, 19.6]       & E        & 35 \\
2  & 000.64$-$00.04	        & 5 & 17 47 18.69   & $-$28 24 25.3  &    49.6 &   6.1 &  17 & 1300~$\times$~760  & [48.2, 52.4]       & C        & 11 \\
3  & 002.53$+$00.19$^{\dagger}$ & 3 &               &                &     3.4 &   7.8 &  34 &  750~$\times$~2100 & [3.0, 19.0]        & E        & 12 \\
4  & 006.18$-$00.35	        & 2 & 18 01 02.17   & $-$23 47 10.8  & $-$30.2 &   7.5 &  18 & 1400~$\times$~1100 & [$-$36.3, $-$29.6] & C        &  4 \\
5  & 006.79$-$00.25	        & 2 & 18 01 57.76   & $-$23 12 34.2  &    26.1 &  24.2 &  72 & 1400~$\times$~640  & [15.0, 30.8]       & E        & 36 \\
6  & 008.68$-$00.36             & 1 & 18 06 23.48   & $-$21 37 10.4  &    43.1 &  21.1 &  25 &  970~$\times$~590  & [40.7, 44.4]       & C        &  9 \\
7  & 008.83$-$00.02             & 1 & 18 05 25.66   & $-$21 19 25.4  &  $-$3.7 &  16.7 &  29 & 1400~$\times$~990  & [$-$5.7, 2.6]      & E        &  7 \\
8  & 009.61$+$00.19             & 6 & 18 06 14.91   & $-$20 31 43.4  &     5.5 &   8.3 &  31 &  490~$\times$~120  & [4.9, 6.9]         & L        & 22 \\
9  & 009.98$-$00.02		& 6 & 18 07 50.12   & $-$20 18 56.5  &    42.4 &  18.9 &  58 & 3000~$\times$~1600 & [40.7, 50.7]       & C        & 32 \\
10 & 010.32$-$00.16             & 4 & 18 09 01.47   & $-$20 05 07.8  &    11.6 &  21.1 &  22 &  490~$\times$~620  & [4.1, 14.2]        & C        & 19 \\
11 & 011.49$-$01.48             & 4 & 18 16 22.13   & $-$19 41 27.2  &     6.3 &  28.8 &  68 &  290~$\times$~710  & [4.5, 17.1]        & C        & 42 \\
12 & 011.90$-$00.14             & 4 & 18 12 11.45   & $-$18 41 28.8  &    43.1 &  38.3 &  29 & 1300~$\times$~480  & [39.6, 44.2]       & P        & 41 \\
13 & 012.02$-$00.03             & 3 & 18 12 01.86   & $-$18 31 55.9  &   108.3 &  22.8 &  25 &  520~$\times$~1200 & [107.1, 109.0]     & A        & 18 \\
14 & 012.68$-$00.18$^{\dagger}$ & 3 &               &                &    58.4 &  11.5 &  20 &  310~$\times$~650  & [52.0, 60.1]       & C        &  1 \\
15 & 012.88$+$00.48             & 5 & 18 11 51.39   & $-$17 31 30.1  &    39.2 &  18.0 &  67 & 2600~$\times$~3800 & [29.9, 40.1]       & C        & 21 \\
16 & 014.10$+$00.08$^{\dagger}$ & 5 &               &                &         &       &     &                    &                    &               \\
17 & 020.23$+$00.06 SW	        & 5 & 18 27 44.56   & $-$11 14 54.1  &    71.5 &   3.2 &  12 &  160~$\times$~280  & [71.1, 73.6]       & P        & 14 \\
   & 020.23$+$00.06 NE          &   & 18 27 44.95   & $-$11 14 47.8  &    60.9 &   1.1 &  14 &  100~$\times$~10   & [60.2, 71.1]       & A        & 24 \\
18 & 023.43$-$00.18 MM1         & 1 & 18 34 39.19   & $-$08 31 25.3  &    96.6 &   5.0 &  10 &  110~$\times$~110  & [96.3, 98.4]       & C        &  7 \\
   & 023.43$-$00.18 MM2	        &   & 18 34 39.27   & $-$08 31 39.3  &   103.0 &   8.1 &  28 & 1600~$\times$~550  & [101.4, 107.9]     & P        &  6 \\
19 & 025.65$+$01.05$^{\ast}$    & 1 &               &                &    41.8 &  26.3 &   8 &   90~$\times$~20   & [41.5, 42.2]       & L        & 15 \\
20 & 025.71$+$00.04             & 1 & 18 38 03.15   & $-$06 24 15.0  &    95.5 &  34.1 &  13 &  930~$\times$~1500 & [89.8, 96.2]       & C        &  4 \\
21 & 025.82$-$00.17             & 6 & 18 39 03.63   & $-$06 24 09.9  &    91.6 &   8.2 &  53 & 1500~$\times$~1500 & [90.5, 99.5]       & E        & 29 \\
22 & 028.83$-$00.25$^{\ast}$    & 2 &               &                &    83.5 &  18.4 &  30 & 1700~$\times$~2000 & [81.2, 92.1]       & A        & 15 \\
23 & 029.86$-$00.04$^{\ast}$    & 6 &               &                &   101.7 &  22.1 &  53 & 1400~$\times$~3200 & [99.5, 105.1]      & A        & 56 \\
24 & 030.70$-$00.06$^{\ast}$    & 2 &               &                &    88.3 &  71.7 &  14 & 4100~$\times$~2500 & [85.7, 89.4]       & P        & 28 \\
25 & 030.76$-$00.05$^{\ast}$    & 5 &               &                &    91.7 &  10.6 &  15 &  160~$\times$~20   & [90.5, 92.6]       & P        & 29 \\
26 & 030.91$+$00.14$^{\ast}$    & 2 &               &                &   101.9 &   3.5 &  10 &  150~$\times$~320  & [100.1, 103.0]     & L        &  2 \\
27 & 031.28$+$00.06$^{\ast}$    & 4 &               &                &   110.5 &   8.2 &  38 & 3300~$\times$~2700 & [104.3, 112.4]     & C        & 11 \\
28 & 032.03$+$00.06$^{\ast}$    & 3 &               &                &    92.7 &  27.8 &  28 & 2100~$\times$~3100 & [92.1, 101.4]      & P        & 25 \\
29 & 037.40$+$01.52$^{\ast}$    & 3 &               &                &    41.1 &  24.0 &   4 &   60~$\times$~20   & [41.0, 41.5]       & L        &  4 \\
30 & 049.49$-$00.38             & 3 & 19 23 43.93   & $+$14 30 35.1  &    59.3 & 134.9 &  21 & 2000~$\times$~1700 & [51.7, 60.2]       & C        & 13 \\
31 & 232.62$+$00.99             & 2 & 07 32 09.78   & $-$16 58 12.4  &    22.9 &  64.4 &  11 &   40~$\times$~120  & [21.9, 23.4]       & P        & 29 \\
32 & 351.77$-$00.53             & 1 & 17 26 42.54   & $-$36 09 17.6  &     1.7 &   4.4 &  18 &   40~$\times$~40   & [$-$2.7, 2.1]      & E        &  1 \\
33 & 352.63$-$01.06             & 1 & 17 31 13.93   & $-$35 44 08.5  &  $-$2.9 & 140.4 &  32 &  330~$\times$~260  & [-5.6, -2.0]       & L        & 58 \\
34 & 353.41$-$00.36             & 2 & 17 30 26.18   & $-$34 41 45.6  & $-$20.5 &  72.7 &  16 &  220~$\times$~460  & [$-$21.5, $-$19.6] & C        & 48 \\
35 & 354.61$+$00.47             & 2 & 17 30 17.09   & $-$33 13 55.0  & $-$24.4 &  26.5 &  63 & 1500~$\times$~1700 & [$-$25.8, $-$15.6] & C        &  9 \\
36 & 359.43$-$00.10$^{\dagger}$ & 3 &               &                & $-$46.8 &  11.2 &   5 & 1200~$\times$~370  & [-52.4, -46.6]     & P        &  6 \\
\hline
\multicolumn{12}{l}{\hbox to 0pt{\parbox{168mm}{\footnotesize
Column~1 -- ID number (as listed in Table~\ref{table1});
Column~2 -- Source name;
Column~3 -- Observational session (corresponding to the sessions listed in Table~\ref{table2});
Columns~4, 5 -- Absolute coordinates obtained by fringe-rate mapping
(for specified sources marked by asterisk or dagger, the absolute coordinates listed in Table~\ref{table1} are listed here);
Columns~6, 7 -- Radial velocity and flux density of the peak maser spot located at the origin of each map, respectively;
Column~8 -- Number of detected maser spots;
Columns~9, and 10 -- Spatial scale (in RA and Dec coordinates), and velocity range of the detected component, respectively;
Column~11 -- Spatial morphology (E: Elliptical; A: Arched; L: Linear; P: Paired; C: Complex).
Column~12 -- Ratio of integrated fluxes of correlated- to total-power spectra.
\\
$\ast$: sources not suitable for fringe-rate mapping because of the equatorial location.\\
$\dagger$: sources not suitable for fringe-rate mapping because of the weak flux or resolved-out.
}\hss}}
\end{tabular}
\end{center}
\end{table*}

\section{Comments on individual sources}
\label{comments}

{\it 000.54$-$00.85}

The main cluster (SE) of the 6.7~GHz methanol maser spots of this source are distributed over 540~mas~$\times$~630~mas,
corresponding to 3900~AU~$\times$~4500~AU, at a near kinematic distance of 7.2~kpc (Figure~\ref{fig1}).
This source has the largest extention of 4900~AU of all imaged sources.
The distance is determined from the radial velocity and the H\emissiontype{I} self-absorption (H\emissiontype{I}SA) spectrum.
Isolated spots (NW) located at 3~arcsec ($2 \times 10^{4}$~AU) northwest are probably associated with another excitation source.
The distribution of the maser spots of the main cluster is \textit{Elliptical}, having roughly circular morphology,
while that of NW component is \textit{Complex}.
These maser spots correspond to spots labeled A$-$L in the ATCA image (IRAS~17470$-$2853, Walsh et al.\ 1998),
while the isolated spots in the EAVN image correspond to spot M in the ATCA image.

{\it 000.64$-$00.04}

This source is associated with a famous high-mass star-forming region in
the Galactic center Sgr~B2 located at a trigonometric parallax distance of 7.9~kpc \citep{2009ApJ...705.1548R}.
The wide absorption trough in the spectrum is remarkable (Menten 1991).
The methanol maser spots in this source show \textit{Complex} spatial distribution over 170~mas~$\times$~100~mas,
corresponding to 1300~AU~$\times$~760~AU at the source (Figure~\ref{fig2}).
These maser spots are extended roughly along the east-west direction,
although the extension is not clearly linear. The ATCA image (IRAS~17441$-$2822, Walsh et al.\ 1998)
extends along the northeast-southwest direction.
Apart from the contradicting elongation direction,
the EAVN and ATCA images display similar radial velocity gradient trends,
from red-shifted at the east to blue-shifted at the west.

{\it 002.53$+$00.19}

The kinematic distance of this source is 4.2~kpc.
Although the north-western part is missing, the spatial distribution is {\it Elliptical} (Figure~\ref{fig3}).
The maser spots are distributed over 180~mas~$\times$~500~mas, corresponding to a linear area of 750~AU~$\times$~2100~AU.
The millimeter and submillimeter emissions have been previously detected at this maser position
(Hill et al 2005; Walsh et al. 2003).

{\it 006.18$-$00.35}

This source locates at the same line of sight toward the W~28 supernova remnant field.
The 6.7~GHz methanol maser source seems to be associated with one of several dense
molecular cores previously identified in this region, namely, Core~3 (Nicholas et al. 2011).
The radial velocity ($-$33~km~s$^{-1}$) of Core~3 approximates that of the spectral
peak of the methanol maser but differs largely from those of other molecular cores.
\citet{2008ApJ...683L.143D} suggested that this source locates in the galactic 3~kpc arm
approximately 5~kpc from the Sun. Therefore, Core 3 (and hence the methanol maser source)
is likely not connected to the gas associated with the W 28 field (Nicholas et al. 2011).
We assumed a source distance of 5.1~kpc (Green \& McClure-Griffiths 2011).
The morphology is {\it Complex} (Figure~\ref{fig4}). The maser spots are distributed over 280~mas~$\times$~220~mas,
corresponding to 1400~AU~$\times$~1100~AU.

{\it 006.79$-$00.25}

This source also locates near the supernova remnant W~28 similar to 006.18$-$00.35
and appears to be associated with Core 1, as defined by Nicholas et al. (2011).
The radial velocity of $+$21~km~s$^{-1}$ of the ammonia line of Core 1 indicates
a kinematic distance of 3.8 kpc, although the uncertainty is large.
The spatial distribution of the masers is clearly {\it Elliptical},
with a clockwise radial velocity gradient (Figure~\ref{fig5}).
The size of this ellipse is 400~mas from east to west, corresponding to 1400~AU.

{\it 008.68$-$00.36}

This source locates at a distance of 4.5~kpc (Green \& McClure-Griffiths 2011).
The spatial distribution of the masers is {\it Complex} (Figure~\ref{fig6}). The maser spots are
distributed over 220~mas~$\times$~130~mas, corresponding to 970~AU~$\times$~590~AU.
These maser spots coincide well with those labeled A-D in the ATCA image
(IRAS~18032$-$2137; Walsh et al. 1998). The 6.7~GHz methanol masers are located
at the center of the continuum emission and the foot-point of the $^{12}$CO\,($J = 2-1$) outflow
(where the $^{12}$CO\,($J = 2-1$) and 1.2~mm continuum images were obtained by the SMA; Longmore et al. 2011).
The $^{12}$CO\,($1-0$) and $^{12}$CO\,($2-1$) line profiles show prominent infall signatures (Ren et al. 2012).
The methanol maser is associated with the weaker of two submillimeter continuum cores known
to exist in this region (Walsh et al. 2003).

{\it 008.83$-$00.02}

This source locates at a distance of 5.2~kpc (Green \& McClure-Griffiths 2011).
The morphology is {\it Elliptical} but displays no simple velocity gradient (Figure~\ref{fig7}).
The maser spots are distributed over 270~mas~$\times$~190~mas, corresponding to 1400~AU~$\times$~990~AU.

{\it 009.61+00.19}

Trigonometric parallax measurements established the distance of this source as 5.2~kpc \citep{2009ApJ...706..464S}.
Various H\emissiontype{II} regions at different evolutionary phases exist in this region, labeled A$-$E by \citet{1993ApJ...418..368G}.
Source 009.61$+$00.19 is associated with D, the ultracompact H\emissiontype{II} region.
The other component E, located \timeform{10"} north of D,
is a candidate hyper-compact H\emissiontype{II} region \citep{2002RMxAC..12...16K}, which
is well known as the strongest 6.7~GHz methanol maser, G\,9.621$+$0.196.
Within this complex region, we imaged only 009.61$+$00.19,
although the strongest emission in the peak total-power spectrum arises from G\,9.621$+$0.196.
The EAVN image reveals a \textit{Linear} distribution of the methanol maser spots,
oriented toward the east-west direction with a linear velocity gradient (Figure~\ref{fig8}). 
The scale distributed over 100~mas~$\times$~20~mas,
corresponding to 490~AU~$\times$~120~AU at the source. Only two spots appear in the ATCA image
(IRAS~18032$-$2032, Walsh et al.\ 1998; \cite{1998MNRAS.300.1131P}).

{\it 009.98$-$00.02}

This source locates at a distance of 12.0~kpc \citep{2011MNRAS.417.2500G}.
The 6.7~GHz methanol maser spots show a \textit{Complex} spatial distribution
with no obvious trend in the velocity distribution (Figure~\ref{fig9}). The scale coverage is 250~mas~$\times$~130~mas,
corresponding to 3000~AU~$\times$~1600~AU at the source.
The maser spots correspond to those labeled A$-$F in the ATCA image (IRAS~18048$-$2019, Walsh et al. 1998).
We detected new maser spots at $V_{\mathrm{lsr}} = 49.98-50.68$~km~s$^{-1}$,
located at ($\Delta \alpha,~\Delta \delta $) = ($+$130, $+$45) and ($+$180, $-$60)~mas
from the brightest spot at the origin of the image.
The flux density of these new spots is 1.5~Jy, which is similar to that observed
in the single-dish spectrum obtained by Parkes 64~m \citep{2010MNRAS.409..913G}.

{\it 010.32$-$00.16}

The distance of this source is assumed as 2.39~kpc,
the measured distance of W31~North \citep{2011MNRAS.411..705M},
with which it is associated.
It should be noted that this source distance is disputed among
the literatures (\cite{2011MNRAS.417.2500G}, references therein).
The 6.7~GHz methanol maser spots show \textit{Complex} spatial distribution over
200~mas~$\times$~260~mas, corresponding to 490~AU~$\times$~620~AU at the source (Figure~\ref{fig10}).
The maser spots clustered in the northern part of the image correspond
to those labeled E$-$H in the ATCA image (IRAS~18060$-$2005, Walsh et al.\ 1998).
We detected new maser spots at $V_{\mathrm{lsr}} = 4.14 - 6.95$~km~s$^{-1}$,
located 220~mas south from the main cluster.
The origin of the image is not coincide with the position of the strongest component,
but a relatively strong, compact component at  $V_{\mathrm{lsr}} = 6.3$~km~s$^{-1}$ is selected for the origin of the image.

{\it 011.49$-$01.48}

This source displays \textit{Complex} spatial and velocity distributions of its methanol masers (Figure~\ref{fig11}).
The overall distribution is elongated along the north-south direction over 180~mas~$\times$~440~mas,
corresponding to 290~AU~$\times$~710~AU at a kinematic distance of 1.6~kpc.
These maser spots coincide with those labeled A$-$M in the ATCA image
(IRAS~18134$-$1942, Walsh et al. 1998).
New maser spots at $V_{\mathrm{lsr}} = 6.40 - 6.76$ and 6.93$-$7.63~km~s$^{-1}$ are detected
at ($\Delta \alpha,~\Delta \delta $) = ($+$18, $+$409)
and ($-$40, $+$360)~mas from the brightest spot, respectively.

{\it 011.90$-$00.14}

The observed maser spots of this source display a \textit{Paired} distribution (Figure~\ref{fig12}),
dividing into two spatially discrete clusters separated by 340~mas,
corresponding to 1300~AU at a near kinematic distance of 4.0~kpc \citep{2011MNRAS.417.2500G}.

{\it 012.02$-$00.03}

This source locates at a distance of 11.1~kpc (Green \& McClure-Griffiths 2011).
The maser spots show an {\it Arched} spatial distribution over 50~mas~$\times$~110~mas,
corresponding to 520~AU~$\times$~1200~AU (Figure~\ref{fig13}). A single maser spot was detected in the ATCA image
(IRAS~18090$-$1832; Walsh et al. 1998), associated with the denser of two submillimeter
continuum cores known to exist in this region (Walsh et al. 2003).

{\it 012.68$-$00.18}

This source, associated with W33B, is located at 2.40~kpc,
determined from trigonometric parallax measurements of 22~GHz water masers (Immer et al. 2013).
Although the maser spots are roughly distributed along the northwest-southeast direction,
we classified the morphology of this source as {\it Complex} (Figure~\ref{fig14}).
The masers are distributed over 130~mas~$\times$~270~mas, corresponding to 310~AU~$\times$~650~AU.
The associated submillimeter continuum emission is extended along the northeast-southwest direction,
perpendicular to the methanol maser distribution (Walsh et al. 2003).

{\it 012.88+00.48}

This source locates at the trigonometric parallax distance of 2.34~kpc \citep{2011ApJ...733...25X}.
The flux of its methanol maser is known to show a periodic variation with
 a period of 29.5~days \citep{2009MNRAS.398..995G}.
The 6.7~GHz methanol maser spots are distributed as a main cluster and two separated clusters,
constituting \textit{Complex} morphology (Figure~\ref{fig15}). The main cluster is seen in the ATCA image
(IRAS~18089$-$1732, Walsh et al. 1998) as A$-$E, H, and J. The cluster located at
1.5 arcsec in the northeast direction corresponds to spot F in the ATCA image.
While spot G in the ATCA image is absent, EAVN detected new maser spots at 1.9 arcsec
northeast from the main cluster, whose $V_{\mathrm{lsr}} = 32.53 - 32.88$~km~s$^{-1}$.
Additional new maser spots, with $V_{\mathrm{lsr}} = 34.29 - 34.99$~km~s$^{-1}$, appear around the origin.
The masers are distributed over 1100~mas~$\times$~1600~mas, corresponding to 2600~AU~$\times$~3800~AU.

{\it 014.10+00.08}

This source locates at a distance of 5.4~kpc (Green \& McClure-Griffiths 2011).
As mentioned in Section 3, 014.10$+$00.08 was detected only by a short baseline
between Mizusawa$-$Hitachi. Therefore, an image of this source is not made.

{\it 020.23+00.06}

This source locates at a kinematic distance of 4.4~kpc \citep{2011MNRAS.417.2500G}.
Two isolated maser clusters are separated by 8.5~arcsec (equivalently, by  $3.7 \times 10^{4}$~AU, Figure~\ref{fig16}).
The maser spots in each cluster are probably excited by different high-mass YSOs.
In the southwestern maser cluster (SW), two isolated clusters are separated by 70~mas (320~AU)
at the source. Therefore, the morphology of this cluster is {\it Paired}.
The velocities of maser spots in the northeastern maser cluster (NE) vary widely (by 11~km~s$^{-1}$),
although the spatial coverage is only 100~AU. The morphology of this cluster is {\it Arched}.

{\it 023.43$-$00.18}

Trigonometric parallax measurements of 12~GHz methanol masers identified this
source at 5.9~kpc (Brunthaler et al. 2009). The masers form two clusters separated
by approximately 14 arcsec, corresponding to $8.2 \times 10^{4}$~AU (Figure~\ref{fig17}). Two millimeter dust
continuum cores have been reported to the north (MM1) and south (MM2) of this region (Ren et al. 2011).
Each methanol maser cluster is associated with each mm dust core,
suggesting that these separated clusters locate in different high-mass star-forming regions.
Both maser clusters appear in the ATCA image (IRAS~18319$-$0834; Walsh et al. 1998).
The maser spots in the MM2 and MM1 regions coincide with those labeled C$-$F and J$-$L
in the ATCA image, respectively. The morphology of the northern cluster MM1 is {\it Complex}.
The maser spots are distributed over 18~mas~$\times$~18~mas, corresponding to 110~AU~$\times$~110~AU.
The southern cluster MM2 is {\it Paired}. The masers in this cluster are distributed over
280~mas~$\times$~90~mas (1600~AU~$\times$~550~AU).

{\it 025.65+01.05}

This source locates at a kinematic distance of 12.5~kpc (Green \& McClure-Griffiths 2011).
The morphology of the methanol maser is {\it Linear} (Figure~\ref{fig18}). The masers are distributed over 7~mas~$\times$~2~mas,
corresponding to 90~AU~$\times$~20~AU. The radial velocity width is quite narrow (less than 1~km~s$^{-1}$).
The ATCA image is elongated approximately 1 arcsec
in the north-south direction (IRAS 18316$-$0602; Walsh et al. 1998).
The spots in the VLBI image coincide with spot B in the ATCA image.

{\it 025.71+00.04}

This source locates at a kinematic distance of 11.8~kpc (Pandian et al. 2008).
The morphology is {\it Complex} (Figure~\ref{fig19}). The maser spots are distributed over 80~mas~$\times$~130~mas,
corresponding to 930~AU~$\times$~1500~AU. These spots coincide with those labeled C, D,
and J in the ATCA image (IRAS~18353$-$0628; Walsh et al. 1998).

{\it 025.82$-$00.17}

This source locates at a kinematic distance of 5.0~kpc \citep{2011MNRAS.417.2500G}.
The 6.7~GHz methanol maser spots are distributed in an \textit{Elliptical} morphology
with a clockwise velocity gradient (Figure~\ref{fig20}).
The scale is 300~mas~$\times$~300~mas, corresponding to 1500~AU~$\times$~1500~AU at the source.
The EAVN and ATCA images of this source are similarly structured (IRAS~18361$-$0627, Walsh et al. 1998).

{\it 028.83$-$00.25}

This source locates at a kinematic distance of 4.6~kpc (Green \& McClure-Griffiths 2011).
The spatial distribution is {\it Arched} (Figure~\ref{fig21}). The maser spots are distributed over 370~mas~$\times$~440~mas,
corresponding to 1700~AU~$\times$~2000~AU. These spots coincide with those labeled B, D, and G
in the ATCA image (IRAS~18421$-$0348; Walsh et al. 1998).
The EAVN image is similar to the JVLA image presented in Cyganowski et al. (2009).

{\it 029.86$-$00.04}

The target source 029.86$-$00.04 is adjacent ($\sim$7~arcmin) to two H\emissiontype{II} regions
designated G\,029.95$-$00.01 and G\,029.97$-$00.04.
From H\emissiontype{I}SA and the formaldehyde absorption line spectrum,
the distances of these H\emissiontype{II} regions have been measured as 9.3 and 9.2~kpc,
respectively (\cite{2009ApJ...690..706A}; \cite{1980A&AS...40..379D}).
Thus, we assume a distance of 9.3~kpc for source 029.86$-$00.04.
The 6.7~GHz methanol maser spots reveal an \textit{Arched} distribution accompanied
by a clear velocity gradient (Figure~\ref{fig22}). The scale of the distribution is 150~mas~$\times$~340~mas,
corresponding to 1400~AU~$\times$~3200~AU at the source.

{\it 030.70$-$00.06}

This source locates in a star-forming region known as the W43 main complex (Motte et al. 2003).
The masers are associated with the millimeter dust continuum MM2 (Motte et al. 2003),
one of the most massive and luminous cores in the complex.
Thus, we assume that 030.70$-$00.06 is located at 5.9~kpc, the measured distance of W43.
The morphology of this source is {\it Paired} (Figure~\ref{fig23}). The maser spots are distributed over 690~mas~$\times$~420~mas,
corresponding to 4100~AU~$\times$~2500~AU.

{\it 030.76$-$00.05}

The kinematic distance of this source is 4.8~kpc \citep{2009ApJ...690..706A}.
The 6.7~GHz methanol maser spots form two distinct clusters isolated by 34~mas,
corresponding to 160~AU; therefore, the morphology of this source is {\it Paired} (Figure~\ref{fig24}).
The maser spots coincide with those labeled D$-$F in the ATCA image (IRAS~18450$-$0200, Walsh et al. 1998).

{\it 030.91+00.14}

This source locates at a kinematic distance of 5.6~kpc (Anderson \& Bania 2009).
The morphology is {\it Linear} with a clear continuous velocity gradient (Figure~\ref{fig25}).
The maser spots are distributed over 30~mas~$\times$~60~mas,
corresponding to 150~AU~$\times$~320~AU, and they are associated
with submillimeter continuum emission (Hill et al. 2005).

{\it 031.28+00.06}

The distance of this source is 5.8~kpc \citep{2009ApJ...690..706A}.
The 6.7~GHz methanol maser spots form a \textit{Complex}  distribution
over 560~mas~$\times$~470~mas, corresponding to 3300~AU~$\times$~2700~AU at the source (Figure~\ref{fig26}).
The maser spots in the VLBI image correspond to those labeled A-K
in the ATCA image (IRAS~18456$-$0129, Walsh et al. 1998).
A VLBI image of this source was also obtained by EVN \citep{2000A&A...362.1093M}.
The overall distributions in the EAVN and EVN images are closely matched.
However, the maser spot at $V_{\mathrm{lsr}} = 109.42$ and 109.87~km~s$^{-1}$ in the EVN image
is absent in the EAVN image, while new maser spots appear at
$V_{\mathrm{lsr}} = 104.34$ and 104.52~km~s$^{-1}$, locating at ($\Delta \alpha,~\Delta \delta $) = ($-$68, $+$215)
from the brightest spot.

{\it 032.03+00.06}

From the NIR extinction-distance relationship, the distance of this source
has been estimated as 7.2~kpc (Stead \& Hoare 2010). The morphology of this source is {\it Paired} (Figure~\ref{fig27}).
The maser spots are distributed over 290~mas~$\times$~430~mas, corresponding to 2100~AU~$\times$~3100~AU.
The masers are associated with millimeter dust continuum and the 4.5~$\mu$m extended
emission observed by Rathborne et al. (2006) and Chambers et al. (2009).

{\it 037.40+01.52} 

The kinematic distance of this source is 2.1~kpc (Fontani et al. 2011).
The morphology is {\it Linear} with a clear velocity gradient (Figure~\ref{fig28}).
The maser spots are distributed over 30~mas~$\times$~10~mas,
corresponding to 60~AU~$\times$~20~AU. The linear spatial distribution, the size, and the velocity gradient
are similar to those of NGC 7538 IRS1 with a linear size of 90~AU (\cite{1998A&A...336L...5M}).

{\it 049.49$-$00.38}

This source locates in the well-known complex high-mass star-forming region W51.
From two reported trigonometric parallax distances, namely 5.1$^{+2.9}_{-1.4}$ \citep{2009ApJ...693..413X}
and 5.41$^{+0.31}_{-0.28}$ \citep{2010ApJ...720.1055S}, we assign a distance of 5.41~kpc to 049.49$-$00.38.
Maser emissions have been reported at several sites within an area of 2~arcmin
enclosing regions W51 Main and IRS2, but here we concentrate
on the target source 049.49$-$00.38 in W51 main.
The 6.7~GHz methanol maser spots are distributed in a \textit{Complex} morphology
over 370~mas~$\times$~310~mas, corresponding to 2000~AU~$\times$~1700~AU at the source (Figure~\ref{fig29}).
From MERLIN data, \citet{2012MNRAS.423..647E} showed that the 6.7~GHz methanol maser spots
of this source are distributed over an area of $3 \times 2$~arcsec$^{2}$.
The southern spots detected by MERLIN were absent in the EAVN image.
The distribution of maser spots of this source was also obtained
by EVN (\cite{2005ASPC..340..342P}; \cite{2012A&A...541A..47S})
and found to be consistent with that in the EAVN image.

{\it 232.62+00.99} 

This source locates at a distance of 1.68~kpc, estimated from trigonometric
parallax measurements of the 12~GHz methanol maser (Reid et al. 2009).
The morphology of this source is {\it Paired} (Figure~\ref{fig30}).
The maser spots are distributed over 20~mas~$\times$~70~mas, corresponding to 40~AU~$\times$~120~AU.
These spots coincide with the spots labeled B and C in the ATCA image (IRAS 07299$-$1651; Walsh et al. 1998).

{\it 351.77$-$00.53}

The distance of this source is 0.4~kpc with a large uncertainty (Green \& McClure-Griffiths 2011).
The morphology of this source is {\it Elliptical} (although the south-western section is missing in the image)
with a counter-clockwise velocity gradient (Figure~\ref{fig31}). The maser spots are distributed over
100~mas~$\times$~100~mas (40~AU~$\times$~40~AU) and coincide with those labeled A$-$D in the ATCA image
(IRAS~17233$-$3606; Walsh et al. 1998). A strong $^{12}$CO bipolar outflow and OH maser
outflow have been reported in this region (Leurini et al. 2008).

{\it 352.63$-$01.06}

This source locates at a kinematic distance of 0.9~kpc, with a large uncertainty
(Green \& McClure-Griffiths 2011). The morphology is {\it Linear}, but the velocity distribution
is complex (Figure~\ref{fig32}). The maser spots are distributed over 370~mas~$\times$~290~mas,
corresponding to 330~AU~$\times$~260~AU. These maser spots coincide with those
labeled D and F in the ATCA image (IRAS~17278$-$3541; Walsh et al. 1998).
New maser spots are detected at ($\Delta \alpha$, $\Delta \delta$) = ($-$210, $-$190)
from the brightest maser spot.
The 6.7~GHz methanol masers locate at a peak of the 1.2~mm dust continuum emission (Fa\'{u}ndez et al. 2004).

{\it 353.41$-$00.36} 

The kinematic distance of this source is 3.8~kpc (Caswell et al. 2011).
The morphology is {\it Complex} (Figure~\ref{fig33}). The maser spots are distributed over 60~mas~$\times$~120~mas,
corresponding to 220~AU~$\times$~460~AU, and they coincide with those labeled A$-$C
in the ATCA image (IRAS~17271$-$3439; Walsh et al. 1998).
The 6.7~GHz methanol maser is associated with a peak of the millimeter
dust continuum and centimeter radio continuum sources (Garay et al. 2006, 2007).
From radio continuum observations, the exciting star is thought to be of spectral
type O9.5. SiO and CH$_{3}$CCH lines have also been detected in this region
(Miettinen et al. 2006).

{\it 354.61+00.47}

This source locates at a kinematic distance of 3.8~kpc (Green \& McClure-Griffiths 2011).
The morphology is {\it Complex},
formed by three clusters (Figure~\ref{fig34}). The scale of the maser distribution is
380~mas~$\times$~440~mas, corresponding to 1500~AU~$\times$~1700~AU.
The 6.7~GHz methanol masers locate close to a peak of the
millimeter dust continuum emission (Fa\'{u}ndez et al. 2004).

{\it 359.43-00.10}

This maser source locates at a kinematic distance of 8.2~kpc,
and the apparent position is approximately 1~arcmin east of Sgr~C (\cite{2003ApJ...587..701F}).
The morphology is {\it Paired} (Figure~\ref{fig35}). The maser spots are distributed
over 140~mas~$\times$~50~mas, corresponding to 1200~AU~$\times$~370~AU.


\section{Discussions}\label{section:discussion}

\subsection{Spatial Distribution of Maser emissions}\label{subsection:5-1}

The distributions of the 6.7~GHz methanol maser spots vary widely in size and structure
similar to the results of previous studies by Bartkiewicz et al. (2009) and Pandian et al. (2011).
The whole size of the spatial distribution is from 9 to 4900~AU.
The 6.7~GHz methanol maser is excited by infrared radiation when its dust
temperature is 100$-$200~K (Cragg et al. 2005).
The spatial scale of the maser distribution can be estimated from
the luminosity of the emitting source. Assuming a luminosity from $10^{3}$ to $10^{5}$~$L_{\solar}$
and a suitable dust temperature of 100~K, the maser is predicted to appear at 500$-$5000~AU
from the excited star, consistent with observation.

On the other hand, the maser distribution of some sources deviate from their expected range.
The three sources 000.54$-$00.85~NW/SE, 020.23+00.06~SW/NE, and 023.43-00.18~MM1/2
cannot be attributed to excitation of a single central source of luminosity  $10^{3}$ to $10^{5}$~$L_{\solar}$
because the separation of their masers exceeds $10^{4}$~AU, indicating
that each source likely involves at least two exciting sources. In the following discussion,
we assume that these largely separated clusters associate with individual exciting sources,
i.e., these are independent sources. Hence we use 38 as the number of the imaged sources, hereafter.

The spatial scale of the maser distribution is plotted as a function of observed
radial velocity range in Figure 37. The spatial scale of each source is defined
as the largest extent of spot distribution in the celestial sphere, while the
velocity range is the difference between the smallest and largest velocities of
the maser spots detected by VLBI (see Table 3). Open circles denote sources with
{\it Linear} morphology, and the filled circles represent other morphologies
({\it Elliptical}, {\it Arched}, {\it Paired}, and {\it Complex}).
A positive correlation is observed between the spatial
scale and radial velocity range, as previously reported by Pandian et al. (2011).
Note, however, that sources of {\it Linear} morphology are the main contributors to this correlation.
The spatial distribution of all {\it Linearly} distributed sources is below 600~AU,
and their radial velocity range is smaller than 4~km~s$^{-1}$, below those of other sources.
If the {\it Linearly} distributed sources are excluded from Figure 37, the correlation between
size and radial velocity range becomes much weaker. This tendency toward small radial
velocity range and spatial extent in sources of {\it Linear} distribution has been similarly
seen in the result by Pandian et al. (2011). As discussed in the following subsection,
we consider that sources of such restricted size and velocity range are the observable
parts of larger structures that are not completely revealed.

\begin{figure}[t]
\begin{center}
\includegraphics[width=77mm,clip]{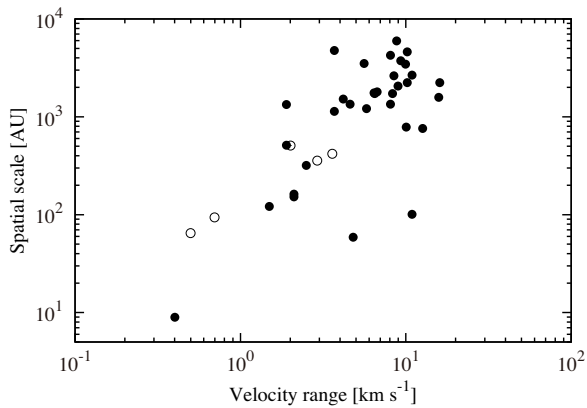}
\end{center}
\caption{{\footnotesize
Spatial scale of maser distribution
as a function of velocity range.
The open circles indicate sources 
with {\it Linear} morphology (see text for detail).
}}
\label{fig5-1}
\end{figure}

\subsection{Morphology and the spreading structure of each spot}\label{subsection:5-2}

Table 4 summarizes the number of sources satisfying each morphology,
as absolute values and as fractions of the total number of sources.
For comparison, 27 sources observed by EVN and classified into the five types are also listed
(these have been previously reported in \cite{2009A&A...502..155B}).
The number of {\it Elliptical} sources, which is expected to trace the
gas disk around YSOs, is six (16{\%}), small as compared with that (33{\%})
in \citet{2009A&A...502..155B}. Our samples contain relatively high numbers of
{\it Paired} (21{\%}) and {\it Complex} (39{\%}) distributions, which contain no apparent
systematic spatial distribution or radial velocity structures.

\begin{figure}[t]
\begin{center}
\includegraphics[width=77mm,clip]{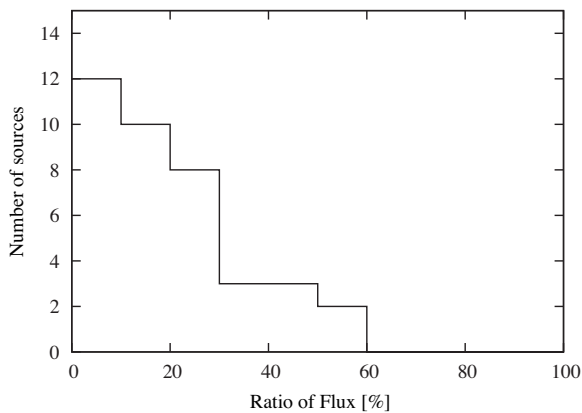}
\end{center}
\caption{{\footnotesize
Histogram of the flux ratio of the correlated to the total-power spectrum.
Bins on the horizontal axis are separated by 10{\%}.
}}
\label{fig5-2}
\end{figure}

\begin{figure}[t]
\begin{center}
\includegraphics[width=77mm,clip]{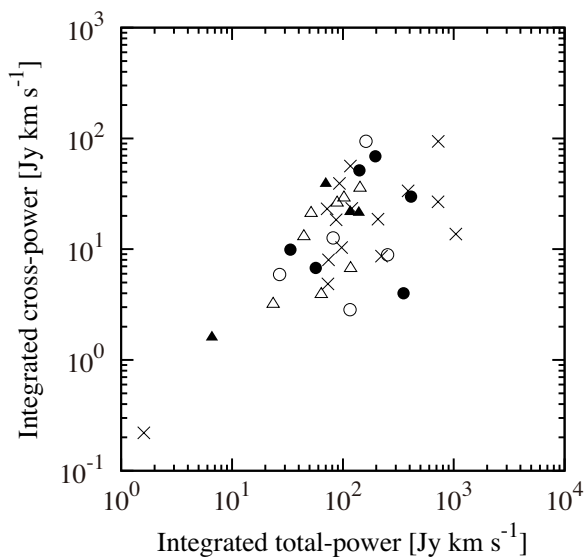}
\end{center}
\caption{{\footnotesize

Distribution of total-power versus correlated flux.
Different symbols denote each morphology: 
The filled circles indicate sources with {\it Elliptical} morphology,
open circles are {\it Linear},
filled triangles are {\it Arched},
open triangles are {\it Paired},
cross are {\it Complex}, respectively.
}}
\label{fig5-3}
\end{figure}

Since structures of low intensity are not detected by VLBI, sources with {\it Paired} or
{\it Complex} morphology may in fact possess low-brightness {\it Elliptical}, or other systematic structures.
The ratio of the integrated flux of CLEAN components to the total flux
is listed in Table~\ref{table3}. To investigate the fraction of maser radiation detectable by VLBI,
a histogram of these ratios is presented in Figure~\ref{fig5-2}. The detectability is from 1{\%} to 58{\%},
averaging around 20{\%}. Similarly, \citet{2009A&A...502..155B} noted that, for most sources,
10$-$30{\%} of the total flux was detected by EVN. Our result suggests that 80{\%}
of the maser emissions spread spatially into a diffuse structure.
Distribution of total-power versus correlated flux is shown in Figure~\ref{fig5-3}
with different symbols for each morphology.
This is for testing if the detectability differs with the morphology,
but no clear tendency of distribution is seen for different morphology.

\citet{2002A&A...383..614M} mentioned that some of the 6.7~GHz methanol maser
emission consists of compact core and spreading halos with sizes of a few (tens)
astronomical units and a few hundred or larger astronomical units, respectively.
\citet{2011ApJ...730...55P} also reported that some maser spots possess no compact core.
The spatial distribution of such dispersed maser emissions cannot be determined by VLBI observation.
Therefore, as discussed for the {\it Linearly} distributed sources, we consider that the exact spatial
distributions or morphological classes of maser emissions are difficult to deduce from VLBI observations alone.

When interpreting the proper motion of the maser spot and its associated site,
any non-observed parts must be identified. To avoid the resolved-out problem,
the diffuse structure should be observed. Assuming that the emission component
spreads into a halo with a size of 300~AU and the distance to the source is 3~kpc,
the angular size of the spread emission is 0.1~arcsec.
To detect this structure, an interferometric observation of spatial resolution
approximately $\geq 1$~arcsec is required. We have observed 24 sources
among the sources in our sample by ATCA, whose typical spatial resolution is $2.0 \times 1.5$~arcsec$^{2}$.
The observed images, which contain the whole emission of the 6.7~GHz masers,
will assist in VLBI image interpretation and obtaining the internal proper motion
of the maser spots. The ATCA results will be published elsewhere.
High-resolution observations by ALMA will be important for determining the gas
and dust distributions within the sources. In a future study,
these distributions could be compared with the distribution of the masers.

\begin{table}[htbp]
\small
\begin{center}
\caption{Spatial morphologies of EAVN and EVN samples.}\label{table4}
\begin{tabular}{lccccc}
\hline\hline
Array         & \multicolumn{5}{c}{Spatial Morphology} \\ \cline{2-6}
              & Elliptical$^{\ast}$ & Arched    & Linear & Paired & Complex    \\ \hline
EAVN          &   6                 &    4      &    5   &   8    &    15      \\
(percentage)  &  (16)               &   (11)    &  (13)  & (21)   &   (39)     \\
EVN           &   9                 &    3      &    5   &   1    &     9      \\
(percentage)  &  (33)               &  (11)     &  (19)  &  (4)   &   (33)     \\ \hline
\multicolumn{6}{l}{\hbox to 0pt{\parbox{77mm}{\footnotesize
The EVN data are taken from \citet{2009A&A...502..155B}.
Note that 27 sources, classified into the five morphologies,
are included in this table.\\
$\ast$ ``Elliptical" is defined as ``Ring" in \citet{2009A&A...502..155B}.
}\hss}}
\end{tabular}
\end{center}
\end{table}

\subsection{Future Prospect for Proper Motion Measurements}\label{subsection5-3}

Despite being limited by the resolved-out problem, VLBI detects 1$-$58{\%} of the maser emission,
implying that the emission is concentrated in sufficiently compact spots.
The number of detected spots in each source is from 3 to 72, with an average of 29.
In 34 of the sources, the spot number is larger than or equal to 10, which is sufficient for investigating
the three-dimensional velocity field. The relatively long lifetime of the 6.7~GHz maser spot
(Goedhart et al. 2004, Ellingsen 2007) has enabled us to observe spots at three epochs
over 2 years.
The relative position
of the maser spot in each map was measured at an accuracy higher than 0.1~mas,
although in practice, the accuracy depends on the signal-to-noise ratio of each component.
Three repeats of this monitoring observation, planned over two years,
will allow us to determine the internal proper motion to an accuracy of $1 \sigma = 0.03$~mas~yr$^{-1}$.
At a typical distance of 3~kpc, this accuracy of proper motion corresponds to a tangential
velocity of 1.5~km~s$^{-1}$. Since the average radial velocity width of the maser is 6.5~km~s$^{-1}$,
the internal proper motion is detectable at sufficiently high signal-to-noise ratio.
In fact, from these previous JVN observations, we measured the internal proper motions
of the spots for four sources (Matsumoto et al. 2011; Sugiyama et al. 2011, 2013; Sawada-Satoh et al. 2013).


\section{Conclusion}\label{section:conclusion}

To study the associated site of the 6.7~GHz methanol maser and the gas dynamics around high-mass YSOs,
we have monitored the internal proper motion of the maser.
For these purposes, 36 selected sources were studied by multi-epoch VLBI observations using EAVN.
We present 35 VLBI images successfully obtained from the first epoch observation.
Three sources contain two separated star-forming regions in each image,
yielding 38 imaged sources.
The distribution of the detected maser spots was from 9 to 4900~AU and displayed
a range of morphologies. The flux detected by VLBI was 1$-$58{\%} of the total
flux, suggesting that a large fraction of the radiation is dispersed into
an extended structure invisible to VLBI. To investigate the associated site and motion of the maser,
shorter baselines are required to recover
the distribution of this extended emission.
Of the 38 imaged sources, we detected 10 or more spots in 34.
The accuracy of the spot position was approximately 0.1~mas.
Therefore, the internal proper motions could be measured with sufficient accuracy
following 2 years of monitoring observation.
From these results, we can statistically investigate the three-dimensional
velocity field around high-mass YSOs.
Since most of the observed sources are located in the southern hemisphere,
they can be observed with the Atacama Large Millimeter/Submillimeter Array (ALMA) in future.

\bigskip


The authors wish to thank the JVN team
for observational assistance and support.
The JVN project is led by the National Astronomical Observatory
of Japan~(NAOJ) that is a branch of the National Institutes of
Natural Sciences~(NINS), Hokkaido University, Ibaraki University,
The University of Tsukuba,
Gifu University,
Osaka Prefecture University,
Yamaguchi University, and Kagoshima University,
in cooperation with Geospatial Information Authority of Japan~(GSI),
the Japan Aerospace Exploration Agency~(JAXA), and the National
Institute of Information and Communications Technology~(NICT).  
This work was financially supported in part by Grant-in-Aid for
Scientific Research (KAKENHI) from the Japan Society for
the Promotion of Science (JSPS), No. 24340034.
This work is partly supported by China Ministry of Science and Technology
under State Key Development Program for Basic Research (2012CB821800),
the National Natural Science Foundation of China (grants 10625314, 11121062, and 11173046),
the CAS/SAFEA International Partnership Program for Creative Research Teams,
and the Strategic Priority Research Program on Space Science,
the Chinese Academy of Sciences (Grant No. XDA04060700).

\clearpage

\appendix
\section*{Spatial Distribution in Unified Scale}
VLBI images of 38 sources are shown in the figures 40-77 in the same spatial scale
(5310 AU in each RA and Dec) for all sources.
These figures correspond to figures 2-36.
A 500 AU scale bar and the spectrum are shown in the corner of each figure.

\begin{figure*}[htbp]
\begin{center}
\includegraphics[width=160mm,clip]{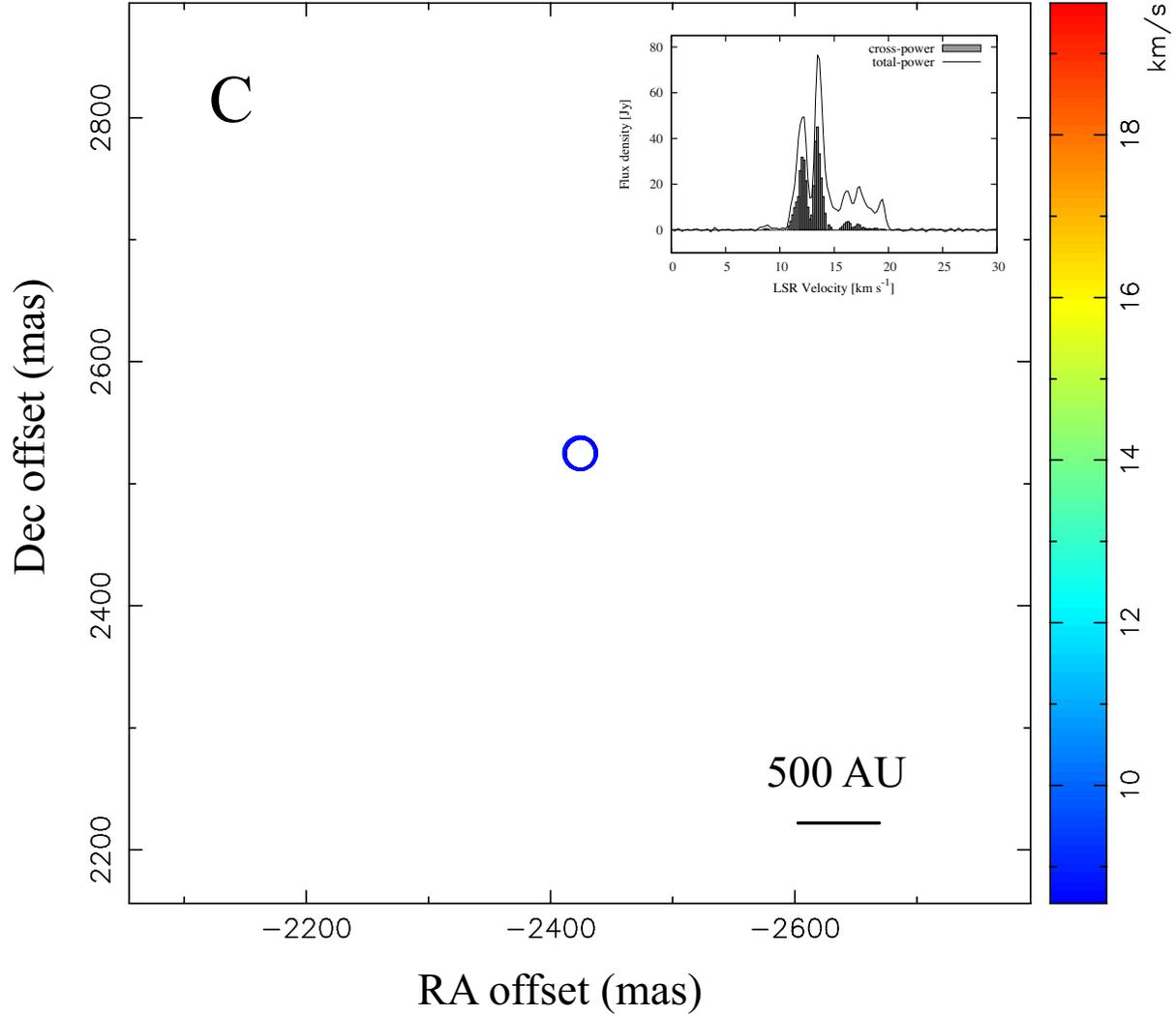}
\end{center}
\caption{
The 6.7~GHz methanol maser emissions of source 000.54$-$00.85~NW.
Spatial distributions of the methanol maser spots obtained
from the EAVN observations. Detail are provided in Section~4 of the text.
\textit{Inset}: Total-power (solid line) and cross-power (hatched box) spectra.
}	
\label{appen-fig1}		
\end{figure*}

\begin{figure*}[htbp]
\begin{center}
\includegraphics[width=160mm,clip]{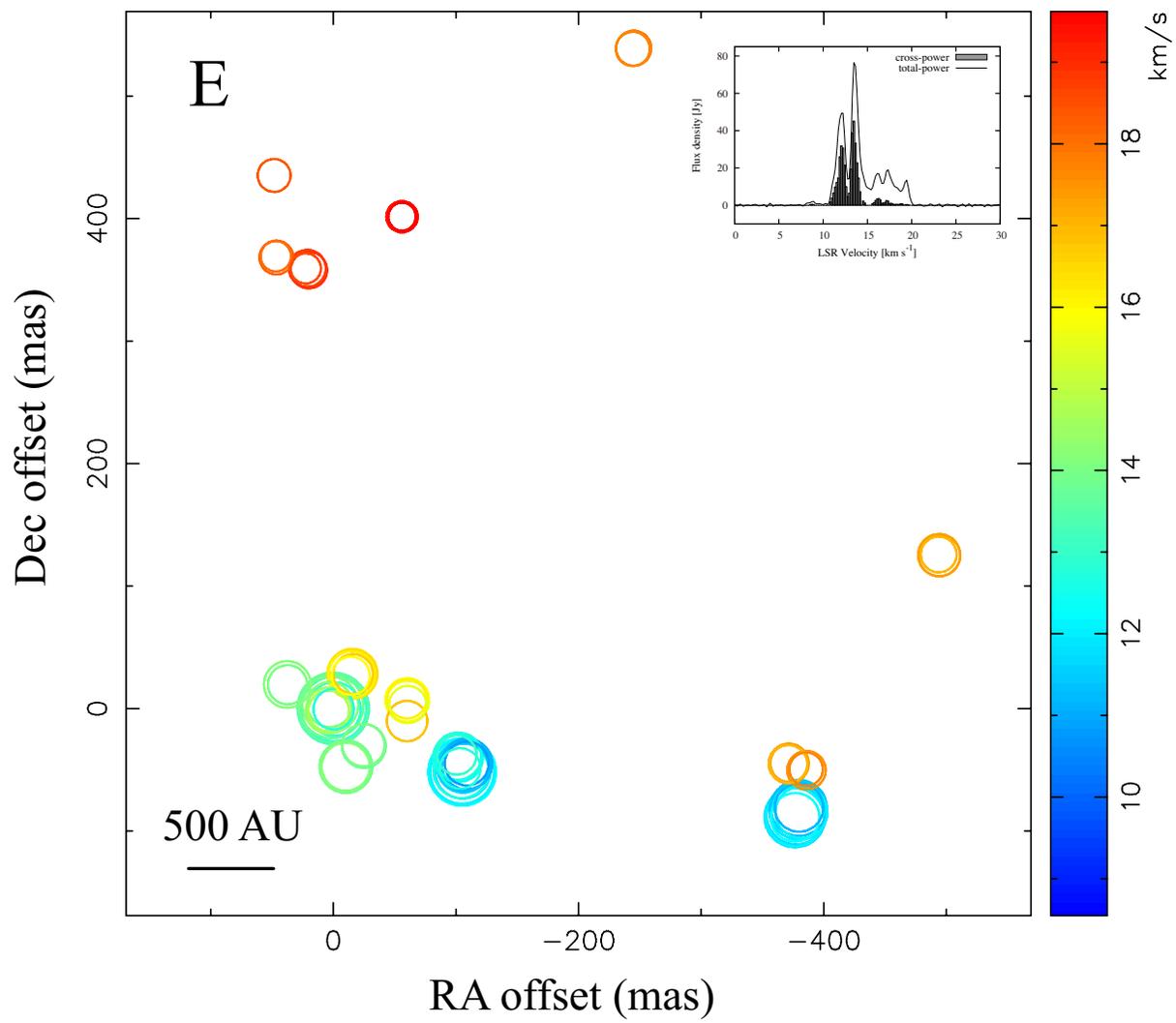}
\end{center}
\caption{
Data for source 000.54$-$00.85~SE, plotted as for Figure~\ref{appen-fig1}.
}	
\label{appen-fig2}		
\end{figure*}

\begin{figure*}[htbp]
\begin{center}
\includegraphics[width=160mm,clip]{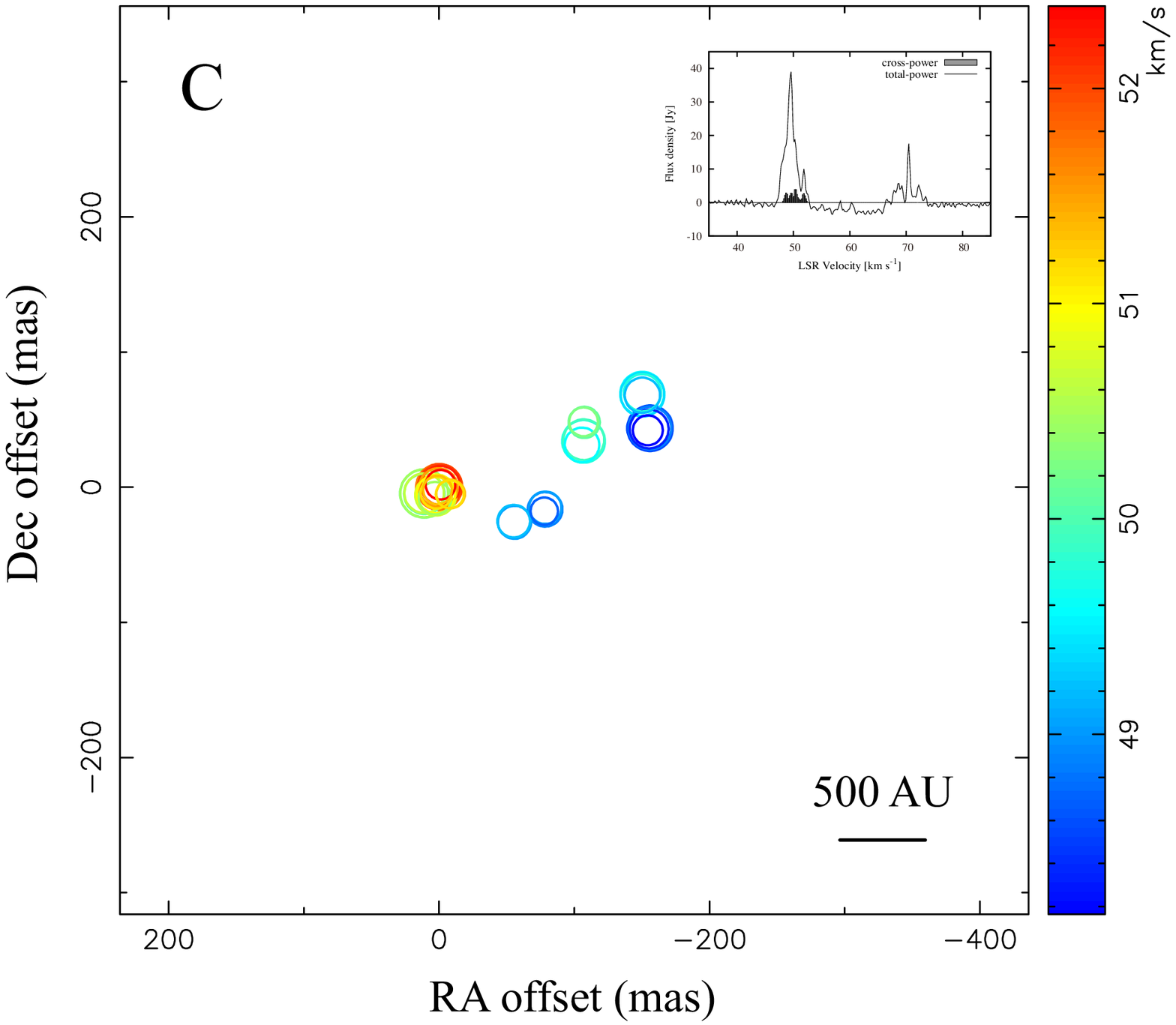}
\end{center}
\caption{
Data for source 000.64$-$00.04, plotted as for Figure~\ref{appen-fig1}.
}	
\label{appen-fig3}		
\end{figure*}

\begin{figure*}[htbp]
\begin{center}
\includegraphics[width=160mm,clip]{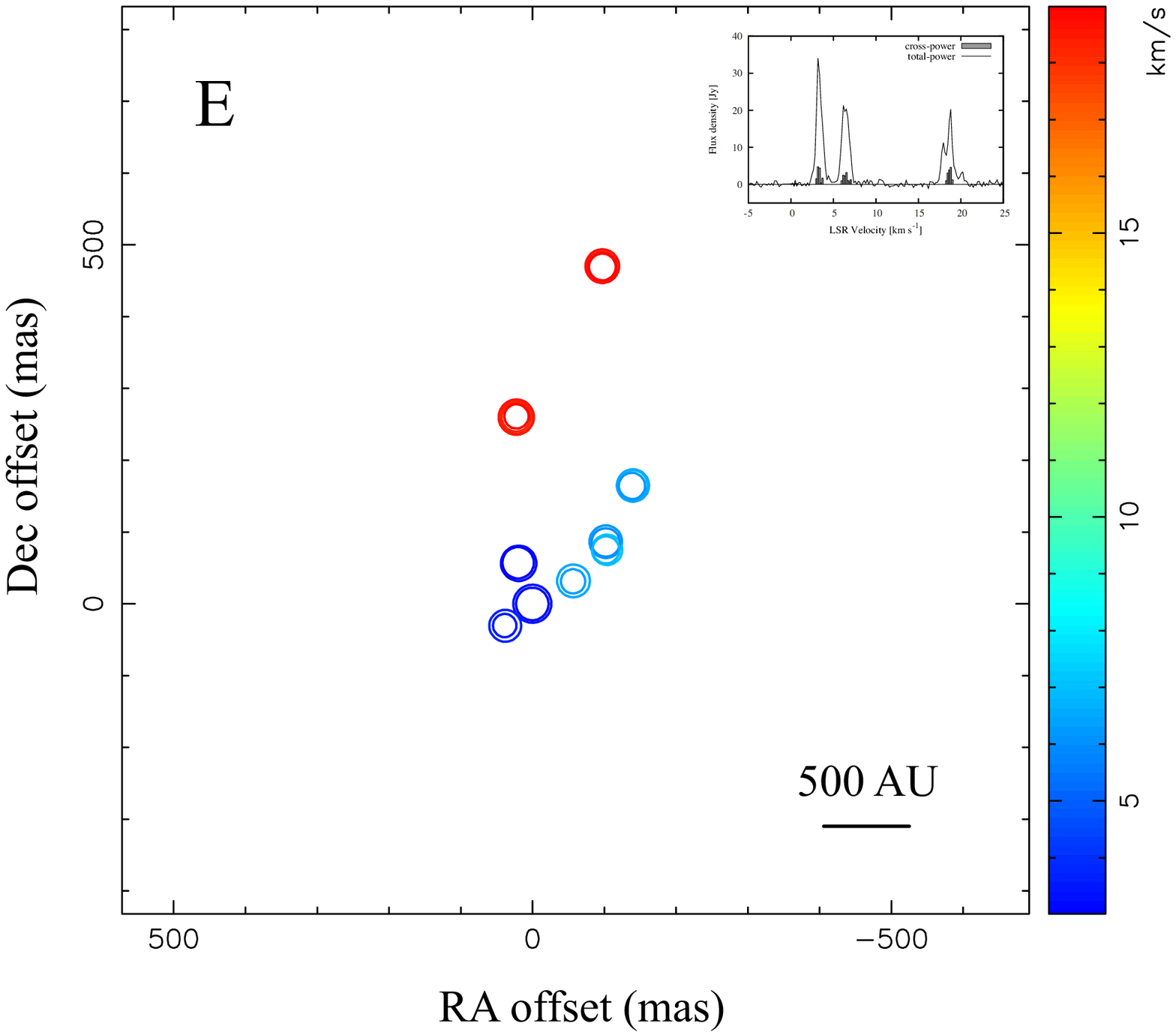}
\end{center}
\caption{
Data for source 002.53$+$00.19, plotted as for Figure~\ref{appen-fig1}.
}	
\label{appen-fig4}		
\end{figure*}

\begin{figure*}[htbp]
\begin{center}
\includegraphics[width=160mm,clip]{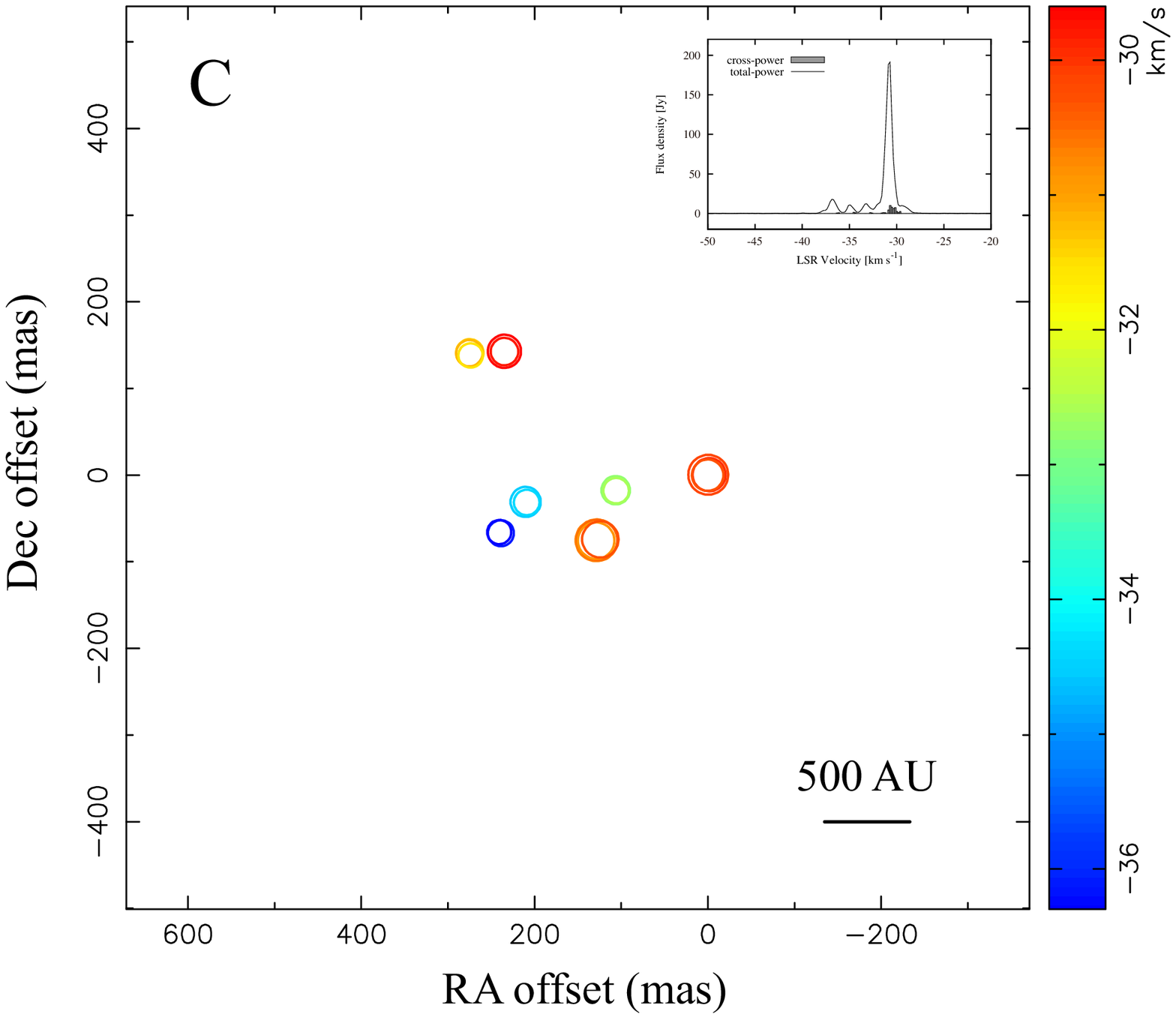}
\end{center}
\caption{
Data for source 006.18$-$00.35, plotted as for Figure~\ref{appen-fig1}.
}	
\label{appen-fig5}		
\end{figure*}

\begin{figure*}[htbp]
\begin{center}
\includegraphics[width=160mm,clip]{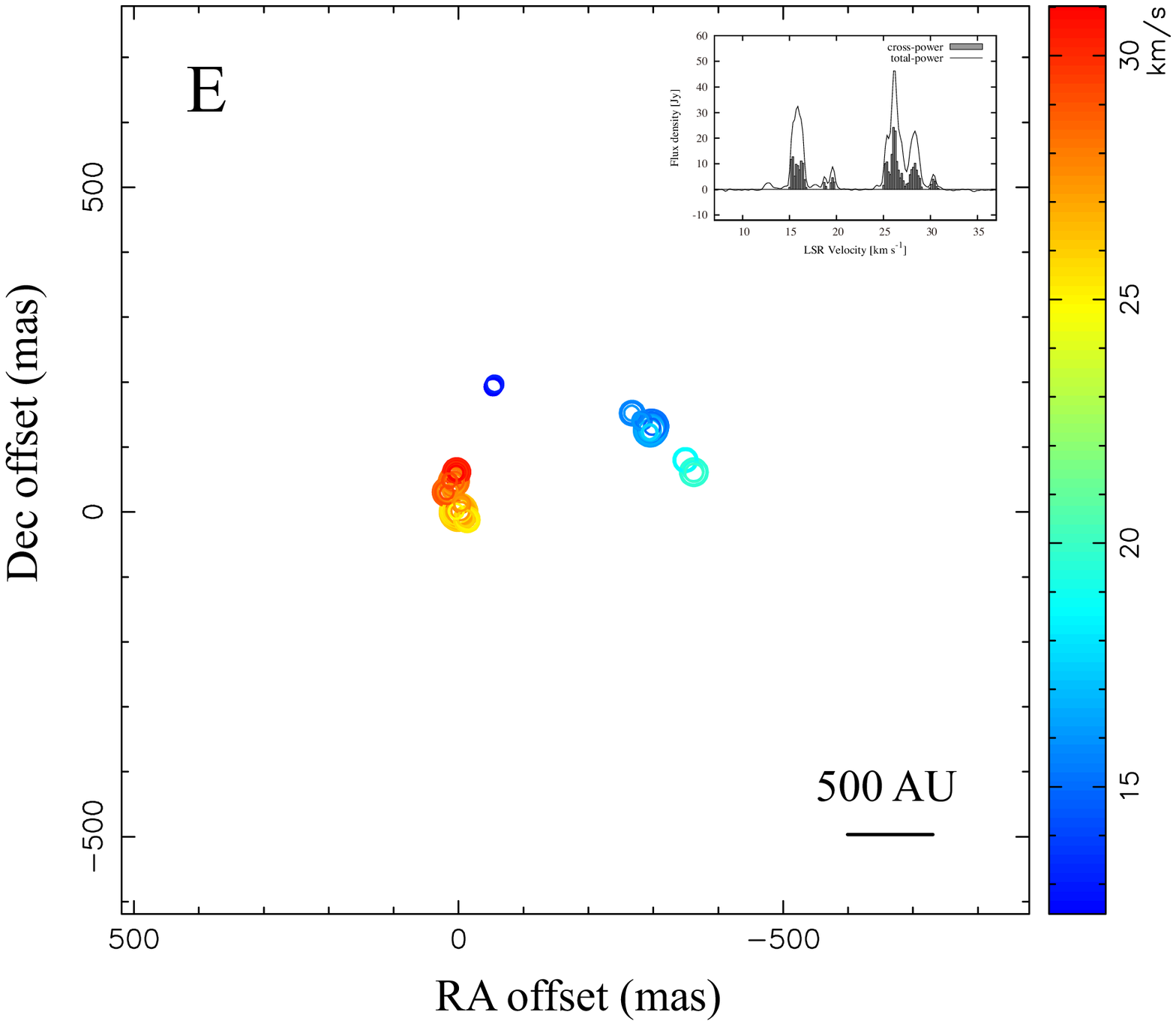}
\end{center}
\caption{
Data for source 006.79$-$00.25, plotted as for Figure~\ref{appen-fig1}.
}	
\label{appen-fig6}		
\end{figure*}

\begin{figure*}[htbp]
\begin{center}
\includegraphics[width=160mm,clip]{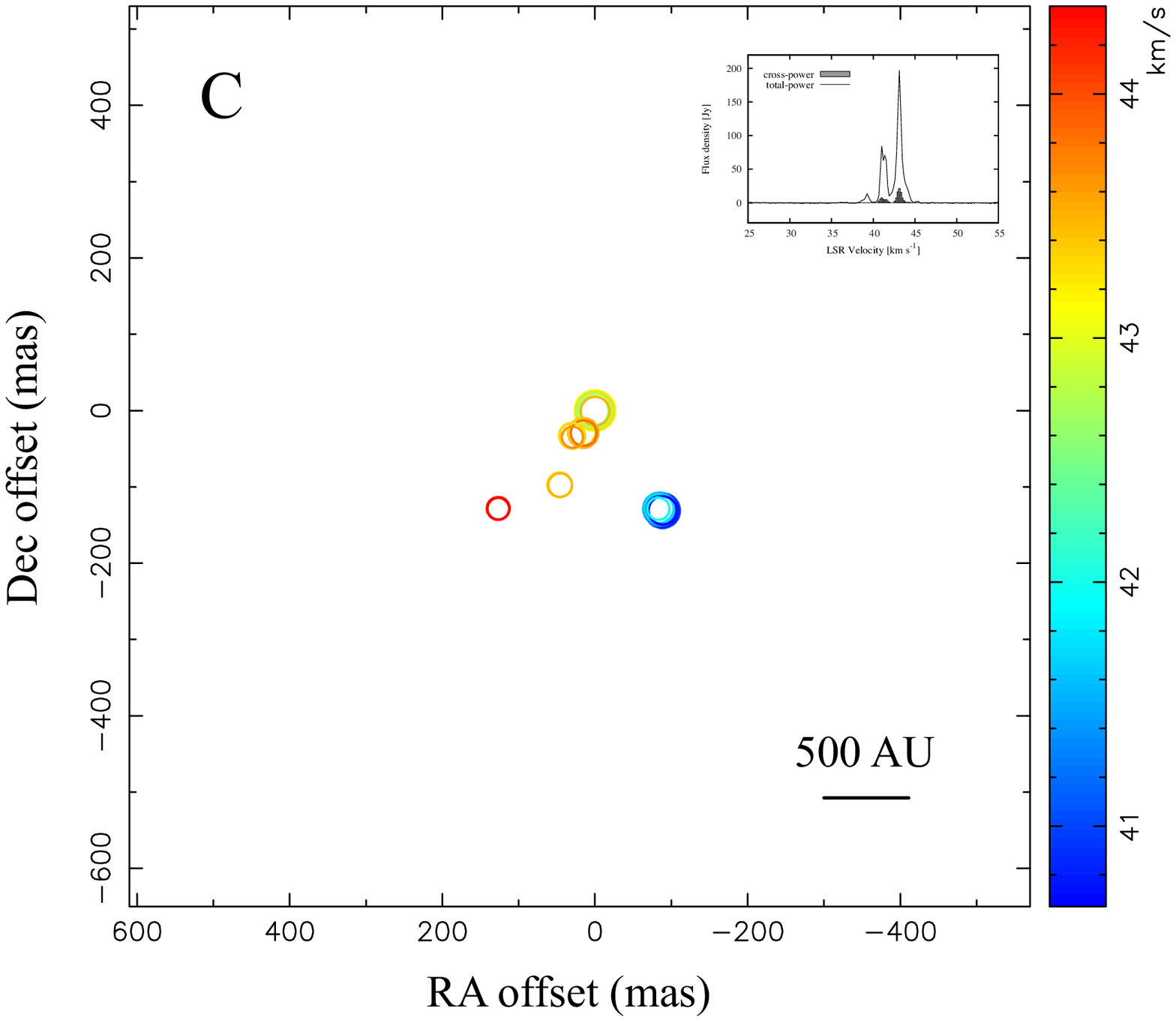}
\end{center}
\caption{
Data for source 008.68$-$00.36, plotted as for Figure~\ref{appen-fig1}.
}	
\label{appen-fig7}		
\end{figure*}

\begin{figure*}[htbp]
\begin{center}
\includegraphics[width=160mm,clip]{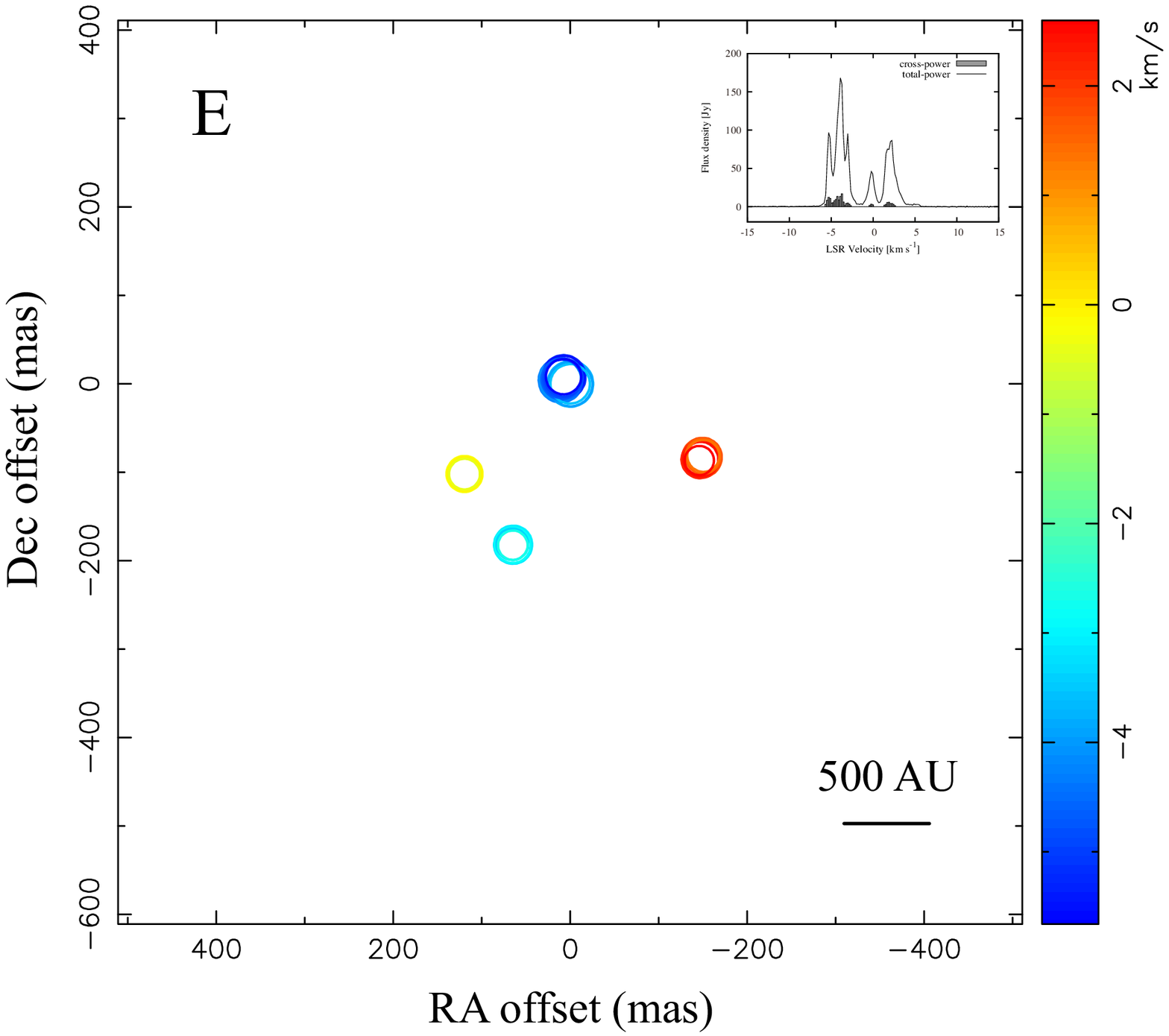}
\end{center}
\caption{
Data for source 008.83$-$00.02, plotted as for Figure~\ref{appen-fig1}.
}	
\label{appen-fig8}		
\end{figure*}

\begin{figure*}[htbp]
\begin{center}
\includegraphics[width=160mm,clip]{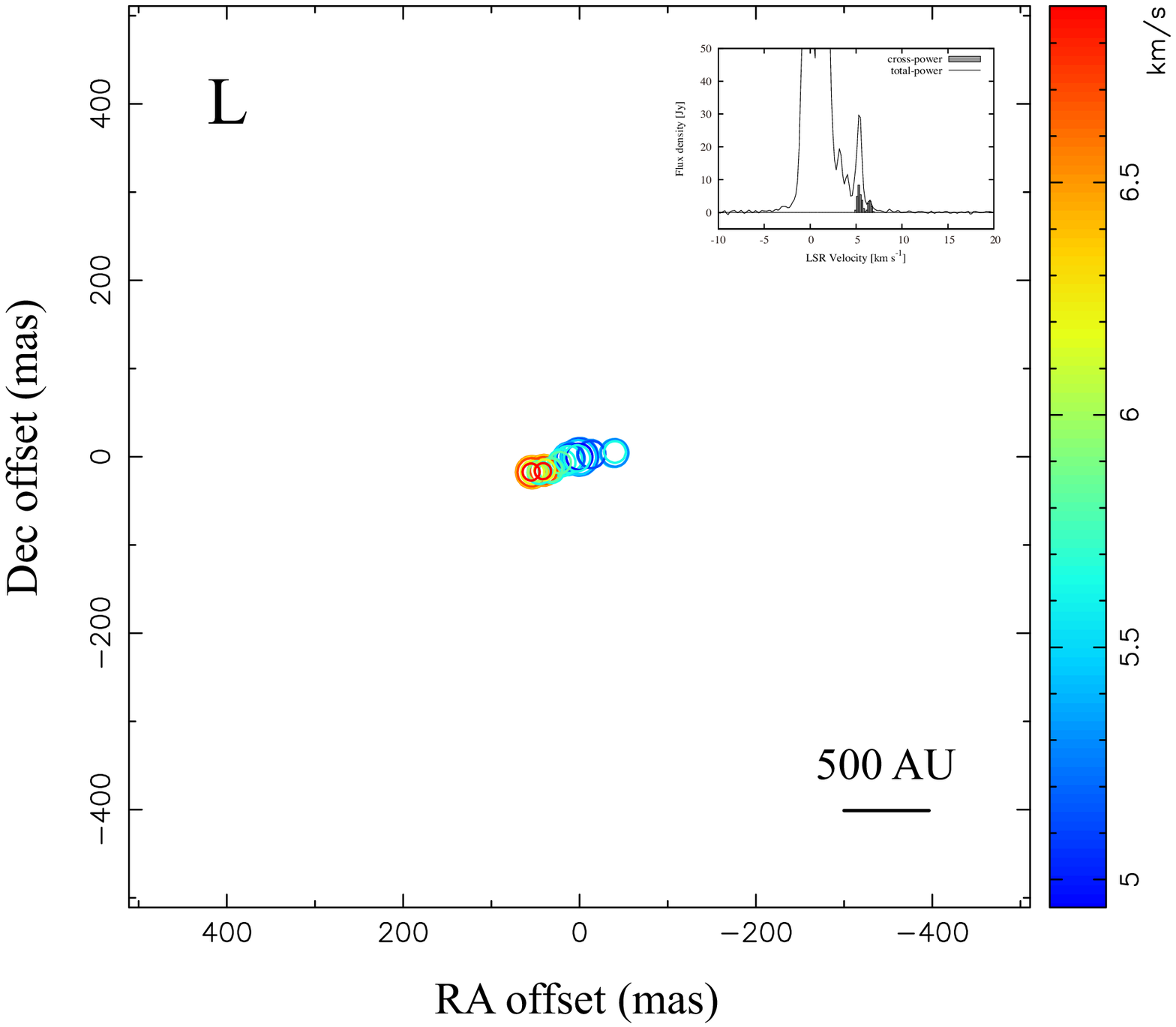}
\end{center}
\caption{
Data for source 009.61$+$00.19, plotted as for Figure~\ref{appen-fig1}.
}	
\label{appen-fig9}		
\end{figure*}

\begin{figure*}[htbp]
\begin{center}
\includegraphics[width=160mm,clip]{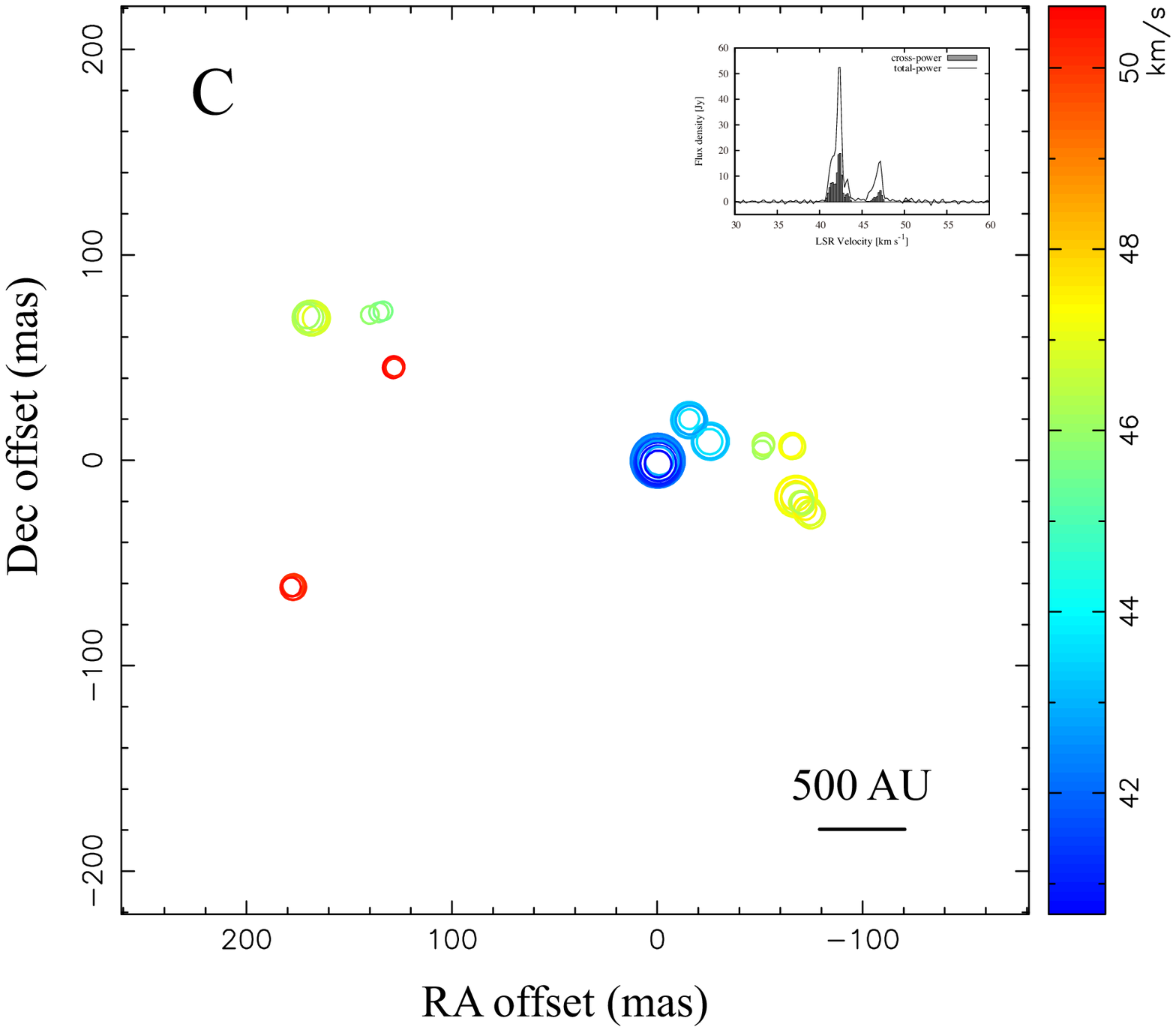}
\end{center}
\caption{
Data for source 009.98$-$00.02, plotted as for Figure~\ref{appen-fig1}.
}	
\label{appen-fig10}		
\end{figure*}

\begin{figure*}[htbp]
\begin{center}
\includegraphics[width=160mm,clip]{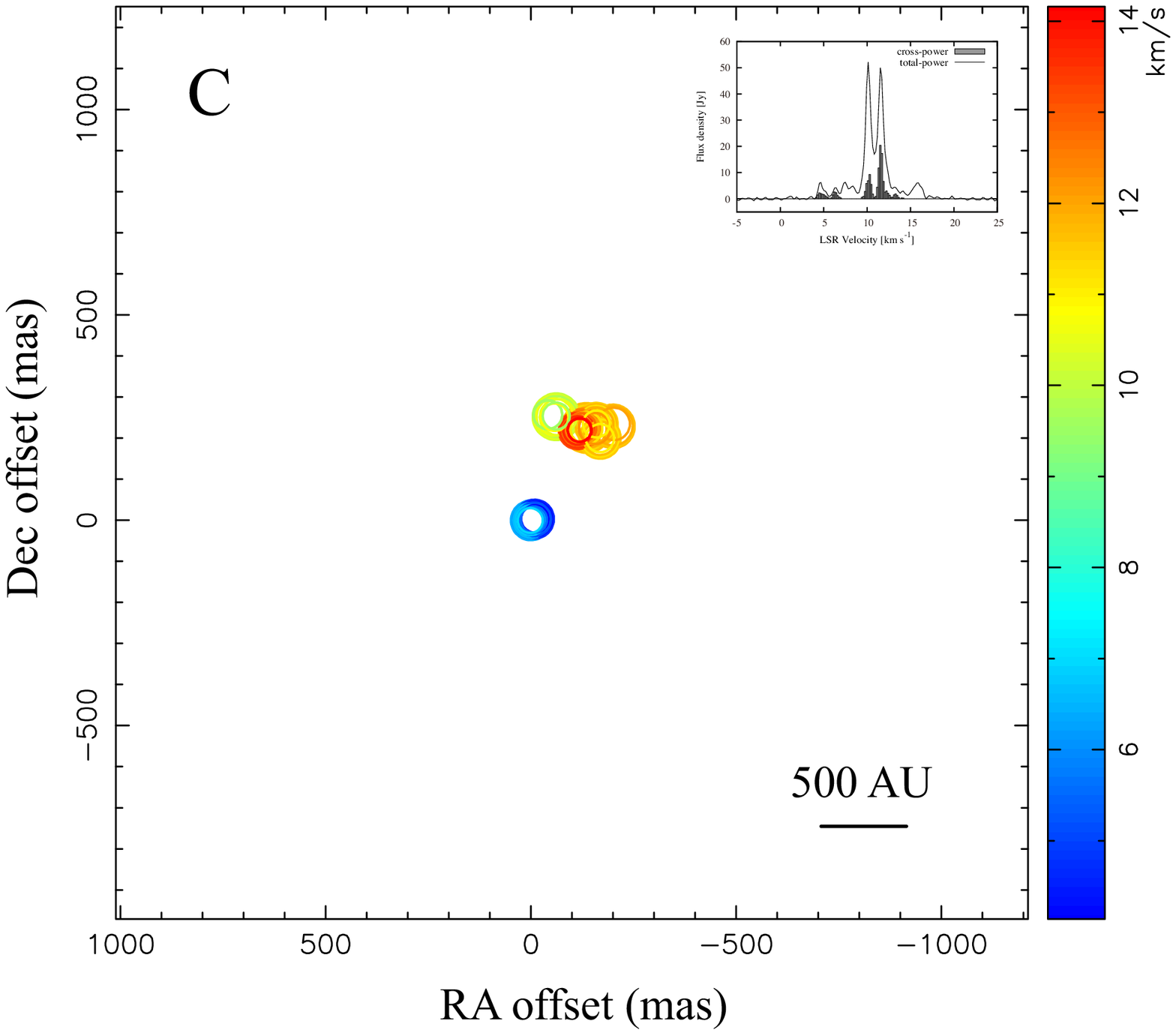}
\end{center}
\caption{
Data for source 010.32$-$00.16, plotted as for Figure~\ref{appen-fig1}.
}	
\label{appen-fig11}		
\end{figure*}

\begin{figure*}[htbp]
\begin{center}
\includegraphics[width=160mm,clip]{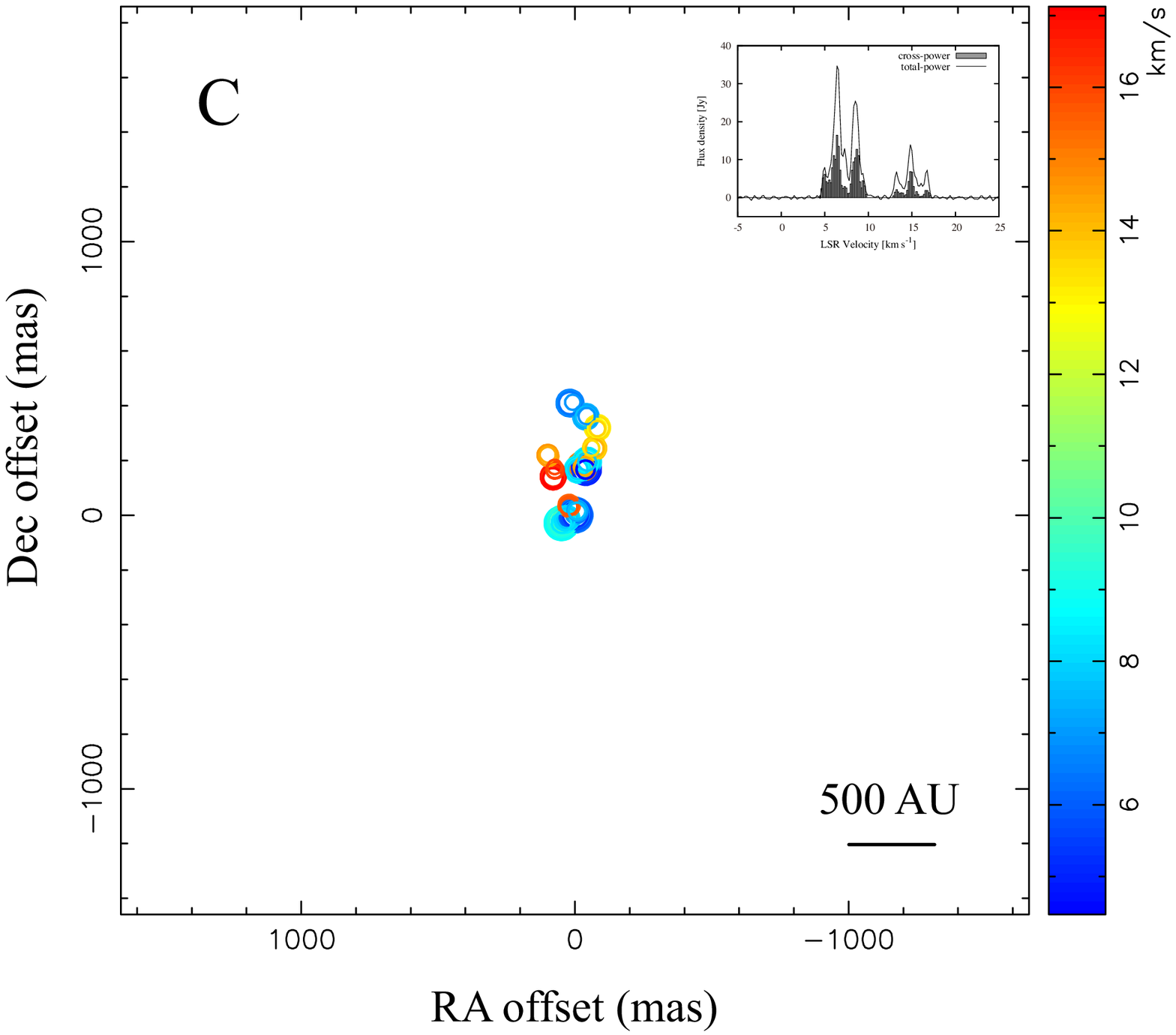}
\end{center}
\caption{
Data for source 011.49$-$01.48, plotted as for Figure~\ref{appen-fig1}.
}	
\label{appen-fig12}		
\end{figure*}

\begin{figure*}[htbp]
\begin{center}
\includegraphics[width=160mm,clip]{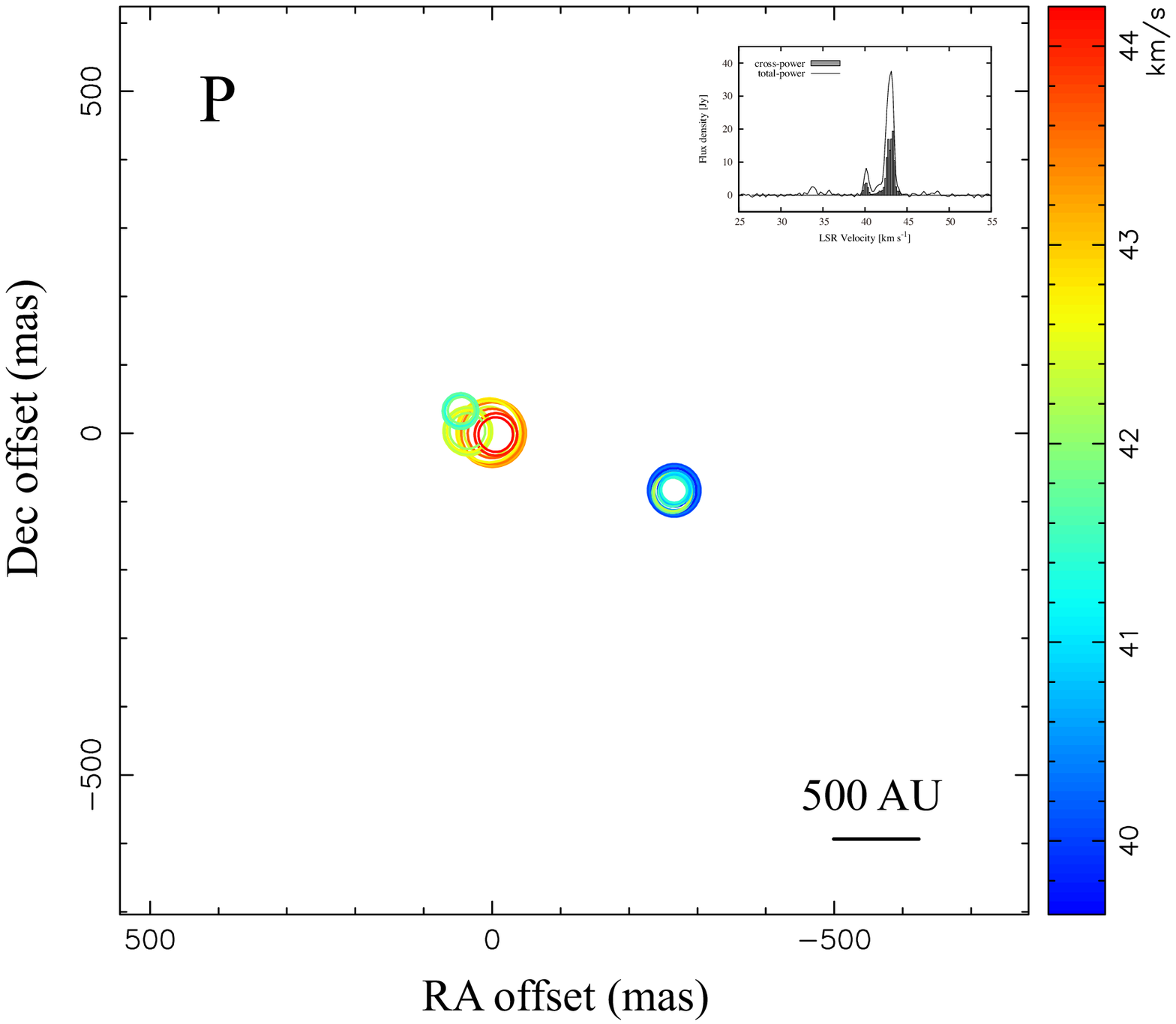}
\end{center}
\caption{
Data for source 011.90$-$00.14, plotted as for Figure~\ref{appen-fig1}.
}	
\label{appen-fig13}		
\end{figure*}

\begin{figure*}[htbp]
\begin{center}
\includegraphics[width=160mm,clip]{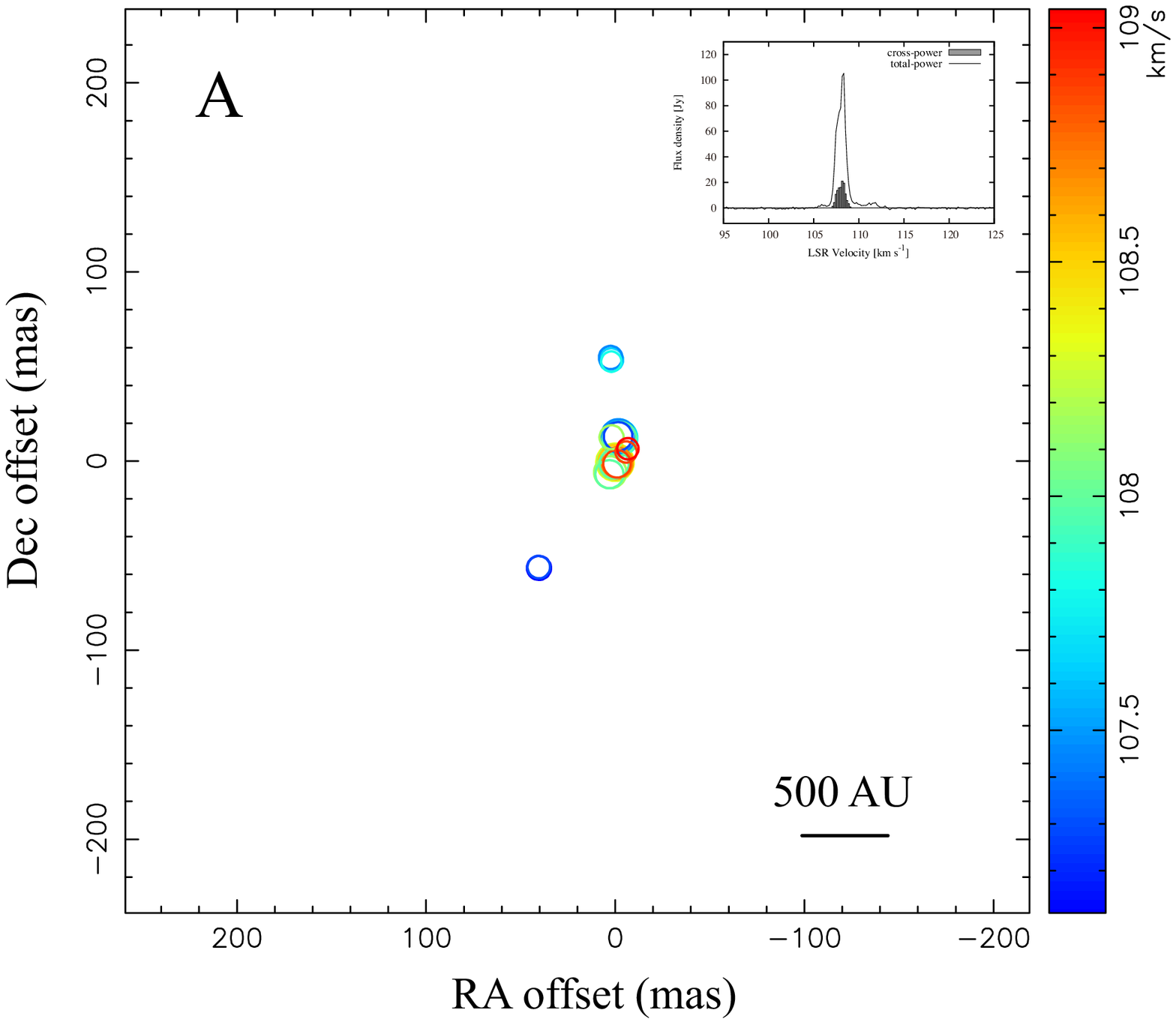}
\end{center}
\caption{
Data for source 012.02$-$00.03, plotted as for Figure~\ref{appen-fig1}.
}	
\label{appen-fig14}		
\end{figure*}

\begin{figure*}[htbp]
\begin{center}
\includegraphics[width=160mm,clip]{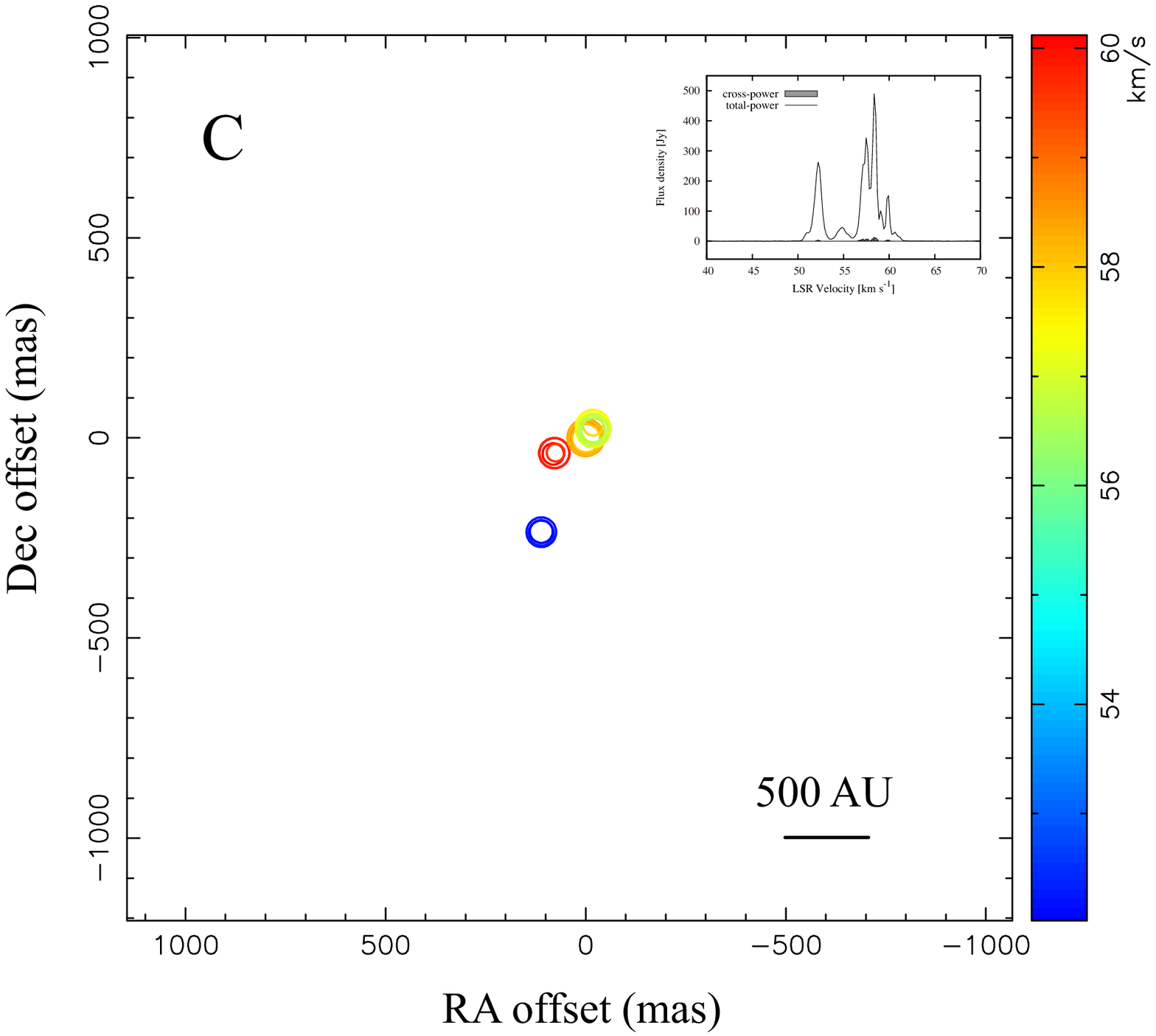}
\end{center}
\caption{
Data for source 012.68$-$00.18, plotted as for Figure~\ref{appen-fig1}.
}	
\label{appen-fig15}		
\end{figure*}

\begin{figure*}[htbp]
\begin{center}
\includegraphics[width=160mm,clip]{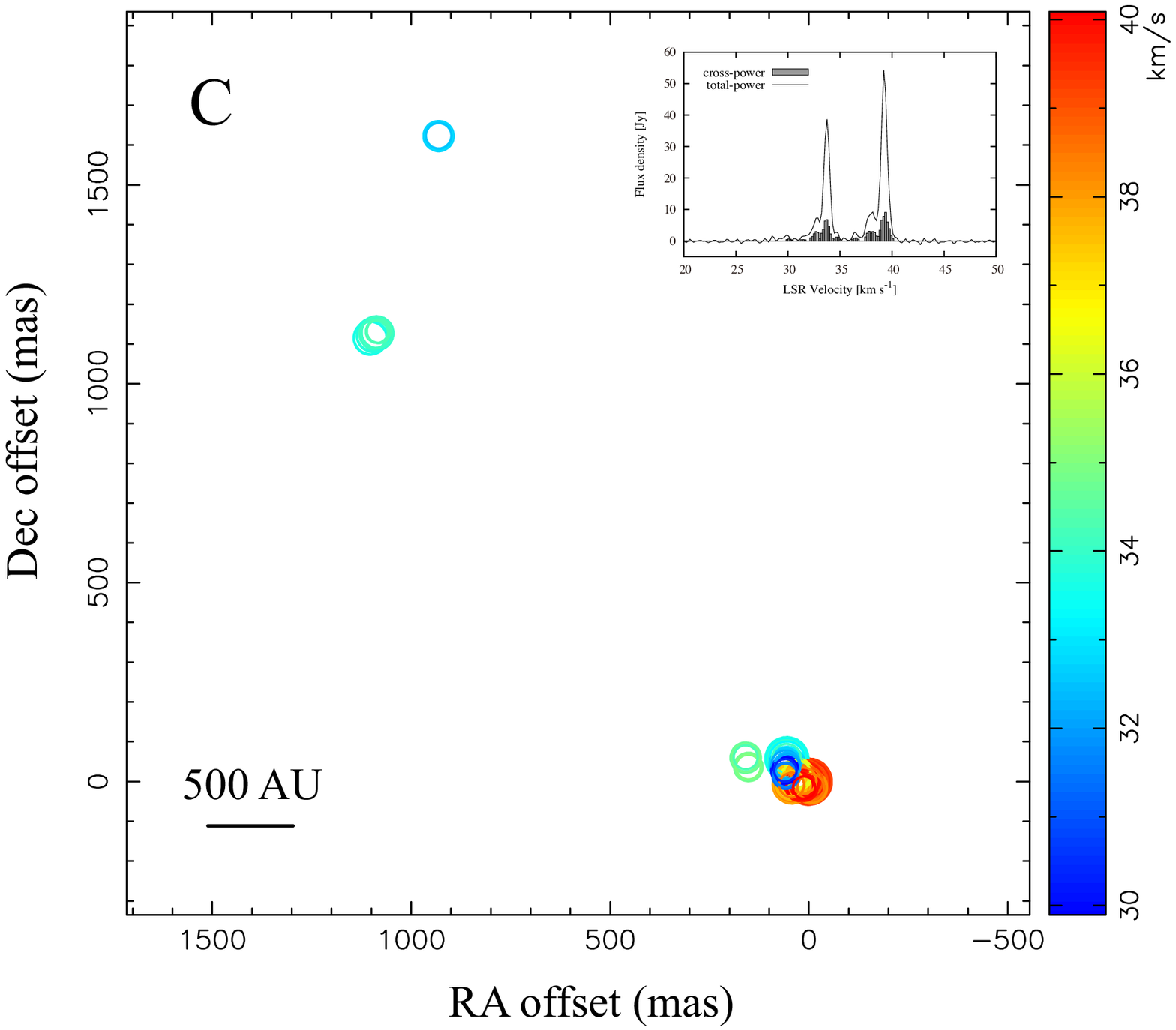}
\end{center}
\caption{
Data for source 012.88$+$00.48, plotted as for Figure~\ref{appen-fig1}.
}	
\label{appen-fig16}		
\end{figure*}

\begin{figure*}[htbp]
\begin{center}
\includegraphics[width=160mm,clip]{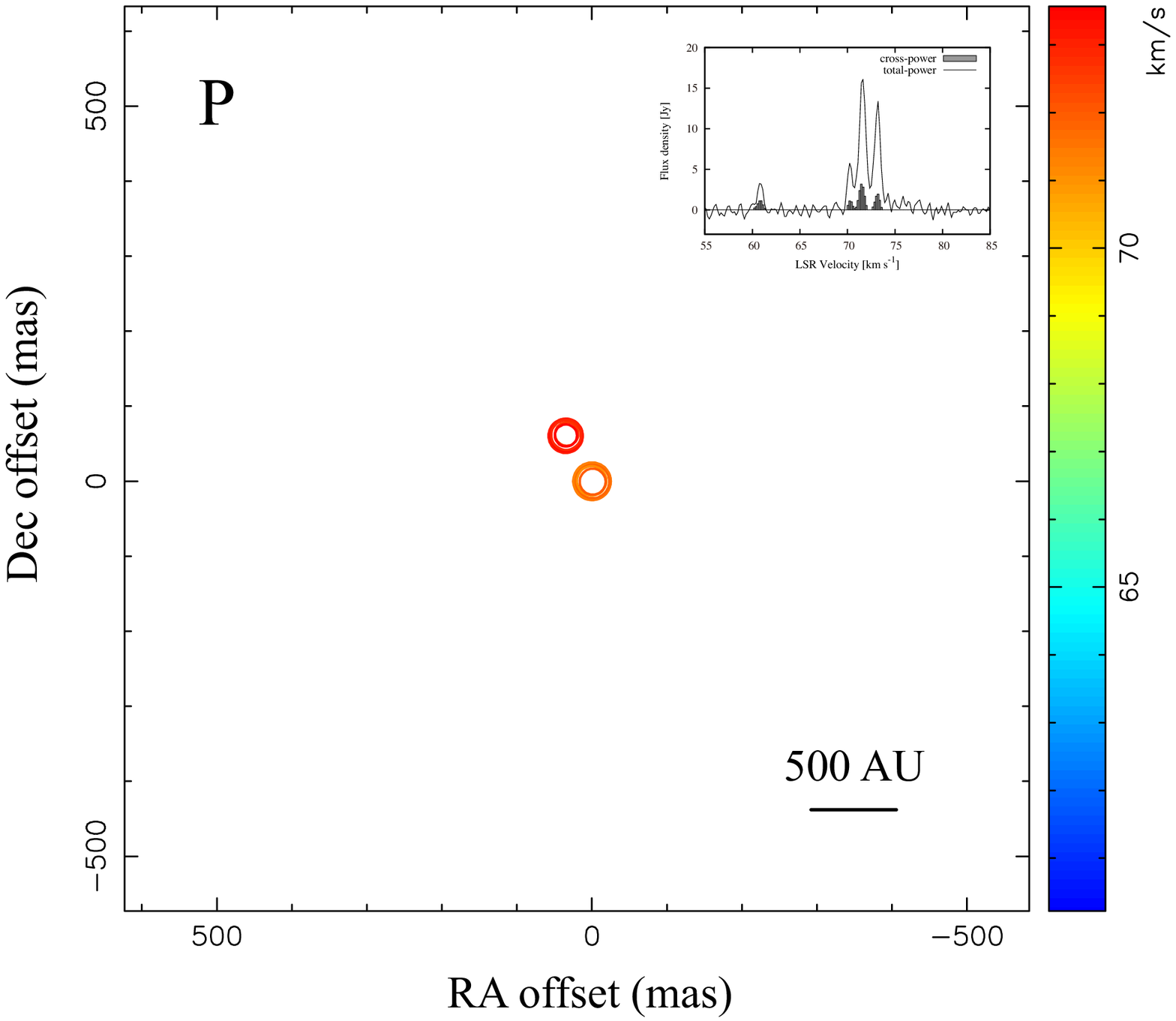}
\end{center}
\caption{
Data for source 020.23$+$00.06~SW, plotted as for Figure~\ref{appen-fig1}.
}	
\label{appen-fig17}		
\end{figure*}

\begin{figure*}[htbp]
\begin{center}
\includegraphics[width=160mm,clip]{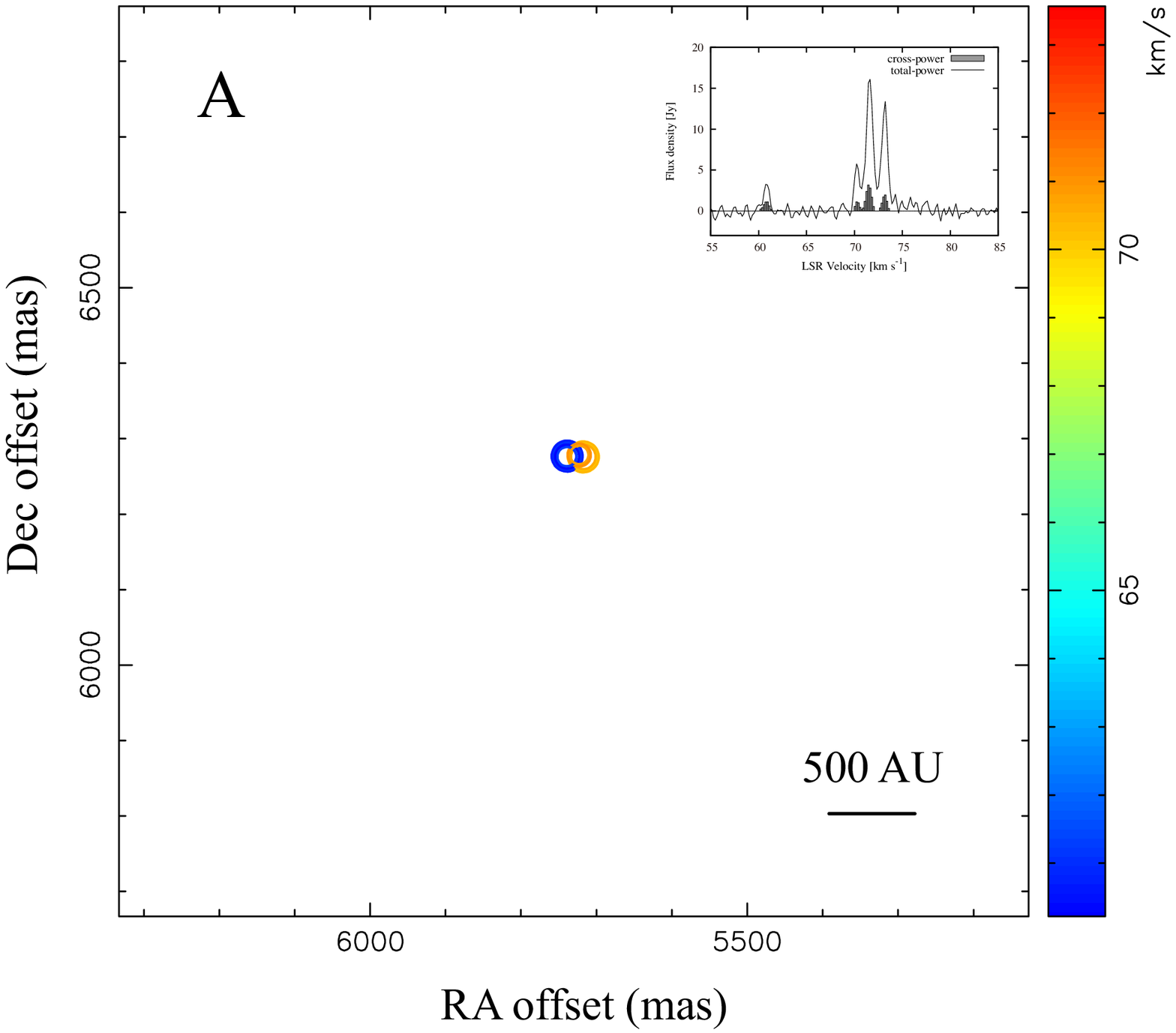}
\end{center}
\caption{
Data for source 020.23$+$00.06~NE, plotted as for Figure~\ref{appen-fig1}.
}	
\label{appen-fig18}		
\end{figure*}

\begin{figure*}[htbp]
\begin{center}
\includegraphics[width=160mm,clip]{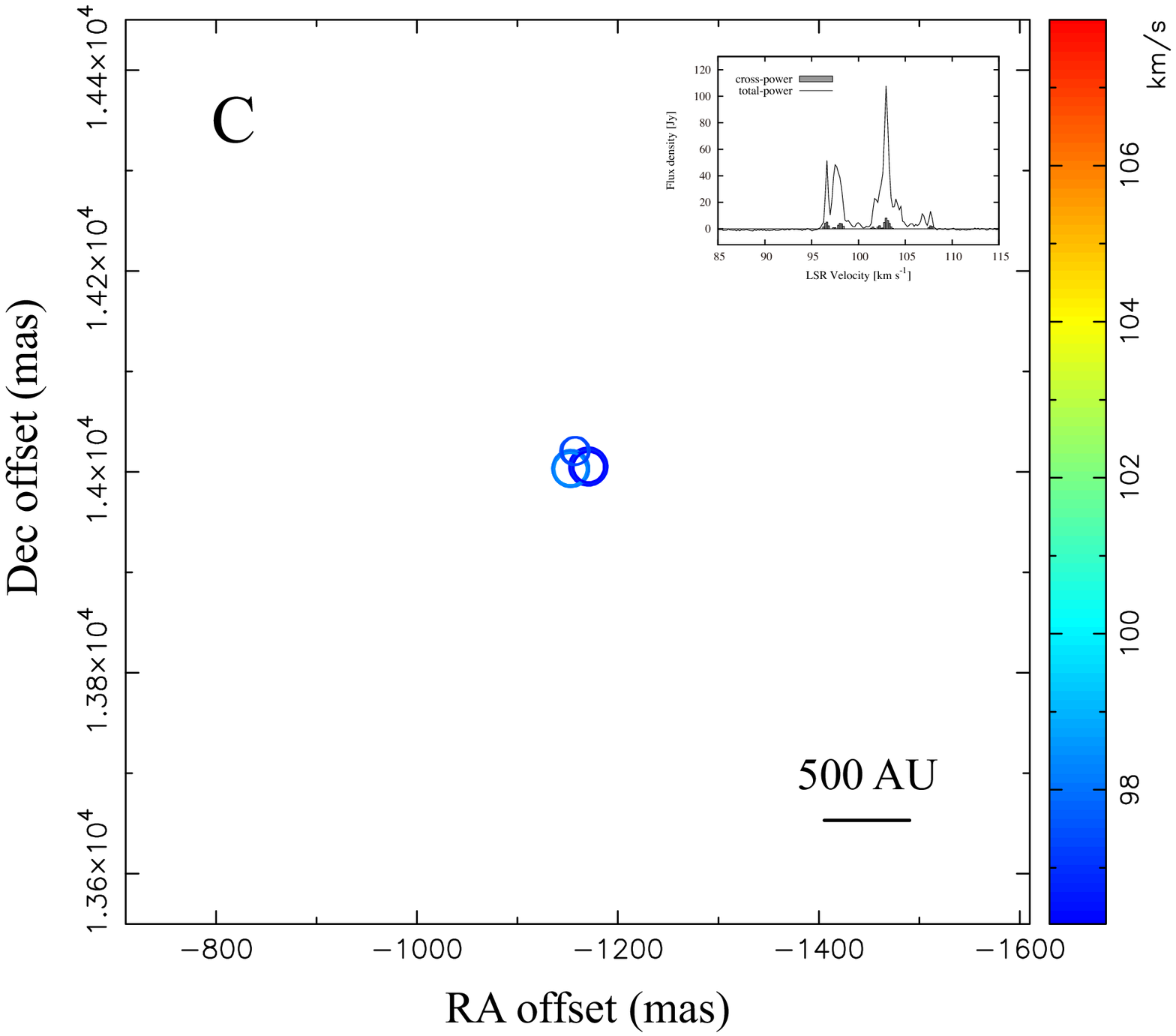}
\end{center}
\caption{
Data for source 023.43$-$00.18~MM1, plotted as for Figure~\ref{appen-fig1}.
}	
\label{appen-fig19}		
\end{figure*}

\begin{figure*}[htbp]
\begin{center}
\includegraphics[width=160mm,clip]{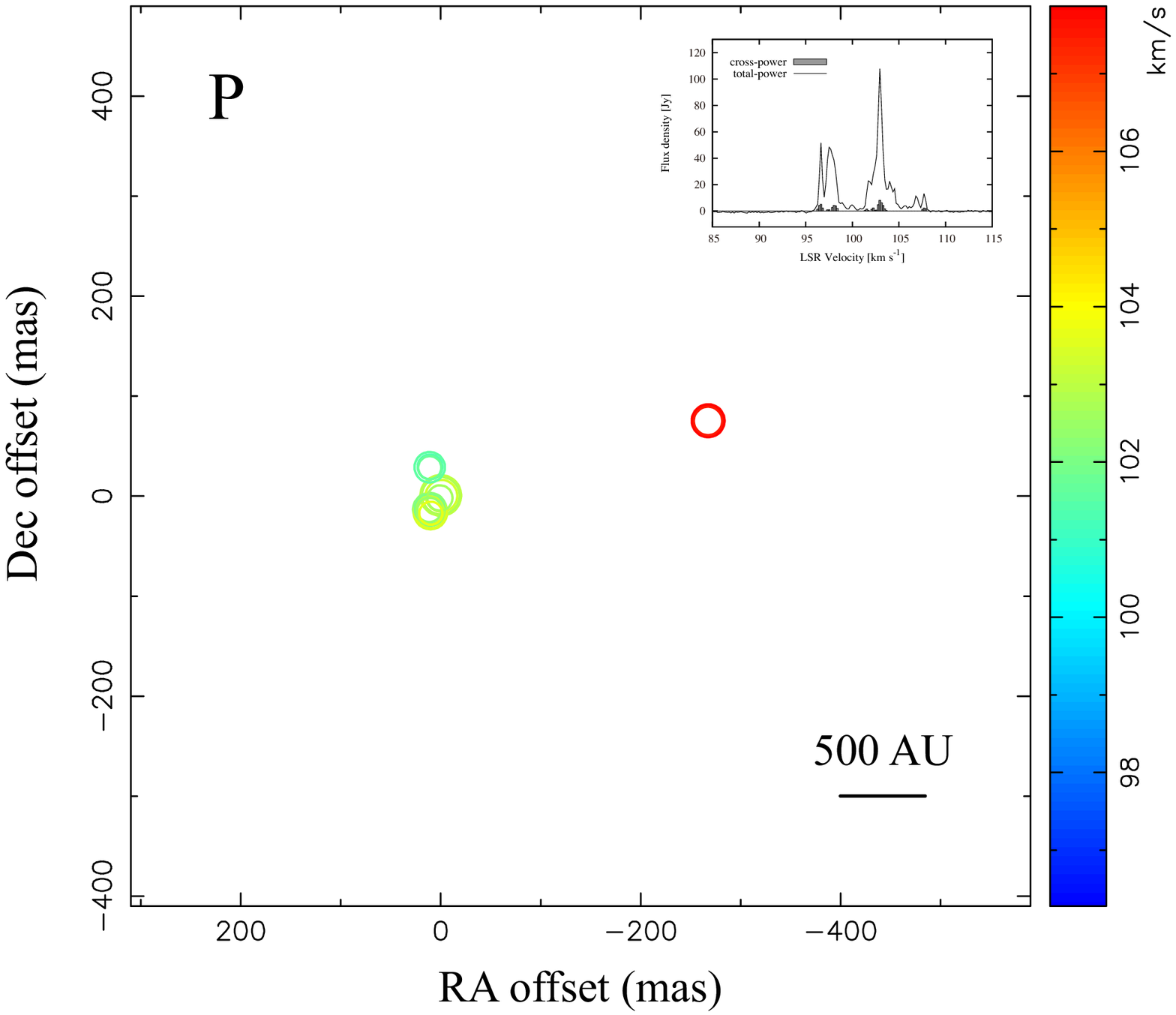}
\end{center}
\caption{
Data for source 023.43$-$00.18~MM2, plotted as for Figure~\ref{appen-fig1}.
}	
\label{appen-fig20}		
\end{figure*}

\begin{figure*}[htbp]
\begin{center}
\includegraphics[width=160mm,clip]{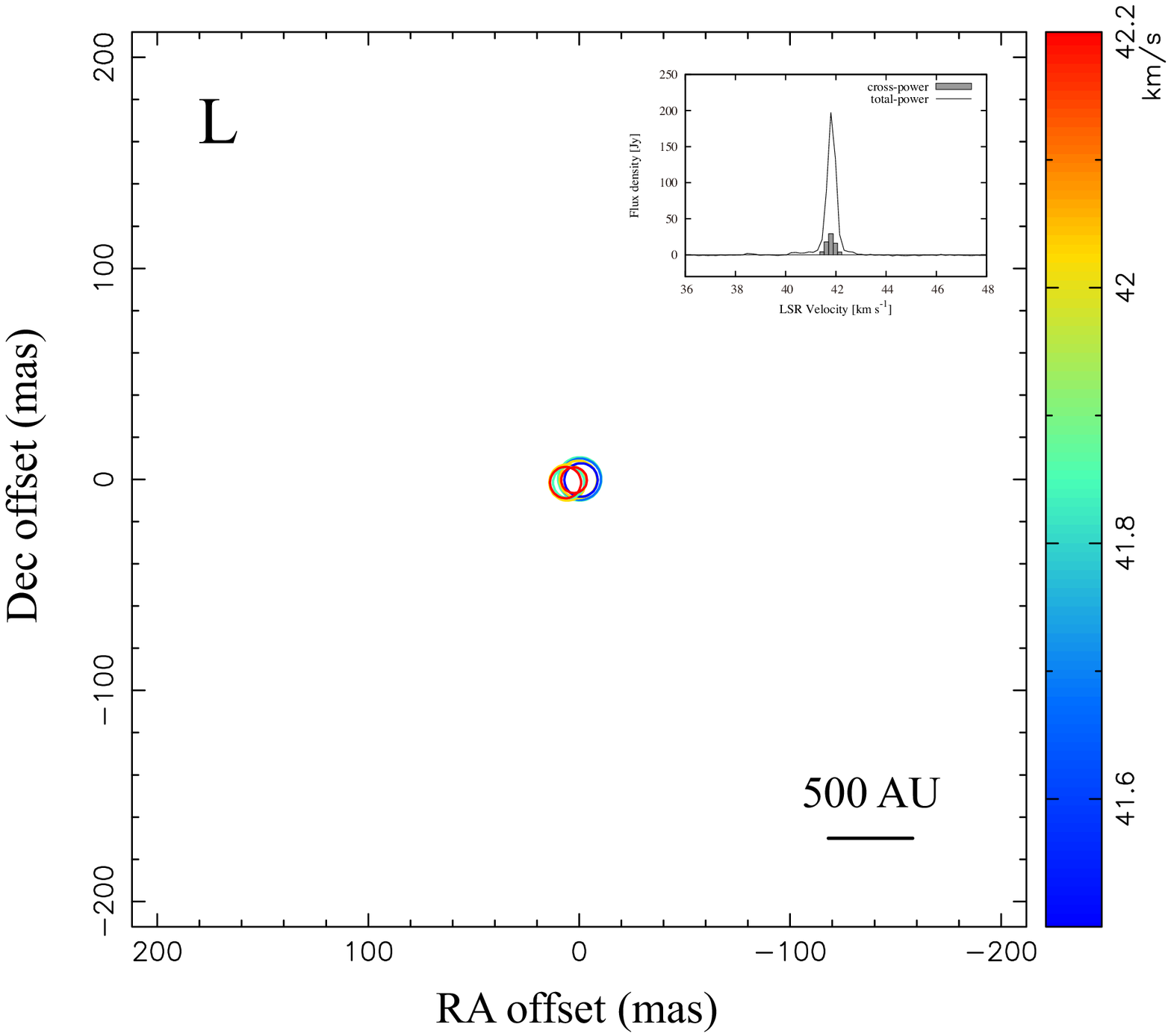}
\end{center}
\caption{
Data for source 025.65$+$01.05, plotted as for Figure~\ref{appen-fig1}.
}	
\label{appen-fig21}		
\end{figure*}

\begin{figure*}[htbp]
\begin{center}
\includegraphics[width=160mm,clip]{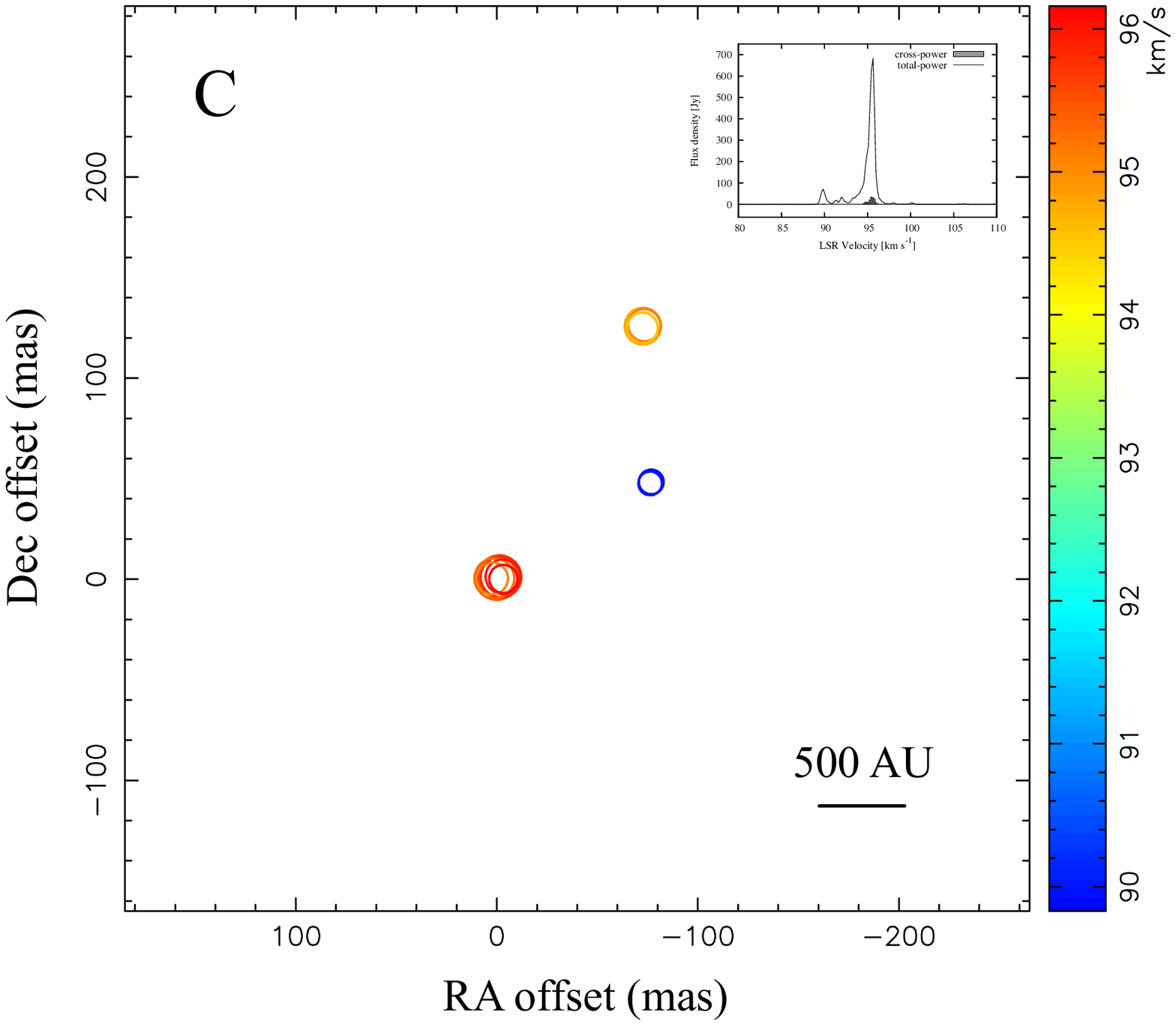}
\end{center}
\caption{
Data for source 025.71$+$00.04, plotted as for Figure~\ref{appen-fig1}.
}	
\label{appen-fig22}		
\end{figure*}

\begin{figure*}[htbp]
\begin{center}
\includegraphics[width=160mm,clip]{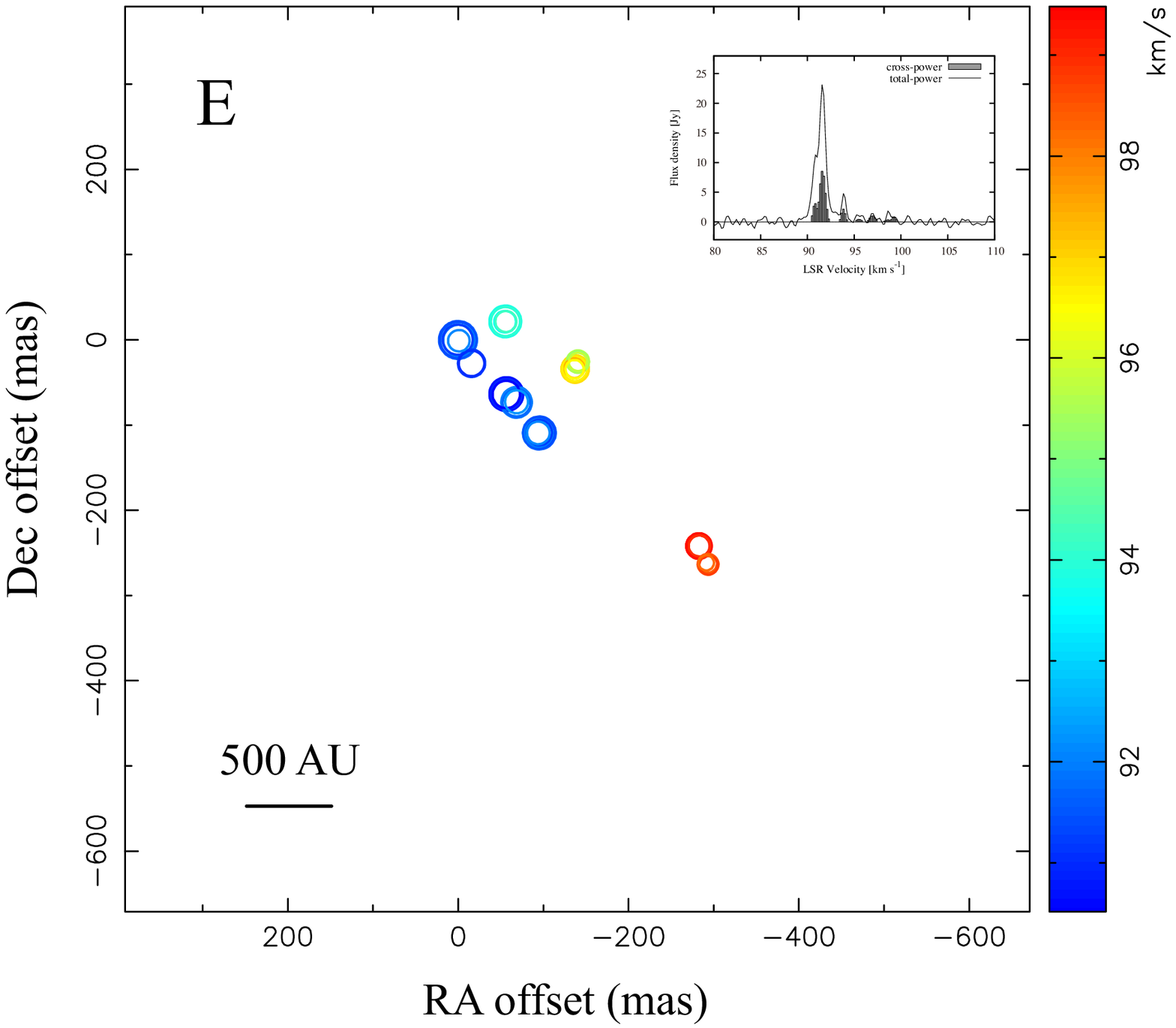}
\end{center}
\caption{
Data for source 025.82$-$00.17, plotted as for Figure~\ref{appen-fig1}.
}	
\label{appen-fig23}		
\end{figure*}

\begin{figure*}[htbp]
\begin{center}
\includegraphics[width=160mm,clip]{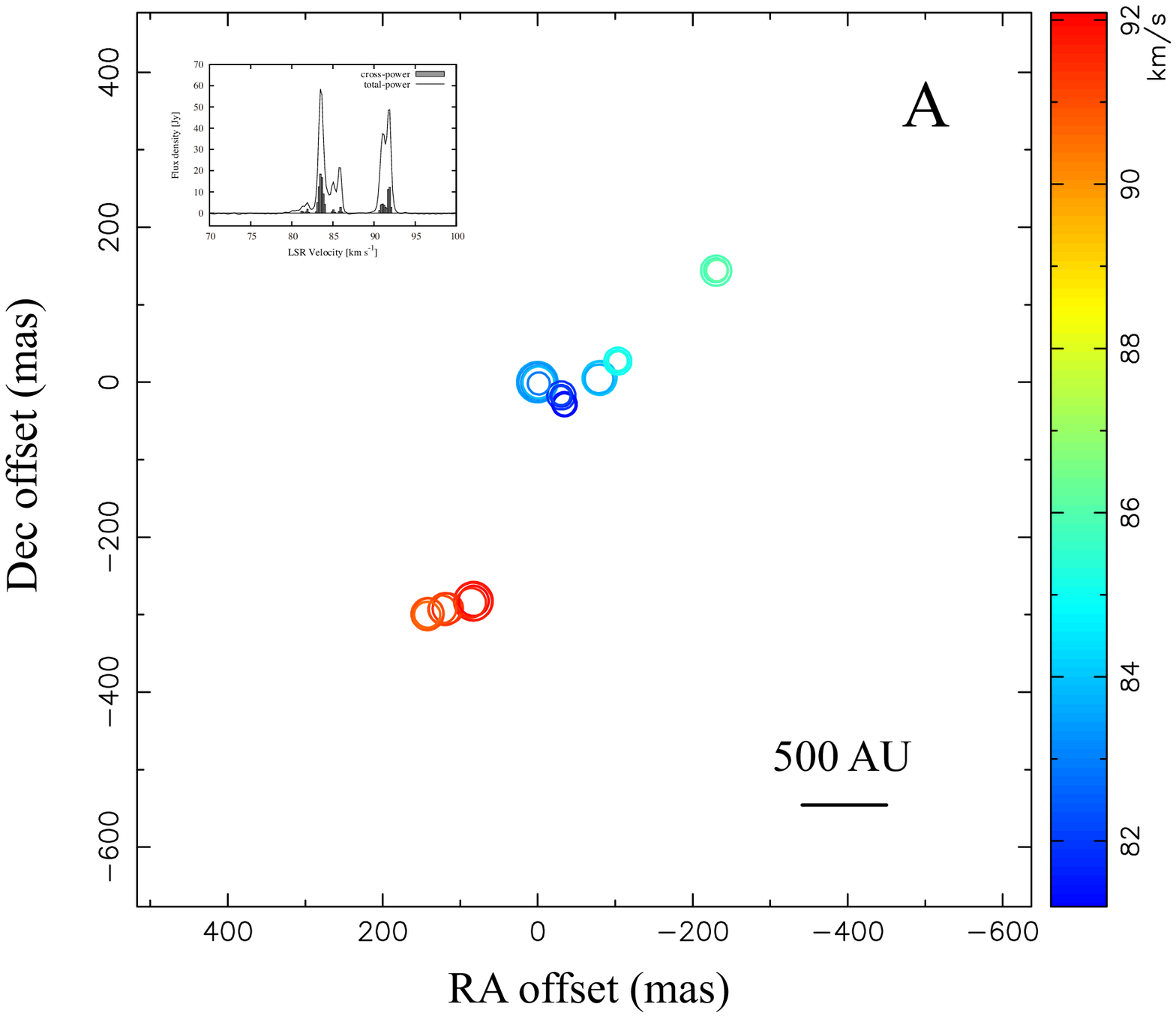}
\end{center}
\caption{
Data for source 028.83$-$00.25, plotted as for Figure~\ref{appen-fig1}.
}	
\label{appen-fig24}		
\end{figure*}

\begin{figure*}[htbp]
\begin{center}
\includegraphics[width=160mm,clip]{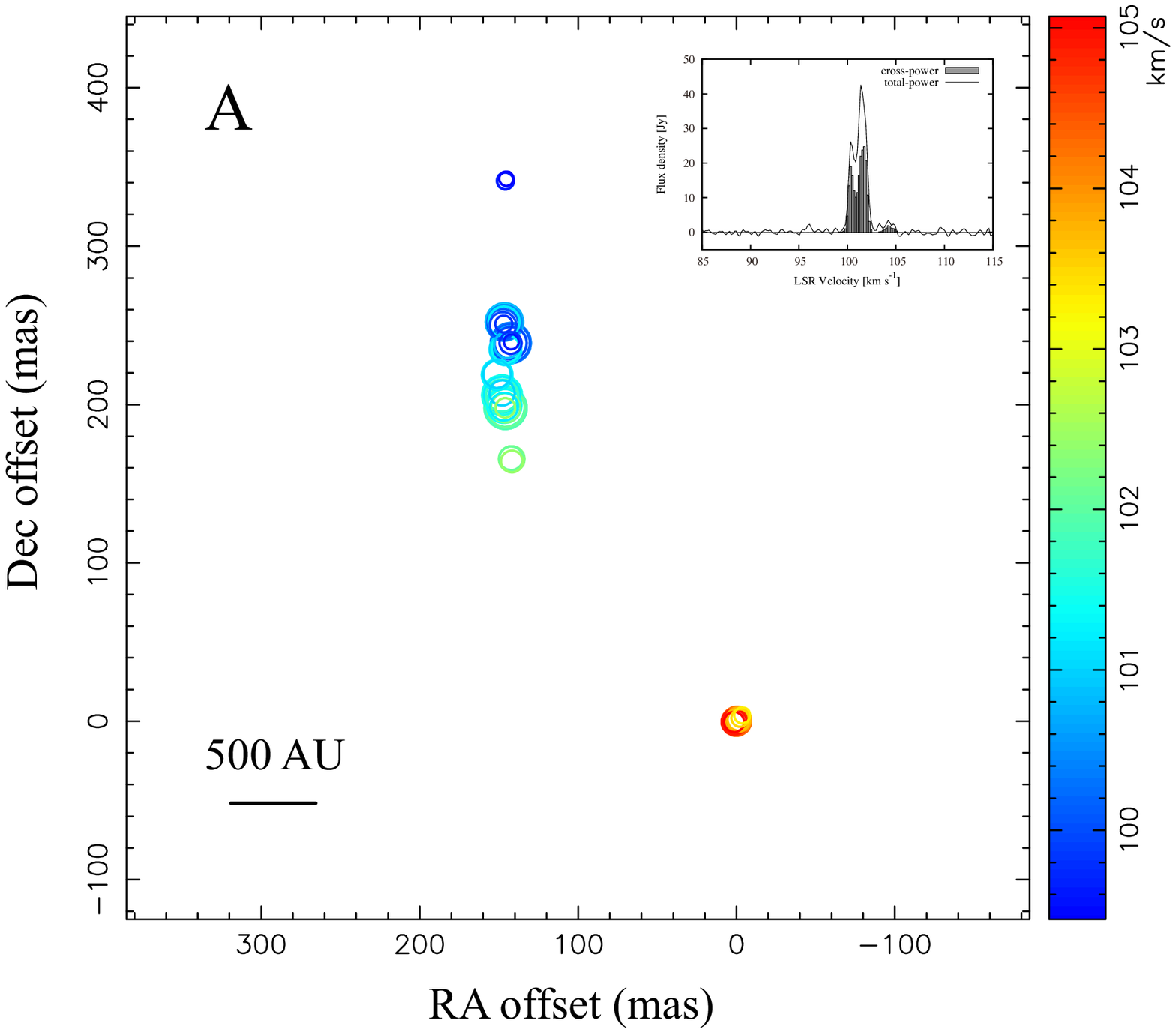}
\end{center}
\caption{
Data for source 029.86$-$00.04, plotted as for Figure~\ref{appen-fig1}.
}	
\label{appen-fig25}		
\end{figure*}

\begin{figure*}[htbp]
\begin{center}
\includegraphics[width=160mm,clip]{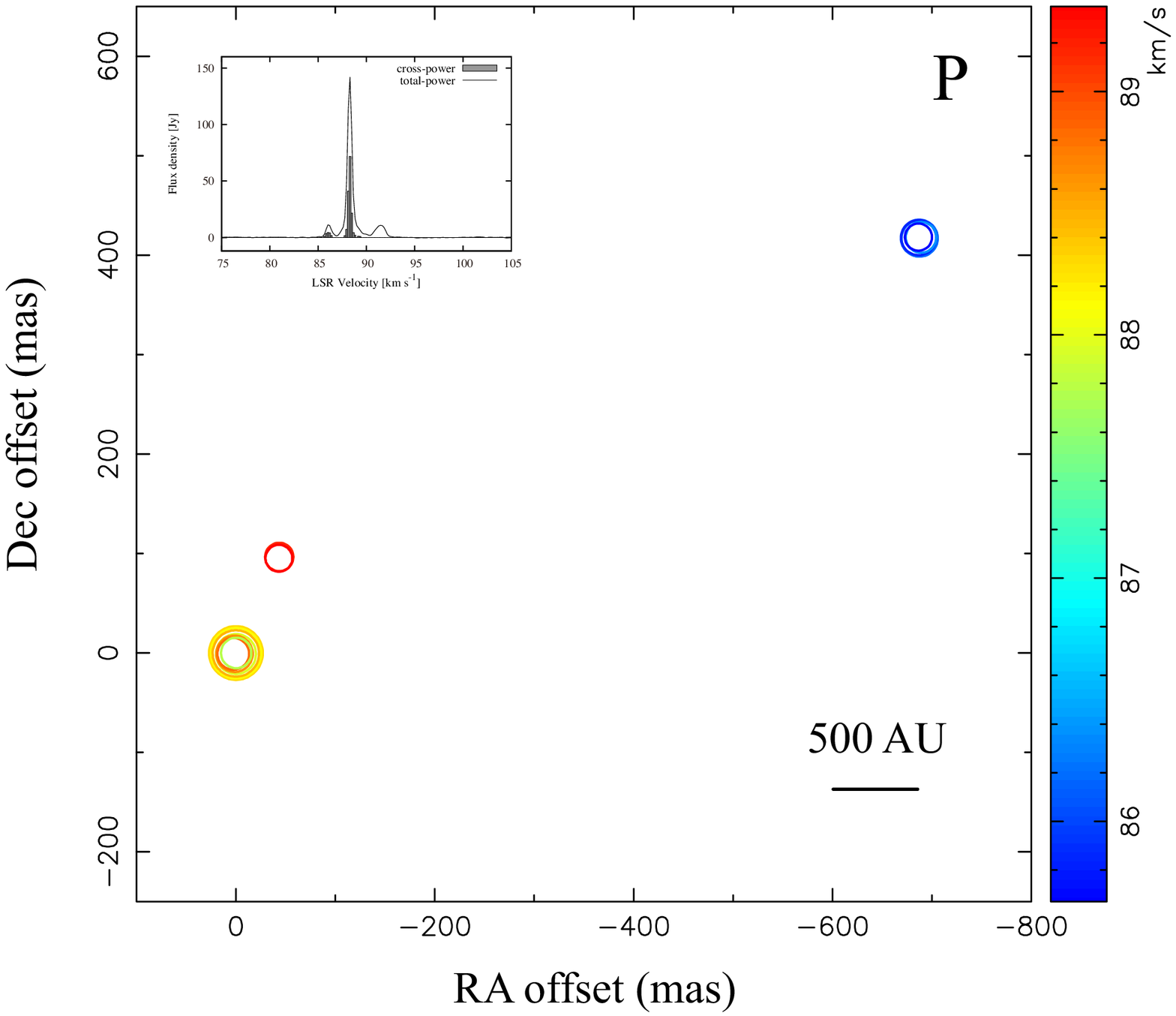}
\end{center}
\caption{
Data for source 030.70$-$00.06, plotted as for Figure~\ref{appen-fig1}.
}	
\label{appen-fig26}		
\end{figure*}

\begin{figure*}[htbp]
\begin{center}
\includegraphics[width=160mm,clip]{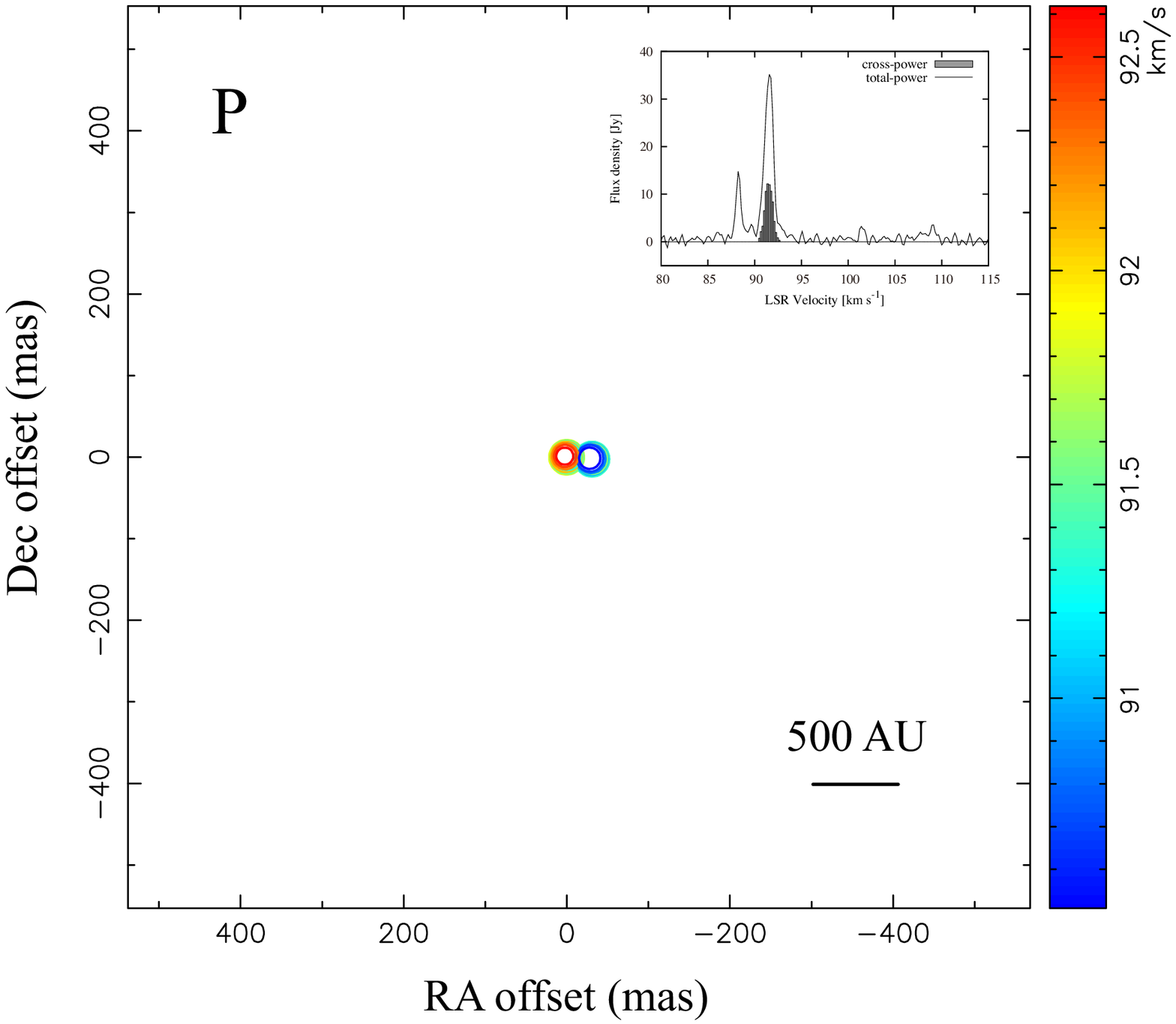}
\end{center}
\caption{
Data for source 030.76$-$00.05, plotted as for Figure~\ref{appen-fig1}.
}	
\label{appen-fig27}		
\end{figure*}

\begin{figure*}[htbp]
\begin{center}
\includegraphics[width=160mm,clip]{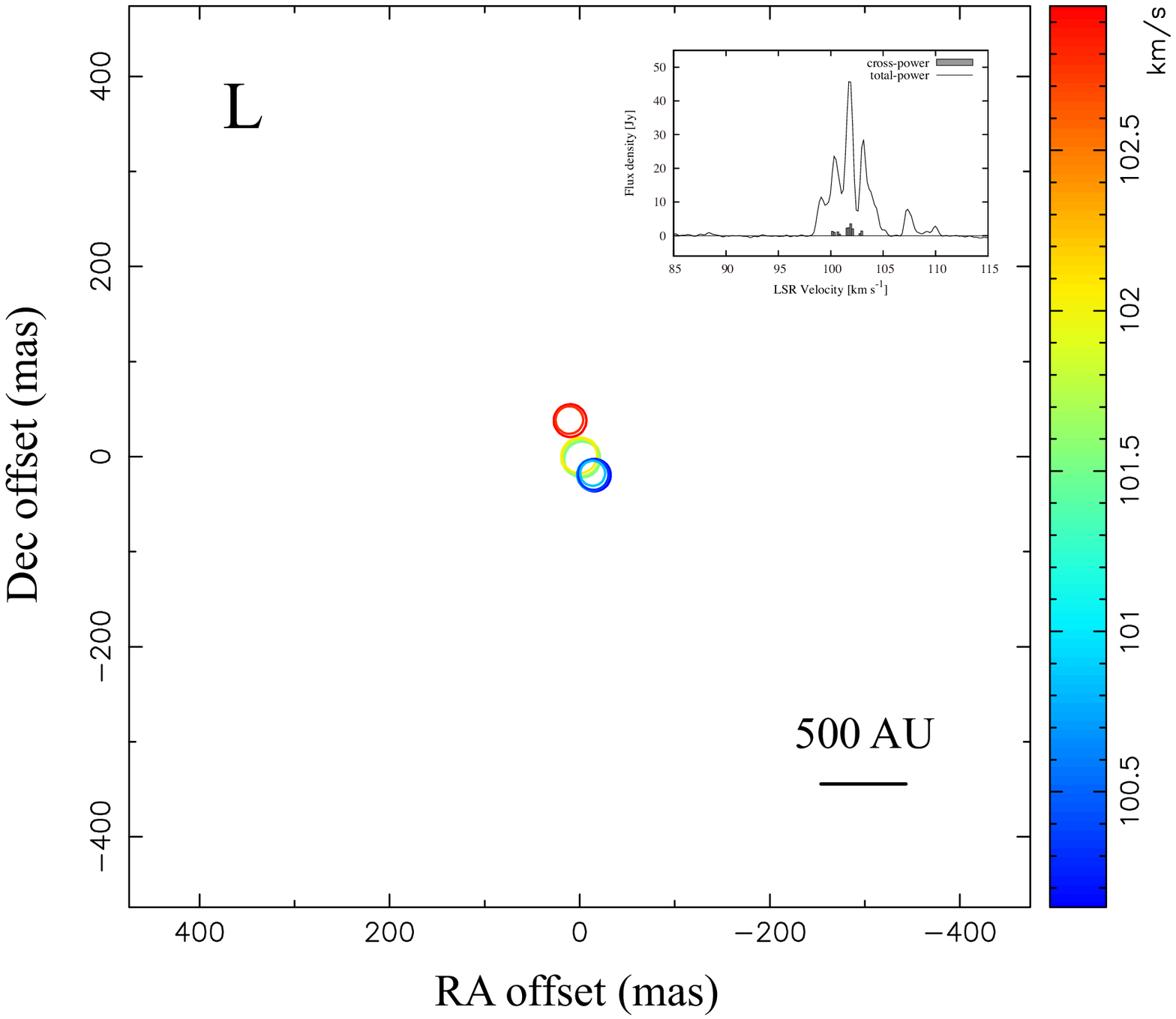}
\end{center}
\caption{
Data for source 030.91$+$00.14, plotted as for Figure~\ref{appen-fig1}.
}	
\label{appen-fig28}		
\end{figure*}

\begin{figure*}[htbp]
\begin{center}
\includegraphics[width=160mm,clip]{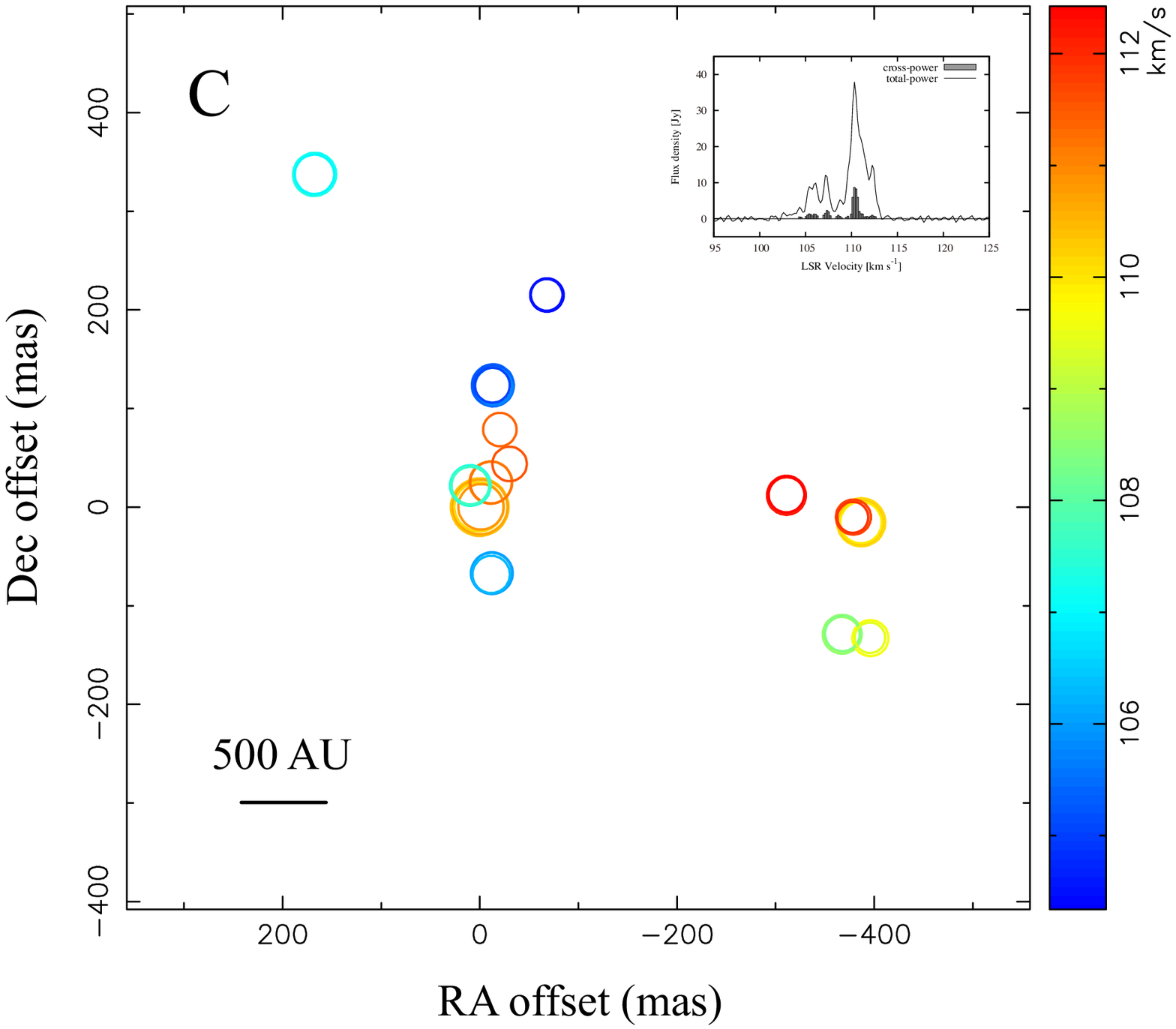}
\end{center}
\caption{
Data for source 031.28$+$00.06, plotted as for Figure~\ref{appen-fig1}.
}	
\label{appen-fig29}		
\end{figure*}

\begin{figure*}[htbp]
\begin{center}
\includegraphics[width=160mm,clip]{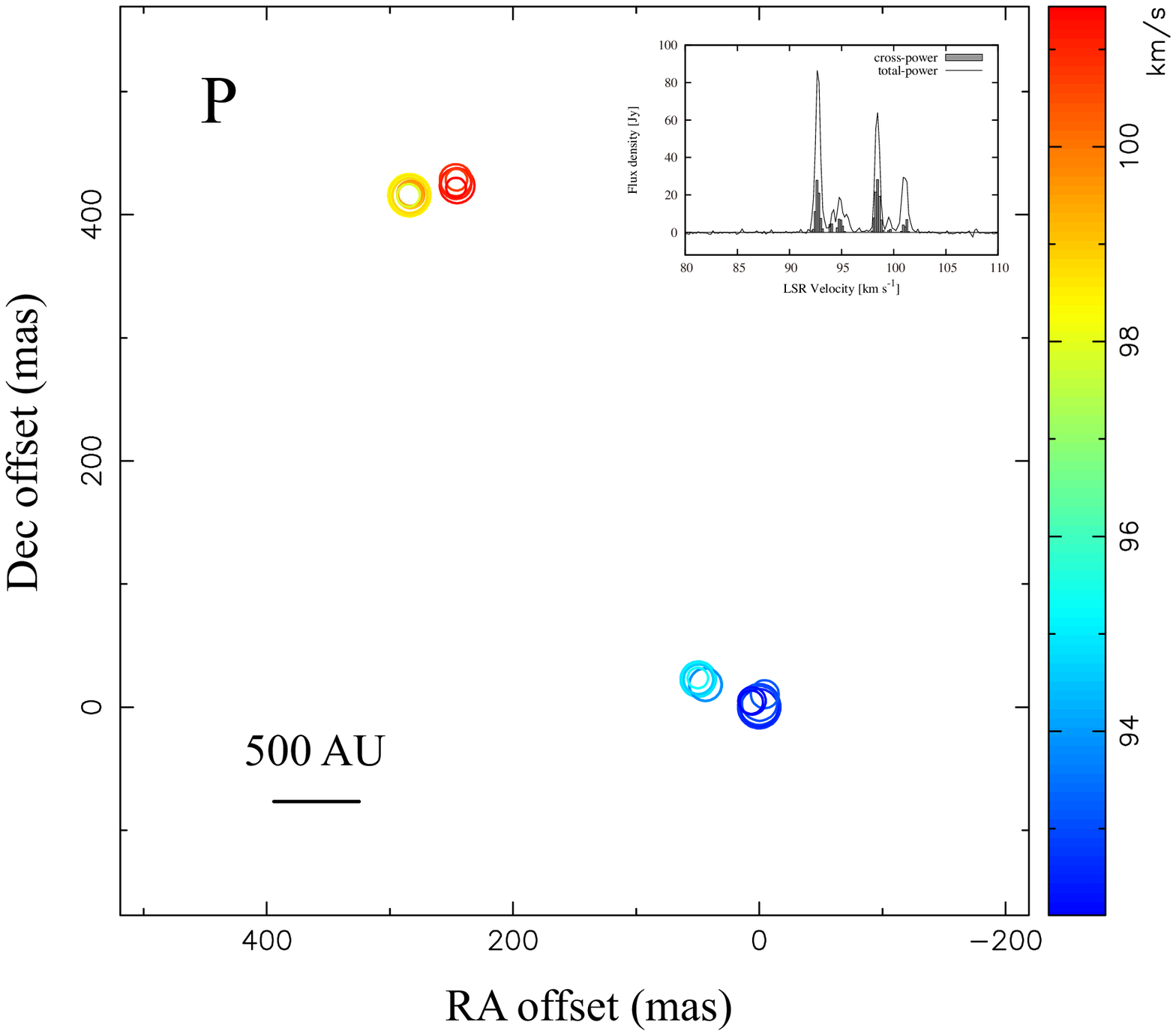}
\end{center}
\caption{
Data for source 032.03$+$00.06, plotted as for Figure~\ref{appen-fig1}.
}	
\label{appen-fig30}		
\end{figure*}

\begin{figure*}[htbp]
\begin{center}
\includegraphics[width=160mm,clip]{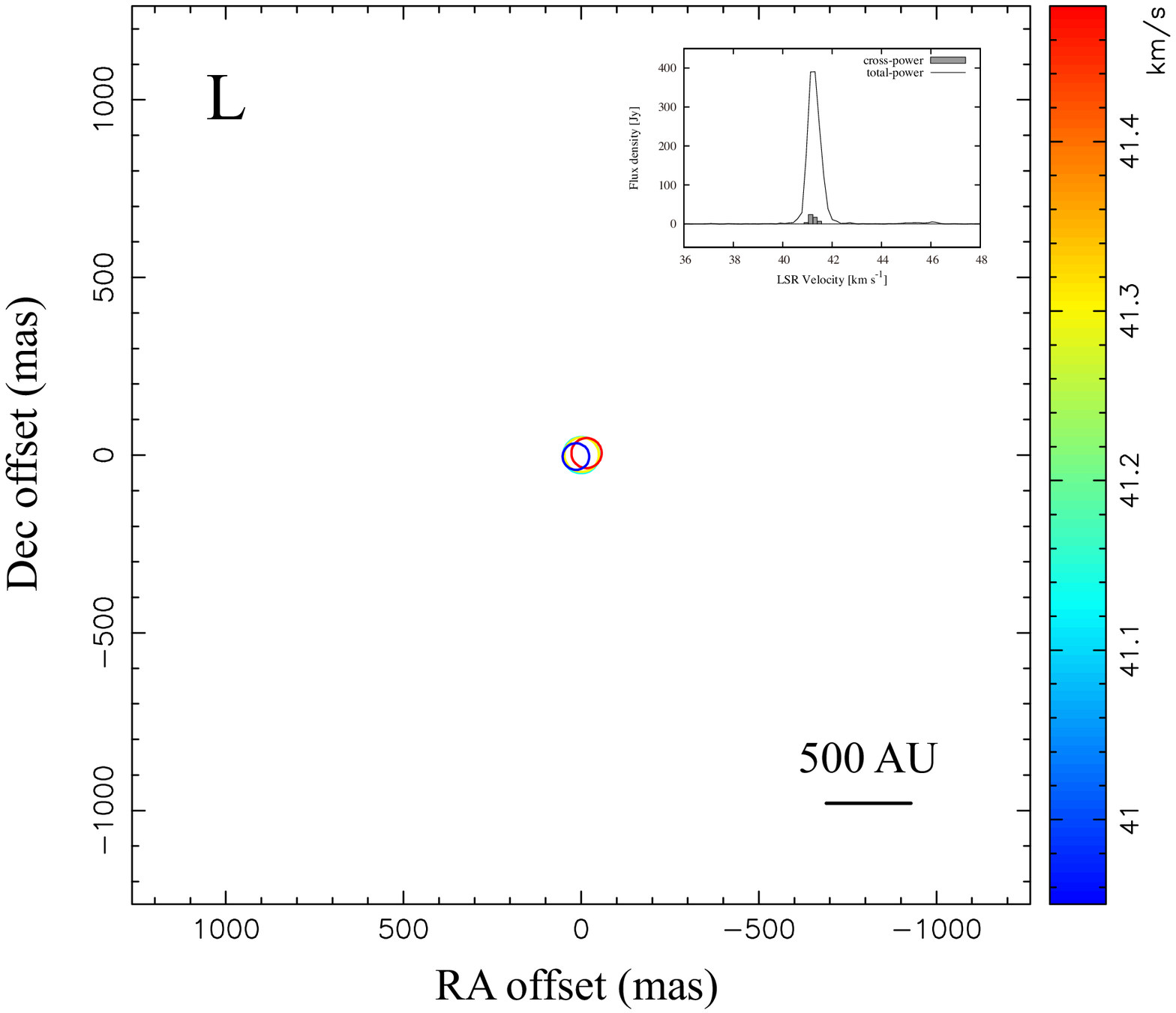}
\end{center}
\caption{
Data for source 037.40$+$01.52, plotted as for Figure~\ref{appen-fig1}.
}	
\label{appen-fig31}		
\end{figure*}

\begin{figure*}[htbp]
\begin{center}
\includegraphics[width=160mm,clip]{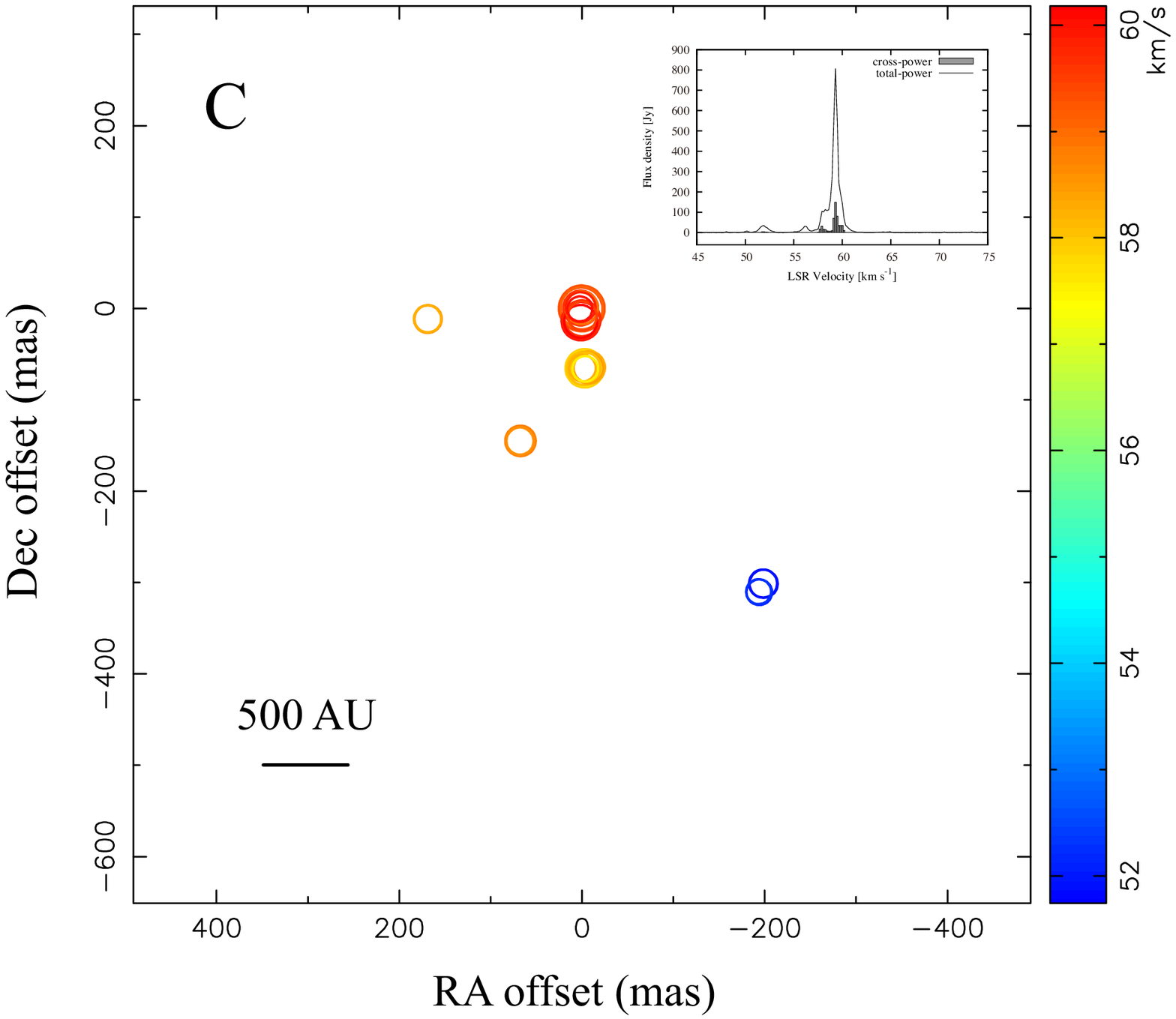}
\end{center}
\caption{
Data for source 049.49$-$00.38, plotted as for Figure~\ref{appen-fig1}.
}	
\label{appen-fig32}		
\end{figure*}

\begin{figure*}[htbp]
\begin{center}
\includegraphics[width=160mm,clip]{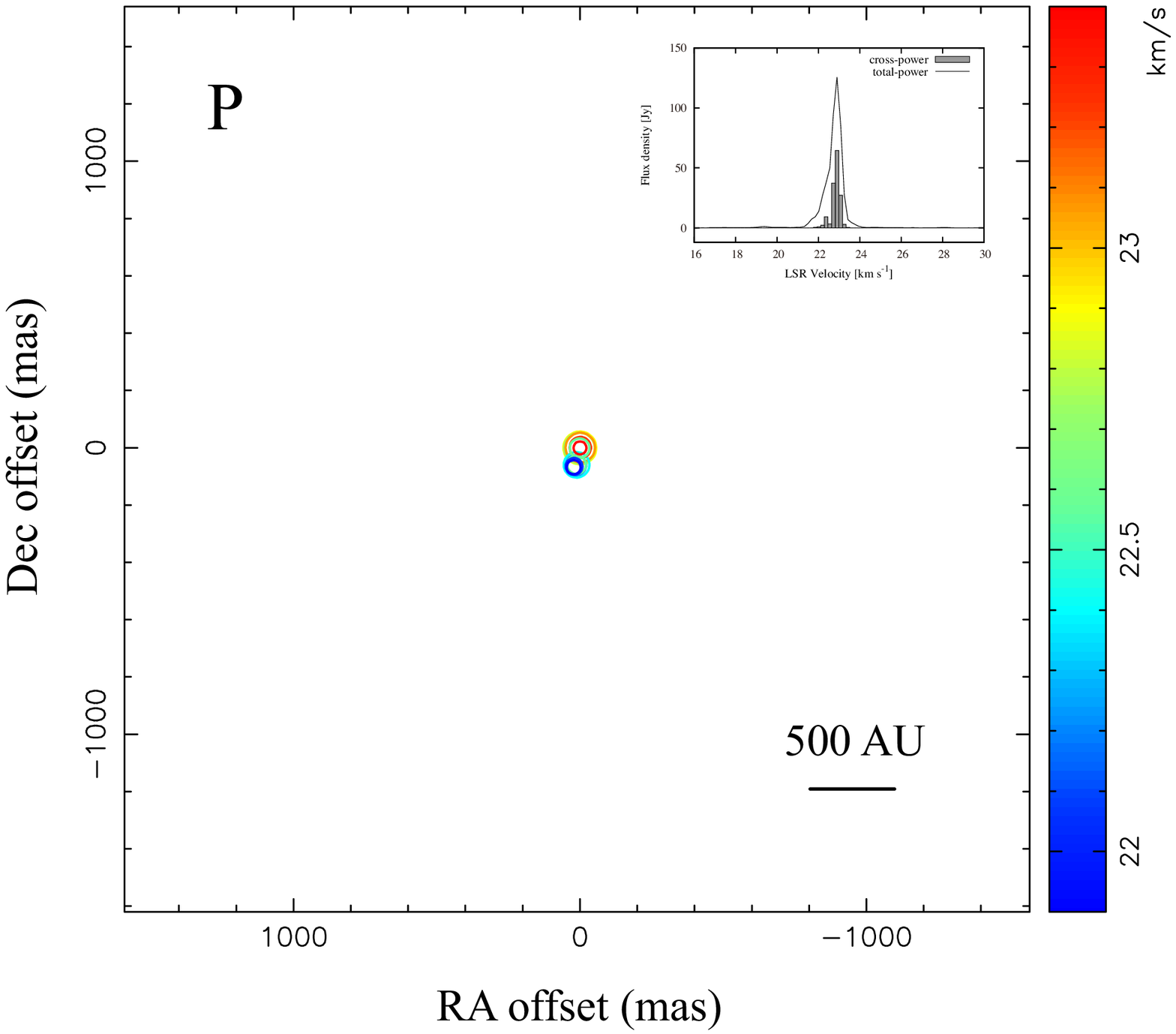}
\end{center}
\caption{
Data for source 232.62$+$00.99, plotted as for Figure~\ref{appen-fig1}.
}	
\label{appen-fig33}		
\end{figure*}

\begin{figure*}[htbp]
\begin{center}
\includegraphics[width=160mm,clip]{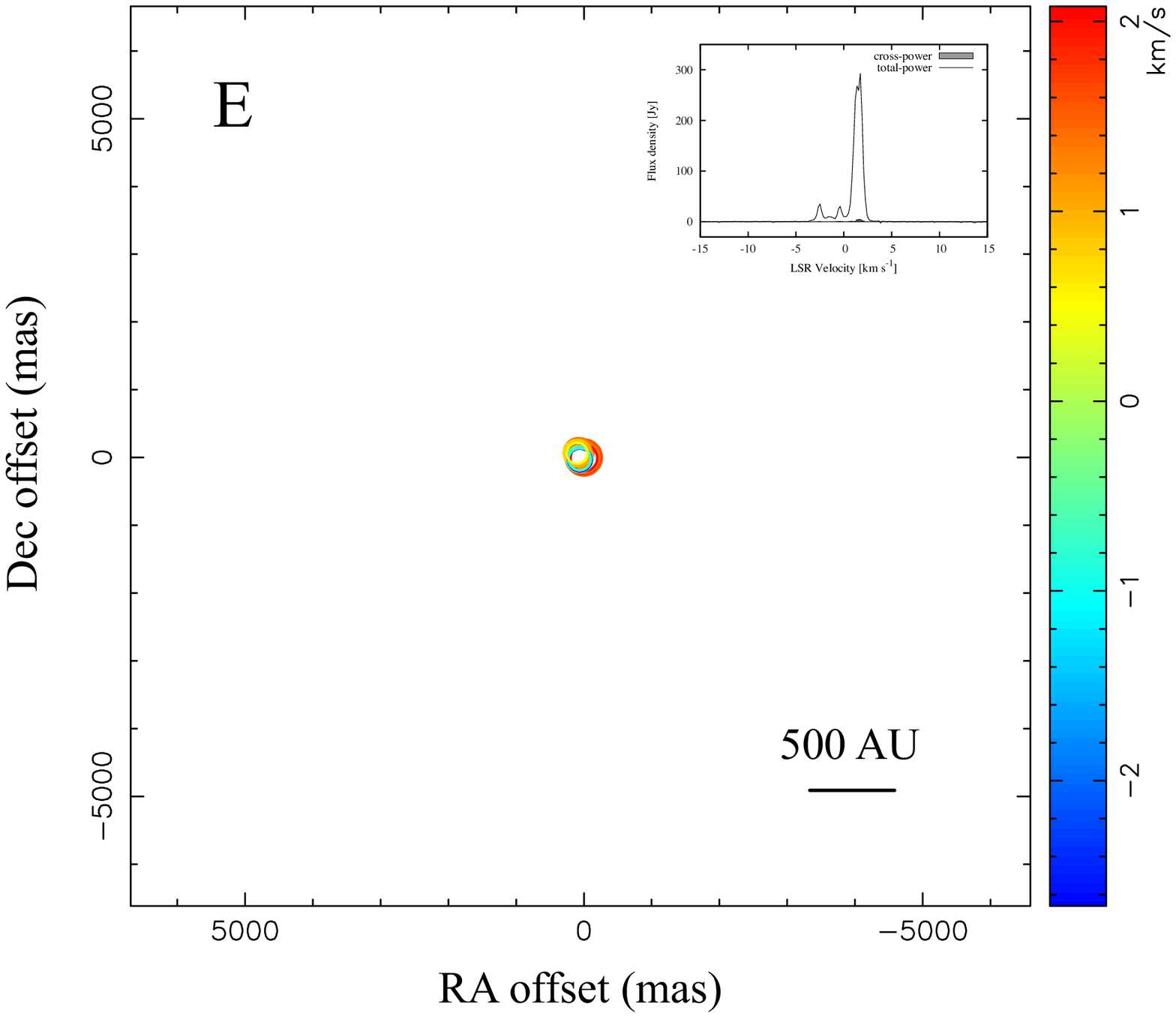}
\end{center}
\caption{
Data for source 351.77$-$00.53, plotted as for Figure~\ref{appen-fig1}.
}	
\label{appen-fig34}		
\end{figure*}

\begin{figure*}[htbp]
\begin{center}
\includegraphics[width=160mm,clip]{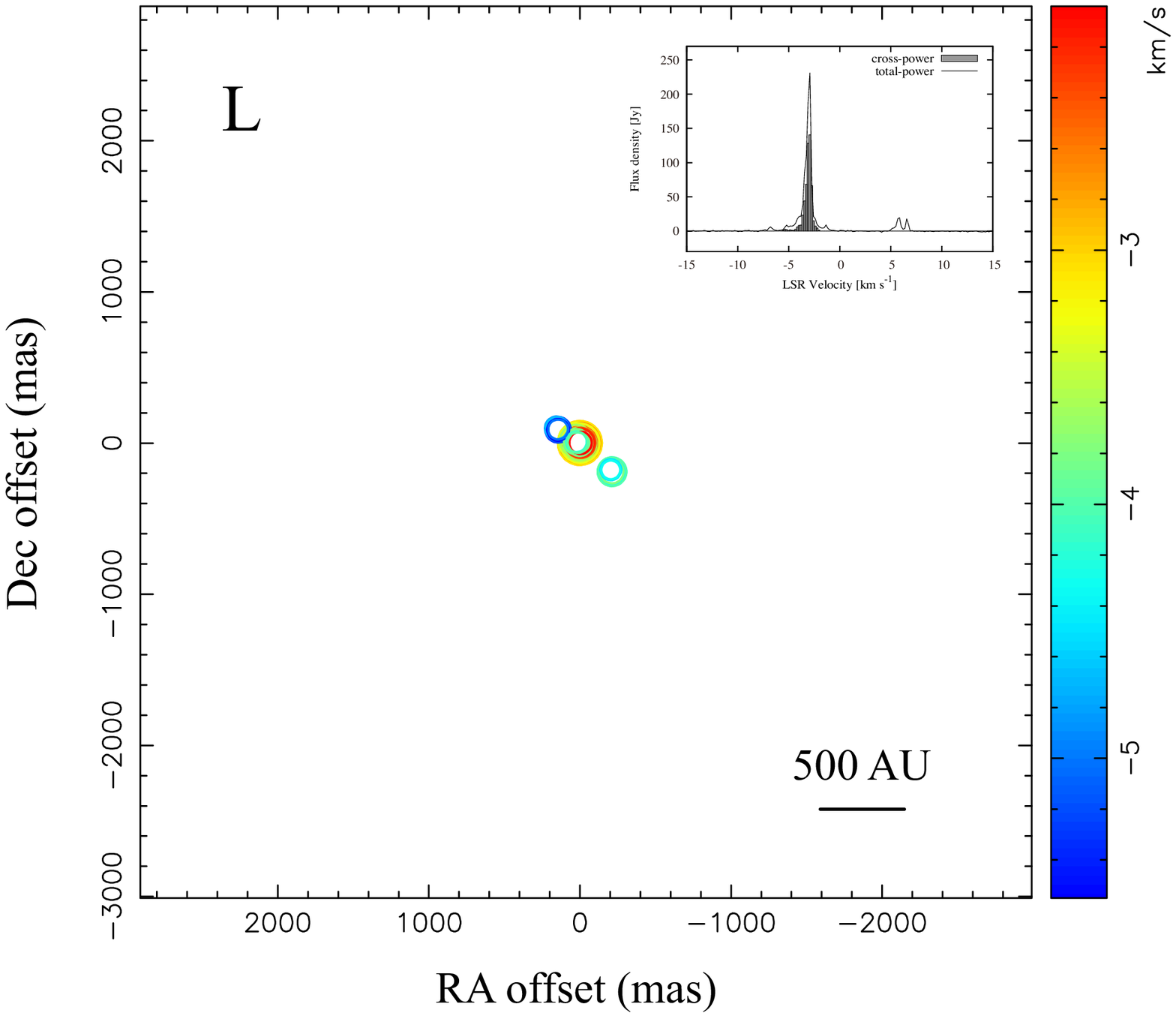}
\end{center}
\caption{
Data for source 352.63$-$01.06, plotted as for Figure~\ref{appen-fig1}.
}	
\label{appen-fig35}		
\end{figure*}

\begin{figure*}[htbp]
\begin{center}
\includegraphics[width=160mm,clip]{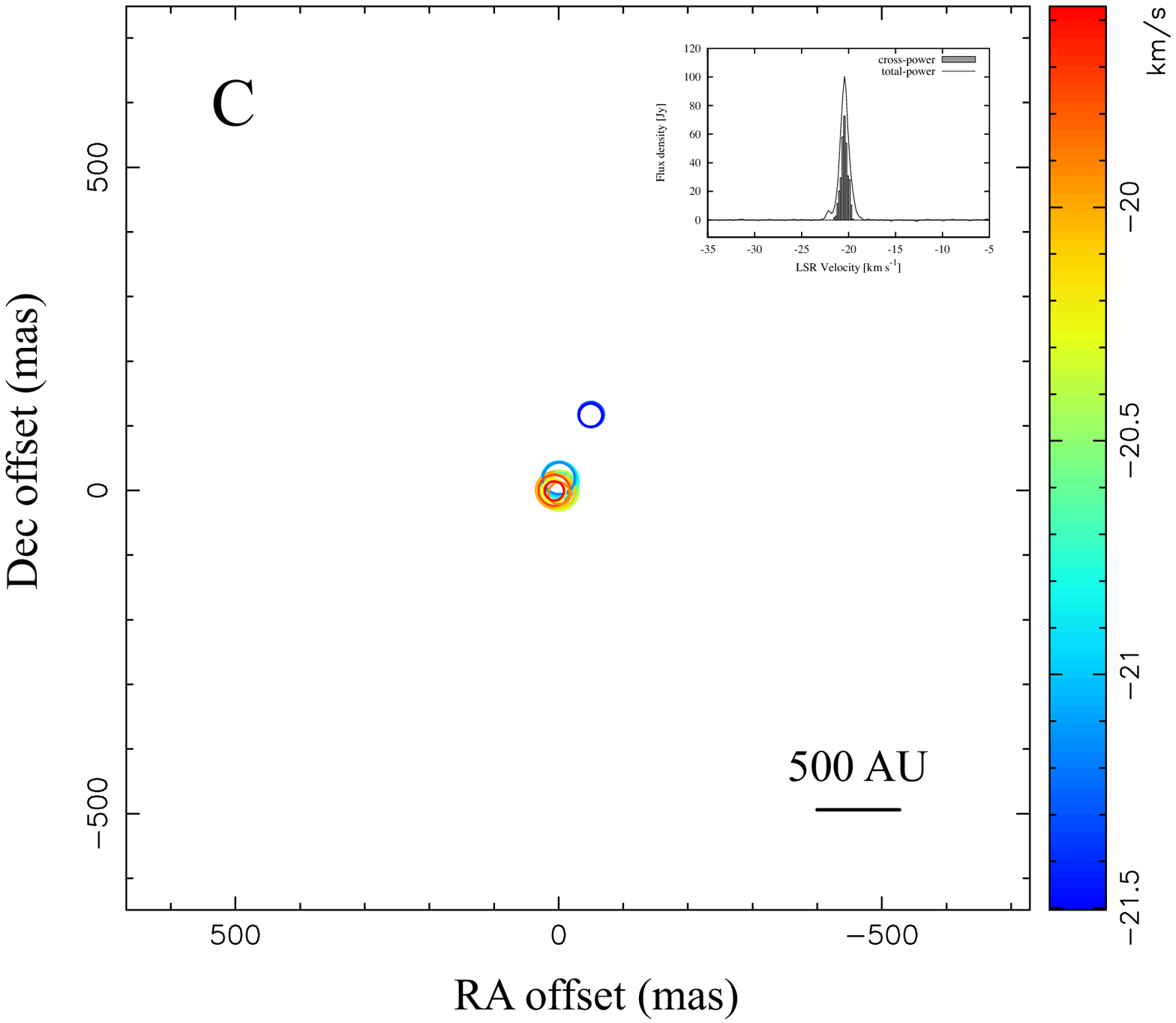}
\end{center}
\caption{
Data for source 353.41$-$00.36, plotted as for Figure~\ref{appen-fig1}.
}	
\label{appen-fig36}		
\end{figure*}

\begin{figure*}[htbp]
\begin{center}
\includegraphics[width=160mm,clip]{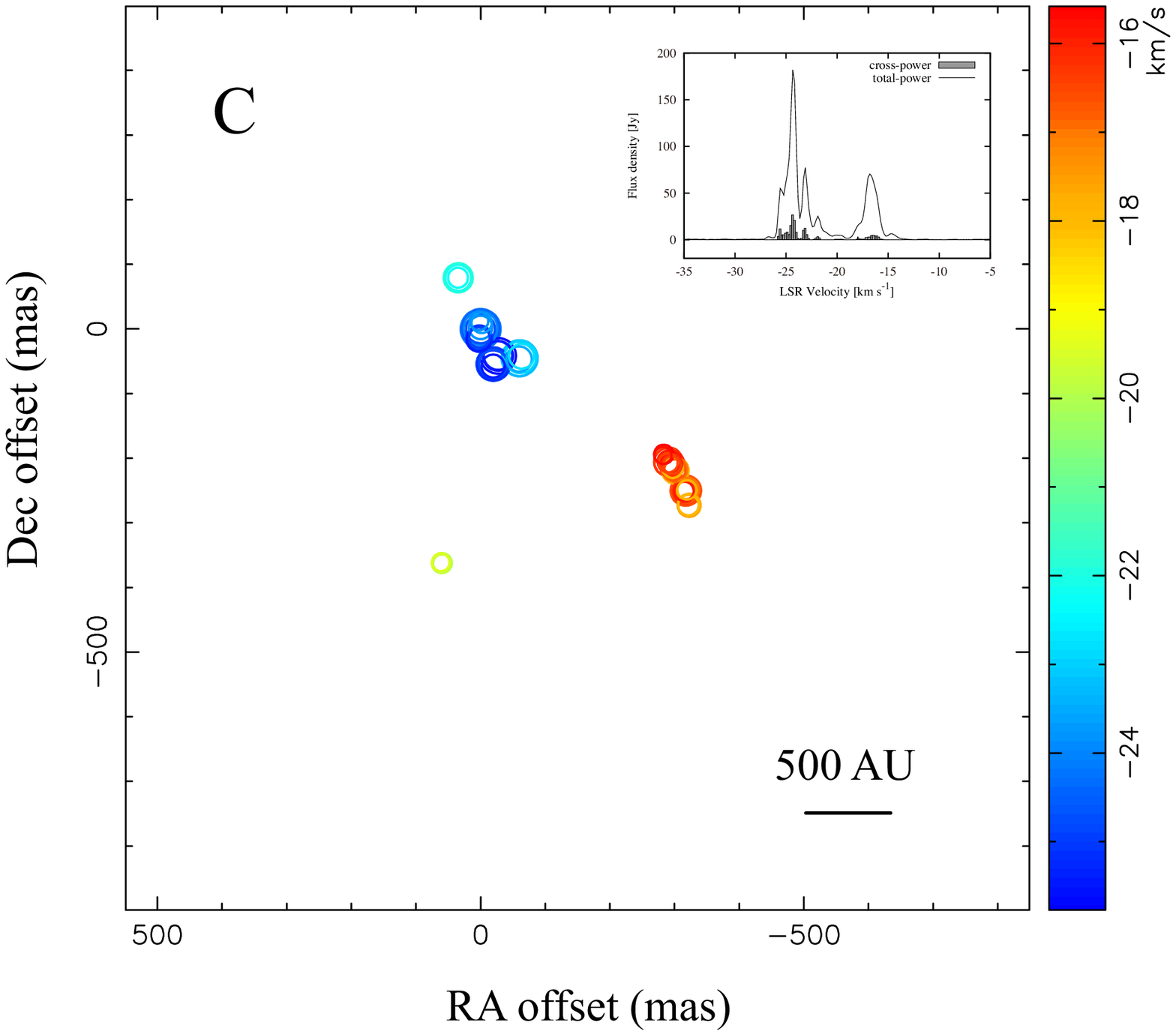}
\end{center}
\caption{
Data for source 354.61$+$00.47, plotted as for Figure~\ref{appen-fig1}.
}	
\label{appen-fig37}		
\end{figure*}

\begin{figure*}[htbp]
\begin{center}
\includegraphics[width=160mm,clip]{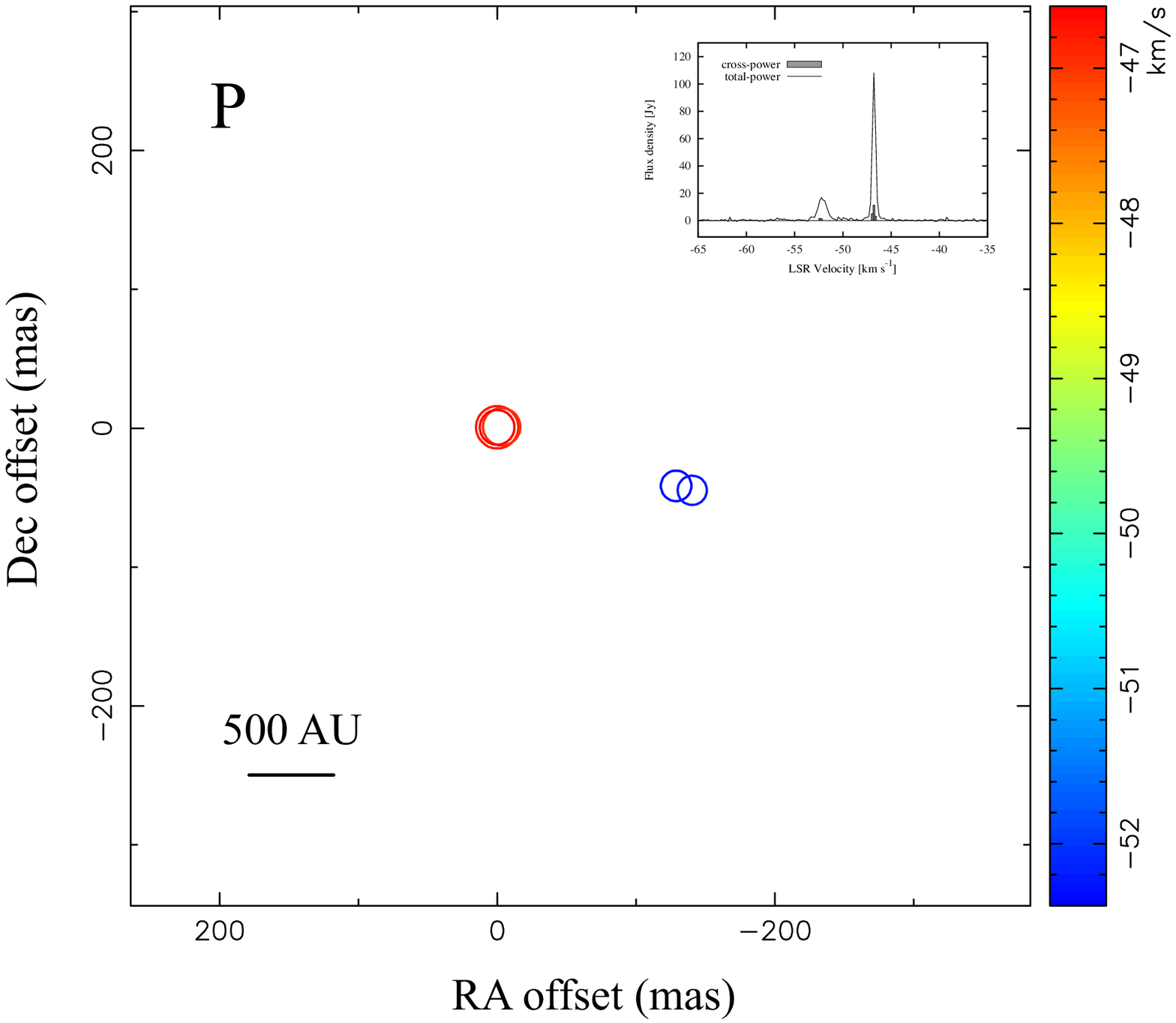}
\end{center}
\caption{
Data for source 359.43$-$00.10, plotted as for Figure~\ref{appen-fig1}.
}	
\label{appen-fig38}		
\end{figure*}


\end{document}